\documentclass[%
 reprint,
 amsmath,amssymb,
 aps,
]{revtex4-2}

\usepackage{graphicx}
\usepackage{dcolumn}
\usepackage{bm}
\usepackage{amsmath}
\usepackage{multirow}

\begin{document}

\preprint{APS/123-QED}

\title{Stability and wave dynamics in polytropic Eddington-inspired Born--Infeld gravitating solar plasmas}

\author{Souvik Das\,\href{https://orcid.org/0000-0002-3906-2773}{\includegraphics[scale=0.1]{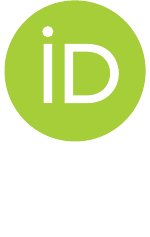}}}
\author{Pralay Kumar Karmakar\,\href{https://orcid.org/0000-0002-3078-9247}{\includegraphics[scale=0.1]{orcid.pdf}}}
\email{pkk@tezu.ernet.in}
\affiliation{
 Department of Physics, Tezpur University, Napaam, Tezpur, Assam 784028, India
}

\date{\today}

\begin{abstract}
We investigate the influence of nonlinear gravity corrections, arising from the Eddington-inspired Born--Infeld (EiBI) theory on wave dynamics, stability, and energy transport processes in polytropic, viscous, and turbulent solar plasmas. Analytical and numerical analyses of the Jeans-normalized quadratic dispersion relation demonstrate that both the EiBI gravity parameter $(\chi)$ and the relative polytropic sound speed $(\beta)$ independently regulate oscillation frequencies, growth rates, phase velocities, perturbation energy partitioning, and outward acoustic energy flux. Positive $\chi$ systematically elevates oscillation frequencies, phase velocities, and outward energy flux level by $\sim$10\% relative to the Newtonian predictions, while larger $\beta$ enhances them by up to 55\%, thereby promoting wave propagation and efficient acoustic transport. Conversely, negative $\chi$ strengthens gravitational binding and increases damping rates by $\sim$40\%, particularly for the \textit{g}-modes. Energy partitioning analyses reveal that the EiBI corrections fundamentally restructure the kinetic-electrostatic-gravitational energy balance. While the Newtonian gravity contributes negligibly ($<$4\%), nonzero $\chi$ channels up to one-third of oscillation energy into gravitational modes. The modal surface flux calculations further confirm that only the \textit{p}-modes drive outward energy transport (amplification for $\chi>0$, suppression for $\chi<0$). A direct comparative analysis with four years of SDO/HMI Doppler velocity observations demonstrate a robust theoretical agreement for $\chi=3\times10^7$ m$^5$\,kg$^{-1}$\,s$^{-2}$, providing the first empirical constraint on the solar EiBI gravity through helioseismology. These findings establish a rigorous framework for testing modified gravity theories and for advancing our understanding about solar plasma stability, helioseismic signatures, and ambient atmospheric energy transport processes.
\end{abstract}

\maketitle

\section{Introduction}
\label{sec:1}
The study of wave dynamics and stability in solar plasmas has long been a central theme in astrophysics, underpinning our understanding of stellar structure, evolution, and oscillations. The solar plasma oscillations, particularly the helioseismic \textit{p}-modes, provide a precise diagnostic tool to probe the internal structure and dynamics of the solar interior \cite{Ulrich1970,Deubner1984,Gough1996, Demarque1999,Christensen-Dalsgaard2002}. Stability analyses of these oscillations, based on perturbative approaches, shed light on the propagation of acoustic and gravity waves, energy transport mechanisms, and the onset of instabilities in the solar interior.\par
Gravitational effects play a fundamental role in maintaining the Sun’s equilibrium and regulating its oscillatory behaviour. Within the standard framework, gravity is described by Einstein’s General Relativity (GR), which has been highly successful in explaining large-scale cosmic phenomena as well as stellar structures. The Einstein--Hilbert action provides the foundation of GR and is expressed as \cite{Misner1973,Carroll2019}  
\begin{equation}
	S_{\mathrm{EH}} = \frac{1}{2 \kappa c} \int \left( R - 2 \Lambda \right) \sqrt{-g} \, d^4x + S_m(g,\text{fields}).
	\label{eq:1}
\end{equation}
Here, $\kappa =  8 \pi G/c^4$ is the Einstein gravitational constant, $R$ is the Ricci scalar, $\Lambda$ denotes the cosmological constant, $g = \det(g_{\mu \nu})$ is the determinant of the metric tensor, and $S_m$ represents the matter action. From this action, the Einstein field equations are derived, which reduce in the weak-field, non-relativistic limit to the classical Newtonian Poisson equation given as \cite{Misner1973,Hartle2003,Carroll2019}  
\begin{equation}
	\nabla^2 \psi = 4 \pi G \rho,
	\label{eq:2}
\end{equation}
where $\psi$ is the self-gravitational potential and $\rho$ denotes the matter density. This Newtonian formulation has been extensively applied in solar modelling and helioseismology, providing the baseline for our understanding of solar plasma oscillation dynamics.\par
Despite its remarkable success, GR encounters conceptual and technical challenges in regimes involving very high densities or nonlinear matter--gravity couplings. These limitations have inspired the development of modified theories of gravity. One such well-studied framework is the Eddington-inspired Born--Infeld (EiBI) theory of gravity \cite{Banados2010}, a Palatini-type extension of GR. Inspired by Born--Infeld electrodynamics, the EiBI theory replaces the Einstein--Hilbert action with a determinant-based structure that regularizes high-curvature behaviour and potentially avoids singularities. The EiBI action is expressed as \cite{Banados2010,Pani2011,Sham2013}

\begin{eqnarray}
S_{\mathrm{EiBI}} &=& \frac{c^3}{8 \pi G \chi} \int d^4x 
\left[ \sqrt{ -\left| g_{\mu \nu} + \chi R_{\mu \nu}(\Gamma^c_{ab}) \right| }
- \lambda \sqrt{-g} \, \right]\nonumber\\
&& + S_m(g,\text{fields}),
\label{eq:3}
\end{eqnarray}

where $R_{\mu \nu}$ is the Ricci tensor constructed from the independent connection $\Gamma^c_{ab}$. The term $\chi$ represents the EiBI gravity parameter. The parameters $\chi$ and $\lambda$ are related to the cosmological constant through $\Lambda = (\lambda - 1)/\chi$. In the limit $\chi \to 0$, the action naturally reduces to the Einstein--Hilbert action of GR \cite{Banados2010,Sham2013}. Unlike GR, the EiBI gravity introduces corrections that become significant in high-density environments while reducing to GR in vacuum or low-density regimes.\par
\begin{figure*}
	\centering
	\begin{tabular}{c c}
		\includegraphics[trim={8cm 0cm 5cm 0cm},clip,width=8cm]{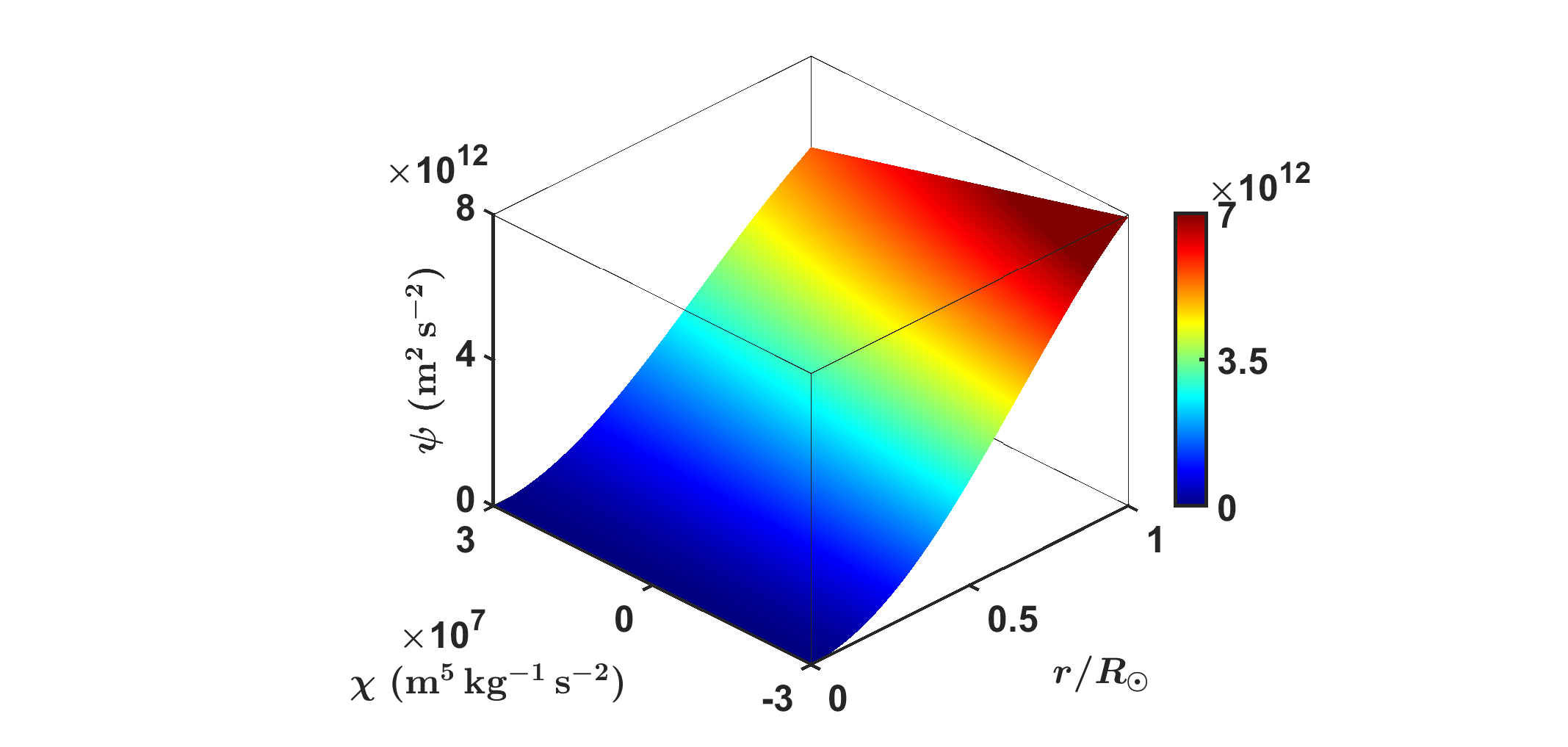}&
		\includegraphics[trim={8cm 0cm 5cm 0cm},clip,width=8cm]{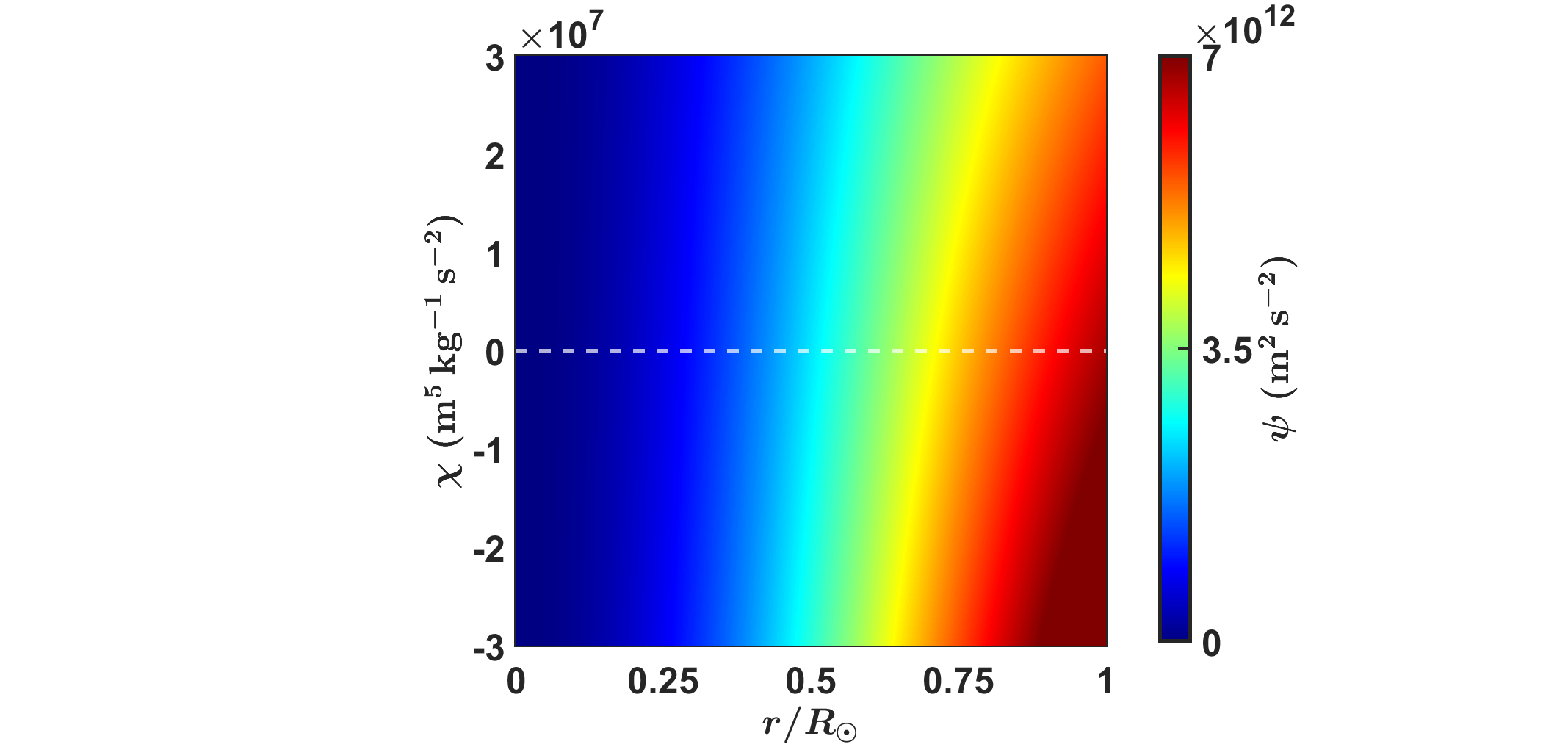}
	\end{tabular}
	\caption{Profile of the EiBI gravity--modified solar self-gravitational potential $(\psi)$ as a function of re-scaled distance $(r/R_\odot)$ and EiBI gravity parameter $(\chi)$.}
	\label{fig:1}
\end{figure*}

Existing studies typically analyse EiBI gravity in the asymptotically flat limit, implemented through the choice  $\lambda=1$, which effectively removes any bare cosmological-constant contribution (i.e., $\Lambda=0$). This formulation is standard in the conventional weak-field analyses employed in stellar and compact-object studies \cite{Pani2011,Pani2012a,Sham2013,Sham2014}. Under this assumption, all deviations from Newtonian gravity arise exclusively from the EiBI gravity parameter $\chi$, which introduces an additional density-gradient contribution to the gravitational potential. Consequently, in the Newtonian limit, the EiBI-modified gravitational Poisson equation takes the form \cite{Banados2010,Avelino2012a,Pani2012a}
\begin{equation}
	\nabla^2 \psi = 4 \pi G \rho + \frac{\chi}{4} \nabla^2 \rho,
	\label{eq:4}
\end{equation}
where the $\chi$-dependent term reflects the nonlinear Born--Infeld--type modification and becomes significant in dense or strongly stratified astrophysical plasmas. Such corrections may influence the equilibrium configuration, oscillatory behaviour, and stability properties of stellar interiors, including the solar plasmas, where even small departures from Newtonian gravity can affect the wave propagation and mode dynamics. Figure~\ref{fig:1} illustrates the dependence of the EiBI gravity--modified solar self-gravitational potential ($\psi$) on the re-scaled distance from the core ($r/R_\odot$) and the EiBI gravity parameter ($\chi$). As $\nabla^2 \rho < 0$ throughout most of the solar interior, a positive $\chi$ weakens the gravitational source and reduces $\psi$, whereas a negative $\chi$ strengthens it and deepens $\psi$.\par

A substantial body of literature has extensively explored the implications of EiBI gravity in compact astrophysical objects \cite{Pani2011,Sham2013,Sham2014,Afonso2019,Afonso2020,Prasetyo2021}, as well as in early-universe cosmology \cite{Rodrigues2008,Delsate2012,Pani2012b} and astrophysical stability analyses \cite{Pani2012a,Kim2014,DeMartino2017,Bessiri2021,Yang2023}. By comparison, relatively little attention has been devoted to the solar context. Astrophysical and cosmological constraints on the EiBI parameter have also been reported \cite{Avelino2012a,Avelino2012b,Casanellas2012,Banerjee2022}. For instance, Casanellas et al.~\cite{Casanellas2012} have examined solar models within the EiBI gravity, demonstrating that both positive and negative $\chi$ significantly modify the solar structure. Previous studies have placed the solar-scale EiBI gravity parameter in the range $\chi \sim 10^5 - 10^7$ m$^5$\,kg$^{-1}$\,s$^{-2}$ \cite{Avelino2012a,Casanellas2012}. These works highlight the astrophysical relevance of EiBI modifications, although most investigations remain focused on compact objects or cosmological applications.\par
Recently, Das and Karmakar~\cite{Das2024} have studied the effect of the EiBI gravity parameter on equilibrium solar plasma properties. However, the influence of EiBI gravity on solar plasma oscillations and wave dynamics remains largely unexplored. Several key issues, such as oscillation frequencies, damping or growth rates, phase velocities, perturbation energy balance, and acoustic energy flux in a self-gravitating solar plasma have not been systematically addressed. This gap motivates the present work.\par
In this study, we incorporate the EiBI gravity into a polytropic, viscous, and turbulent solar plasma framework to systematically analyse oscillation dynamics and stability. Particular attention is given to the EiBI gravity parameter $\chi$, which quantifies departures from Newtonian gravity and may leave observable imprints on helioseismic features. By examining frequency shifts, stability criteria, oscillation energetics, and energy transport, we aim to assess whether nonlinear matter--gravity coupling within EiBI theory can be constrained or probed through solar plasma oscillation dynamics.\par
The remainder of this article is organized as follows. Section~\ref{sec:2} presents the physical model configuration and the underlying assumptions adopted in this study. In Section~\ref{sec:3}, the detailed mathematical formalism and derivations are developed. Section~\ref{sec:4} is devoted to the presentation and discussion of the main results, including a comparative assessment and validation with available astronomical solar observations reported in the literature. Finally, Section~\ref{sec:5} summarizes the major conclusions of our systematic investigation and highlights possible directions for future research on solar surface oscillations under EiBI gravity.

\section{Physical model development}
\label{sec:2}
In the present study, we formulate a quasi-neutral, viscous, turbulent, unmagnetized, and polytropic solar plasma system embedded within the framework of EiBI gravity. For analytical tractability, the system is assumed to be spherically symmetric. The plasma is primarily composed of classical Maxwellian electrons (non-inertial) and isothermal (inertial) ions, consistent with the standard descriptions of the solar interior \cite{Stix2002, Aschwanden2014, Priest2014, Ambastha2020}. Within the solar interior, the thermal plasma pressure $p_t=n k_{\mathrm{B}} T$ dominates over the magnetic pressure $p_m=B^2/(2\mu_0)$, leading to a high plasma beta $\beta_p=p_t/p_m\gg 1$ \cite{Stix2002, Priest2014}. Such a regime implies that magnetic fields do not play a dynamically significant role in the large-scale collective oscillations of the interior plasma \cite{Roberts1986, Das2023}. Moreover, previous works indicate that high-frequency acoustic \textit{p}-mode oscillations are largely insensitive to magnetic effects \cite{Roberts1986, Cally1995}, thereby justifying the neglect of magnetization in our model.\par
To account for dissipative and fluctuating effects, the model incorporates viscous and turbulent contributions. Viscosity introduces resistive damping of plasma fluid motion, while turbulence drives irregular outward fluctuations in pressure and velocity fields \cite{Adams1994, Chandra2013}. The thermodynamic state of the plasma is characterized by a polytropic equation of state (EoS) that relates the effective fluid pressure to the local mass density as \cite{Hendry1993, Hansen2004, Horedt2004}
\begin{equation}
	p_p = K_p \rho^{\Gamma} = K_p \rho^{1 + n_p^{-1}},
	\label{eq:5}
\end{equation}
where $K_p$ is the polytropic constant, $\Gamma = 1 + n_p^{-1}$ denotes the polytropic exponent, and $n_p$ is the polytropic index. For the solar interior, the typical choice $n_p = 3$ yields $\Gamma = 4/3$ \cite{Hendry1993, Hansen2004}.\par
The presence of density gradients in the solar plasmas naturally gives rise to turbulent pressure \cite{Stix2002, Ambastha2020}, particularly in regions of vigorous convective motions. In the near-surface convection zone and overshoot region, turbulence contributes an additional isotropic pressure component beyond the thermal contribution. To model this effect, we adopt the Larson logabarotropic equation of state \cite{Larson1981, Lizano1989, Adams1994}, expressed as
\begin{equation}
	p_{\rm turb} = p_0 \ln \frac{\rho}{\rho_0},
	\label{eq:6}
\end{equation}
which provides a convenient description of turbulent support in stratified, compressible astrophysical fluid. Here, $p_0 = n_0 k_{\mathrm{B}} T_0$ is the mean (equilibrium) isothermal fluid pressure, $\rho$ refers to the inhomogeneous fluid density, and $\rho_0$ denotes the mean solar interior fluid density. Accordingly, the net effective pressure of the polytropic inhomogeneous turbulent ionic fluid is expressed as $p = p_p + p_{\rm turb}$. For mathematical simplicity, we neglect additional effects arising from the heliospheric magnetic field, gyro-fluid interactions, non-ideal plasma corrections, differential rotation, large-scale vorticity, viscoelasticity, and tidal influences. All physical constants used in this work, such as the universal gravitational constant $G$, vacuum speed of light $c$, and Boltzmann constant $k_{\mathrm{B}}$, are retained in symbolic form and assume their standard values as adopted in the astrophysical literature.\par
The model is further embedded in the EiBI gravitational framework, which modifies the standard Newtonian Poisson equation by incorporating nonlinear Born--Infeld-type corrections to spacetime curvature. This departure from the Einstein--Hilbert action of general relativity provides a richer theoretical setting for analysing solar plasma oscillations and stability under nonstandard gravitational responses.\par

\section{Mathematical formalism}
\label{sec:3}
In this work, we systematically investigate the dynamics of self-gravitating solar plasma fluctuations under the framework of EiBI gravity, with the objective of identifying possible deviations from the classical Newtonian description. The governing mathematical framework is developed rigorously and presented below in a structured manner.

\subsection{Governing equations}
The self-gravitating solar interior plasma is modelled as a two-component system consisting of non-inertial thermal electrons (assumed Maxwellian) and an inertial ion fluid. The electron density, following the Maxwell--Boltzmann distribution, can be expressed as \cite{Chen2016}
\begin{equation}
	n_e = n_0 \exp\left(\frac{e\phi}{k_{\mathrm{B}}T_e}\right),
	\label{eq:7}
\end{equation}
where $n_0 = \rho_0 / m_i$ denotes the mean (equilibrium) particle number density in the solar plasma interior.\par
The heavier ions are treated as a viscous turbulent fluid in order to capture their full inertial response. Their collective dynamics are governed by the standard set of hydrodynamic equations: the continuity equation for mass conservation and the momentum equation for force balance. These equations can be expressed using conventional notations respectively as \cite{Landau1987, Choudhuri1998, Clarke2007, Chen2016}
\begin{equation}
	\frac{\partial n_i}{\partial t} + \nabla \cdot (n_i \mathbf{v}_i) = 0,
	\label{eq:8}
\end{equation}
\begin{eqnarray}
	m_i n_i \frac{D \mathbf{v}_i}{D t} &=& q_i n_i \mathbf{E} - \nabla p 
	+ m_i n_i \mathbf{g} + \eta \nabla^2 \mathbf{v}_i\nonumber\\
	&&+ \left( \zeta + \frac{1}{3}\eta \right) \nabla (\nabla \cdot \mathbf{v}_i).
	\label{eq:9}
\end{eqnarray}
Here, $n_i$ and $T_i$ denote the ion number density and temperature, respectively. The material (convective) derivative is defined as $D/Dt \equiv \partial/\partial t + (\mathbf{v}_i \cdot \nabla)$. Equation~\eqref{eq:9} represents the Navier--Stokes equation including both shear ($\eta$) and bulk ($\zeta$) viscosity. According to the Stokes's hypothesis, the bulk viscosity of a Newtonian fluid can be neglected ($\zeta=0$) \cite{Stokes1845, Buresti2015}. Furthermore, using the vector identity $\nabla (\nabla \cdot \mathbf{v}_i ) = \nabla^2 \mathbf{v}_i + \nabla \times(\nabla \times \mathbf{v}_i)$, the viscous term simplifies to $(4/3)\eta \nabla^2 \mathbf{v}_i$ for irrotational flow, where $\nabla \times \mathbf{v}_i = \mathbf{0}$.\par
The dynamic (shear) viscosity coefficient $\eta$ is given as (in kg\,m$^{-1}$\,s$^{-1}$ or Pa\,s)
\begin{equation}
	\eta = 0.96 \, n_i k_{\mathrm{B}} T_i t_r,
	\label{eq:10}
\end{equation}
where $t_r$ is the Coulombic relaxation time for solar ions, expressed as \cite{Braginskii1965, Hollweg1985, Chandra2013}
\begin{equation}
	t_r = \frac{12 \pi^{3/2} \varepsilon_0^2 m_i^{1/2} \left(k_{\mathrm{B}} T_i\right)^{3/2}}
	{e^4 n_i \ln \Lambda}.
	\label{eq:11}
\end{equation}
Here, $\varepsilon_0 = 8.85 \times 10^{-12}$ C$^2$\,kg$^{-1}$\,m$^{-3}$\,s$^2$ is the vacuum permittivity and $\ln \Lambda$ denotes the Coulomb logarithm.\par
The effect of EiBI gravity is incorporated through the modified Poisson equation for self-gravity given with all the usual notations as \cite{Banados2010, Pani2011, Avelino2012a}
\begin{equation}
	\nabla^2 \psi = 4 \pi G \rho + \frac{\chi}{4} \nabla^2 \rho.
	\label{eq:12}
\end{equation}
For a self-gravitationally bounded astrophysical object of radius $R$, the consistency condition requires \cite{Avelino2012a}
\[
\chi < \frac{4}{\pi} G R^2.
\]
Hence, for the solar interior, substituting $R = R_\odot$ yields an upper bound
\[
\chi < \frac{4}{\pi} G R_\odot^2 \sim 4 \times 10^7 \;\; \mathrm{m^5\, kg^{-1}\,s^{-2}}.
\]
Finally, the electrostatic Poisson equation is cast as \cite{Chen2016}
\begin{equation}
	\nabla^2 \phi = - \frac{\rho_q}{\varepsilon_0},
	\label{eq:13}
\end{equation}
where $\phi$ is the electrostatic potential and $\rho_q$ denotes the charge density of the plasma medium. Together, Equations~\eqref{eq:7}--\eqref{eq:13} constitute the governing framework for describing the gravito-electrostatic dynamics of self-gravitating solar plasmas within the EiBI gravity theory.\par

\begin{table*}
	\caption{Adopted normalization scheme.}
	\label{tab:1}
	\begin{ruledtabular}
		\begin{tabular}{lcc}
			Normalized parameter & Normalizing parameter & Magnitude \\
			\hline
			Radial distance ($\xi$) & Jeans length $\left(\lambda_{\mathrm{J}}=\tau_{\mathrm{J}} C_s\right)$ & $3.10 \times 10^{8}~\mathrm{m}$ \\
			Time ($\tau$) & Jeans time $\left(\tau_{\mathrm{J}}=\sqrt{\pi/(G \rho_0)}\right)$ & $10^{3}~\mathrm{s}$ \\
			Population density $\left(N_{e(i)}\right)$ & Mean population density $\left(n_0\sim \lambda_{\mathrm{D}}^{-3}\right)$ & $10^{30}~\mathrm{m^{-3}}$ \\
			Mach number ($M$) & Ion acoustic phase speed $\left(C_s=\sqrt{k_{\mathrm{B}}T_e/m_i}\right)$ & $3.10 \times 10^{5}~\mathrm{m\,s^{-1}}$ \\
			Gravitational potential ($\Psi$) & Acoustic kinetic potential $\left(\psi_{\mathrm{th}}=C_s^2\right)$ & $9.61 \times 10^{10}~\mathrm{m^{2}\,s^{-2}}$ \\
			Electrostatic potential ($\Phi$) & Electron thermal potential $\left(\phi_{\mathrm{th}}=k_{\mathrm{B}} T_e / e\right)$ & $10^{2}~\mathrm{J\,C^{-1}}$ \\
		\end{tabular}
	\end{ruledtabular}
\end{table*}

All quantities are normalized using a standard Jeans-type normalization scheme, as detailed in Table~\ref{tab:1}. Accordingly, the governing hydrodynamic and field equations can be recast into their one-dimensional, Jeans-normalized radial forms using standard notations as
\begin{equation}
	N_e = e^{\Phi},
	\label{eq:14}
\end{equation}
\begin{equation}
	\frac{\partial N_i}{\partial \tau} 
	+ N_i \frac{\partial M}{\partial \xi}
	+ M \frac{\partial N_i}{\partial \xi}
	+ \frac{2}{\xi} N_i M = 0,
	\label{eq:15}
\end{equation}
\begin{eqnarray}
	N_i \left( \frac{\partial M}{\partial \tau} + M \frac{\partial M}{\partial \xi} \right) 
	+ \epsilon_{T0} \frac{\partial P}{\partial \xi} 
	+ N_i \frac{\partial \Phi}{\partial \xi} 
	+ N_i g_s & = \nonumber\\
	\frac{4}{3}\eta^* \left[ 
	\frac{\partial^2 M}{\partial \xi^2} 
	+ \frac{2}{\xi}\frac{\partial M}{\partial \xi} 
	- \frac{2}{\xi^2}M \right],
	\label{eq:16}
\end{eqnarray}
\begin{equation}
	\frac{\partial g_s}{\partial \xi} + \frac{2}{\xi} g_s 
	= N_i + \alpha \chi \left( \frac{\partial^2 N_i}{\partial \xi^2} 
	+ \frac{2}{\xi} \frac{\partial N_i}{\partial \xi} \right),
	\label{eq:17}
\end{equation}
\begin{equation}
	\left( \frac{\lambda_{\mathrm{D}}}{\lambda_{\mathrm{J}}} \right)^2 
	\left( \frac{\partial^2 \Phi}{\partial \xi^2} 
	+ \frac{2}{\xi} \frac{\partial \Phi}{\partial \xi} \right) 
	= N_e - N_i.
	\label{eq:18}
\end{equation}

Equation~\eqref{eq:16} represents the normalized Navier--Stokes equation that includes only the dynamic (shear) viscosity contribution. The normalized viscosity is defined as $\eta^* = \eta/\eta_{\mathrm{J}}$, where $\eta_{\mathrm{J}} = m_i n_0 \omega_{\mathrm{J}} \lambda_{\mathrm{J}}^2$ is the Jeans (critical) viscosity. Here, the corresponding Jeans length and Jeans angular frequency are given, respectively, by
\[
\lambda_{\mathrm{J}} = \frac{2\pi}{k_{\mathrm{J}}}
= \sqrt{\frac{\pi}{G \rho_0}}\, C_s,
\qquad 
\omega_{\mathrm{J}} = C_s k_{\mathrm{J}} = \sqrt{4\pi G \rho_0},
\]
with $k_{\mathrm{J}} = \sqrt{4\pi G \rho_0}/C_s$, the Jeans angular wavenumber.\par
The total normalized pressure is expressed as
\[P = P_p + P_{\rm turb} 
= \frac{\beta}{\Gamma} N_i^\Gamma + \ln N_i,
\]
where $P_p = (\beta/\Gamma) N_i^\Gamma$ is the normalized polytropic contribution and 
$P_{\rm turb} = \ln N_i$ represents the normalized turbulence pressure. The parameter $\beta = c_{p0}^2/c_{t0}^2$ quantifies the relative strength of the equilibrium polytropic and isothermal sound speeds. The electron temperature enters through the normalization $\epsilon_{T0}^{-1} = T_e/T_0$. The normalized electrostatic potential is denoted by $\Phi$, while the Jeans-normalized self-gravitational acceleration is defined as $g_s = \partial_\xi \Psi$, with $\Psi$ representing the Jeans-normalized EiBI gravity--modified self-gravitational potential. In Equation~\eqref{eq:17}, the constant parameter 
\[
\alpha = \frac{1}{16 \pi G \lambda_{\mathrm{J}}^2} \approx 3.1 \times 10^{-9} \;\; \mathrm{m^{-5}\,kg \,s^2}
\]
appears so that the product $\alpha \chi$ becomes dimensionless, thereby ensuring that the EiBI gravity parameter $\chi$ is incorporated consistently within the Jeans-normalized EiBI-modified gravitational Poisson equation.\par
The additional curvature terms proportional to $2/\xi$ arise naturally in the governing equations owing to the adoption of a spherical geometry configuration ($\xi \nrightarrow \infty$) and vanish only in the planar limit ($\xi \rightarrow \infty$). The ratio $\lambda_{\mathrm{D}}/\lambda_{\mathrm{J}}$ appearing in Equation~\eqref{eq:18} involves the equilibrium electron Debye length, defined as
\[
\lambda_{\mathrm{D}} = \frac{C_s}{\omega_p}
= \sqrt{\frac{\varepsilon_0 k_{\mathrm{B}} T_e}{n_0 e^2}},
\]
where $\omega_p = \sqrt{n_0 e^2/(\varepsilon_0 m_e)}$ denotes the equilibrium electron plasma oscillation frequency. Finally, under the physically relevant condition $\lambda_{\mathrm{D}}/\lambda_{\mathrm{J}} \approx 10^{-12} \sim 0$, the electrostatic Poisson equation reduces to the global quasi-neutrality condition $N_e \approx N_i = N$, which significantly simplifies the system and facilitates analytical treatment.

\subsection{Perturbative treatment}
To investigate the dynamical response of the present quasi-neutral, viscous, turbulent, and polytropic EiBI gravitating solar plasmas, we perform a linear stability analysis of the governing system (Equations~\eqref{eq:14}--\eqref{eq:17}). Small-amplitude perturbations ($\delta F$) are introduced systematically around the homogeneous equilibrium state $(F_0)$ defined by
\[
N_0 = 1, \quad M_0 = 0, \quad \Phi_0 = 0, \quad {g_s}_0 = 0.
\]
Accordingly, each physical variable is decomposed into equilibrium and perturbed parts as
\[
F(\xi,\tau) = F_0 + \delta F(\xi,\tau).
\]
We now adopt a spherical (radial) outgoing wave ansatz for the perturbed fields:
\[
\delta F(\xi,\tau) \equiv F_1(\xi,\tau) 
= \frac{F_{10}}{\xi} \exp\!\left[-i\left(\Omega \tau - K \xi \right)\right],
\]
which incorporates the spherical geometric spreading factor $(1/\xi)$ and a plane-wave-like radial phase $K\xi$. Here, $F_{10} \ll 1$ is the perturbation amplitude, while $\Omega = \omega/\omega_{\mathrm{J}}$ and $K = k/k_{\mathrm{J}}$ denote the Jeans-normalized angular frequency and angular wavenumber, respectively. Such perturbations correspond to the Eulerian perturbation at a fixed spatiotemporal point \cite{Christensen-Dalsgaard2002}.\par
The spherical ansatz transforms derivatives from real space $(\xi,\tau)$ into Fourier space $(K,\Omega)$ via
\[
\frac{\partial}{\partial \tau} \;\rightarrow\; -i\Omega, 
\quad 
\frac{\partial}{\partial \xi} \;\rightarrow\; iK - \frac{1}{\xi}, 
\quad 
\frac{\partial^2}{\partial \xi^2} \;\rightarrow\; \frac{2}{\xi^2} - K^2 - \frac{2iK}{\xi}.
\]
Substituting these perturbations into Equations~\eqref{eq:14}--\eqref{eq:17} and retaining only the linear terms, we obtain the following set of coupled relations:
\begin{equation}
	N_1 = \Phi_1, 
	\label{eq:19}
\end{equation}
\begin{equation}
	M_1 = \frac{i\Omega}{\,iK + 1/\xi\,} \, N_1,
	\label{eq:20}
\end{equation}
\begin{eqnarray}
	\left[i\Omega - \tfrac{4}{3}\eta^* \left(K^2 + \tfrac{2}{\xi^2}\right)\right] M_1 &=& \nonumber\\
	\big(1 + \epsilon_{T0} + \beta \epsilon_{T0}\big)\left(iK - \tfrac{1}{\xi}\right)\Phi_1 &+& {g_s}_1,
	\label{eq:21}
\end{eqnarray}
\begin{equation}
	{g_s}_1 = \frac{N_1 \left(1 - \alpha \chi K^2 \right)}{\,iK + 1/\xi\,}.
	\label{eq:22}
\end{equation}
By systematically eliminating the perturbation variables i.e., $\Phi_1$, $N_1$, $M_1$, and ${g_s}_1$, the generalized quadratic dispersion relation is obtained as
\begin{equation}
	A_2 \Omega^2 + A_1 \Omega + A_0 = 0,
	\label{eq:23}
\end{equation}
with coefficients
\begin{eqnarray}
	A_2 &=& 1, 
	\label{eq:24} \\[4pt]
	A_1 &=& i \tfrac{4}{3}\eta^* \left(K^2 + \tfrac{2}{\xi^2}\right),
	\label{eq:325} \\[4pt]
	A_0 &=& 1 - \alpha \chi K^2 - \big(1 + \epsilon_{T0} + \beta \epsilon_{T0}\big)\left(K^2 + \tfrac{1}{\xi^2}\right).
	\label{eq:26}
\end{eqnarray}
Equation~\eqref{eq:23} governs the stability dynamics of polytropic EiBI--gravitating solar plasmas. The spherical geometry contributes explicitly through the $1/\xi^2$ curvature terms. The effective EiBI correction enters via
$\mathcal{G}(\alpha,\chi,K) = - \alpha \chi K^2$,
which encapsulates the modified gravitational coupling. The viscous dissipation ($\eta^*$, via $A_1$) and the relative polytropic sound speed ($\beta$, via $A_0$) further influence the modal wave behaviour and stability features of the system. The full parameter dependence of dispersion and growth/damping characteristics can be explored through numerical analysis of Equation~\eqref{eq:23}.

\section{Results and discussions}
\label{sec:4}
The dispersion relation derived in Equation~\eqref{eq:23} forms the fundamental basis for investigating the stability characteristics of the quasi-neutral, viscous, turbulent, and polytropic solar plasmas under the EiBI gravity framework. To elucidate the dynamic behaviour and the influence of key physical parameters on the system, we numerically solve and analyse this relation across a comprehensive range of relevant parameter values. The Jeans-normalized complex angular frequency is expressed as
$\Omega = \Omega_r + i \Omega_i$, with $\Omega_r$ denoting the oscillatory component of the perturbation and $\Omega_i$ determining temporal stability. Specifically, $\Omega_i > 0$ indicates exponential growth (instability), while $\Omega_i < 0$ corresponds to exponential decay (damping).\par
\begin{figure*}
	\centering
	\begin{tabular}{c c}
		\includegraphics[trim={5cm 0cm 4cm 0cm},clip,width=8cm]{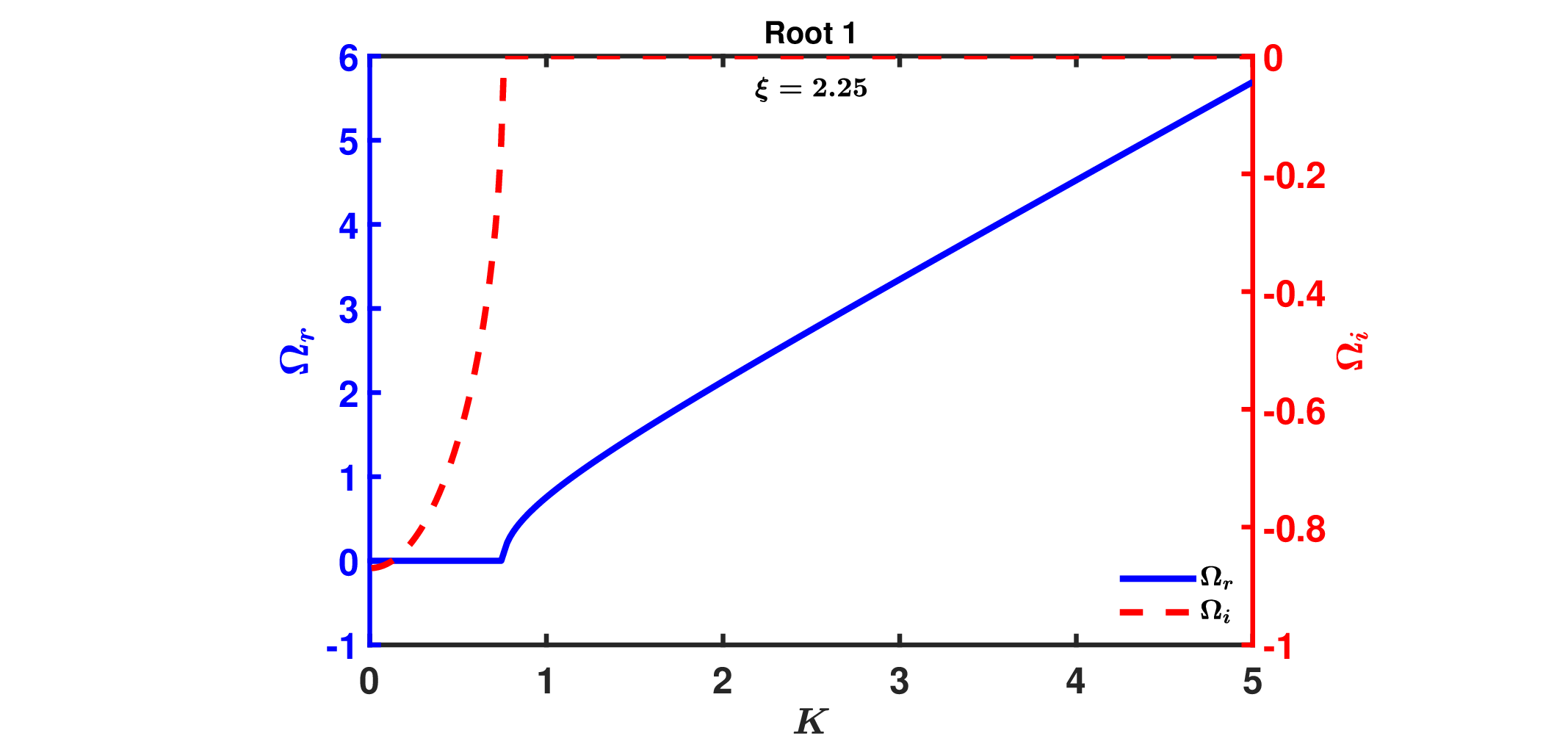}&
		\includegraphics[trim={5cm 0cm 4cm 0cm},clip,width=8cm]{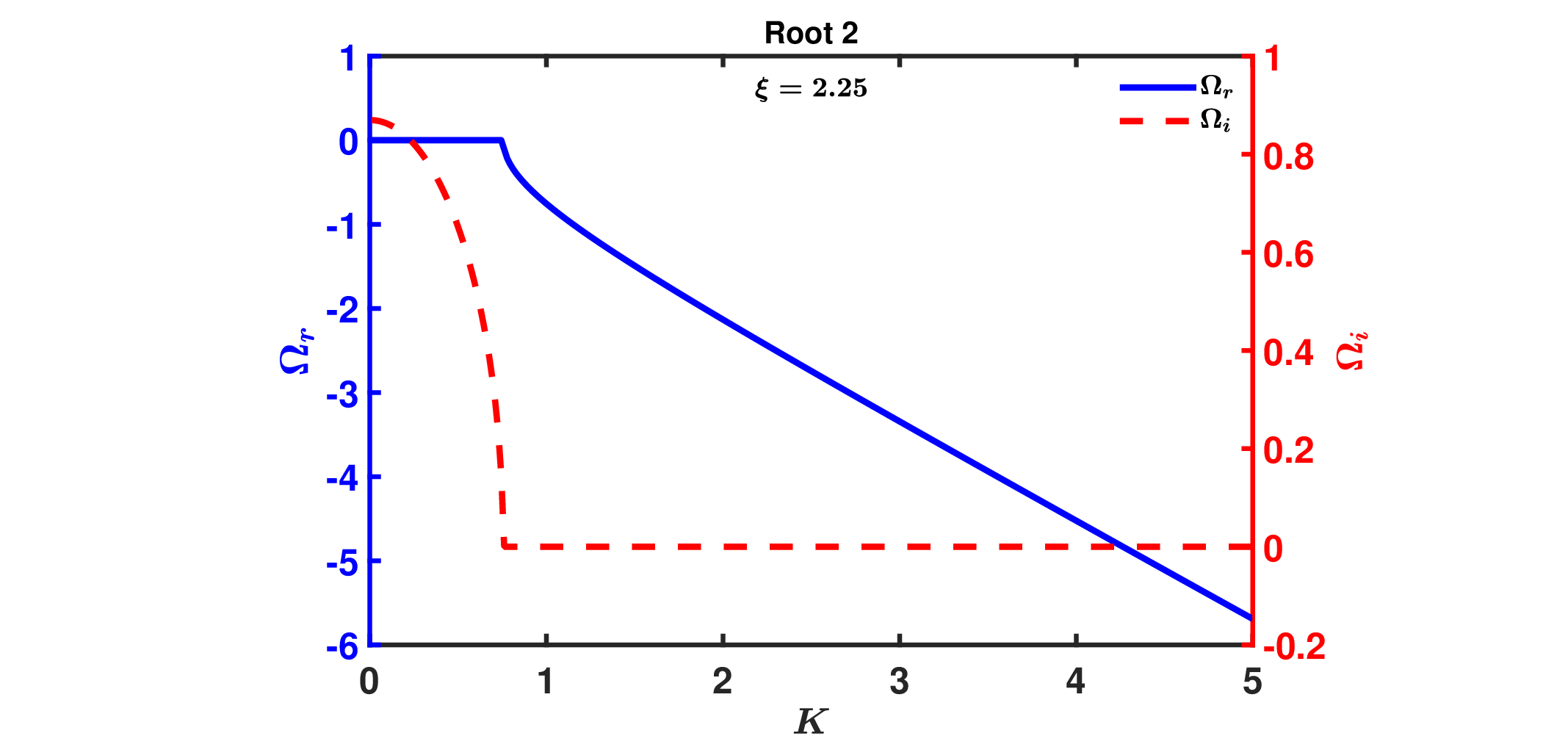}\\
		\includegraphics[trim={5cm 0cm 4cm 0cm},clip,width=8cm]{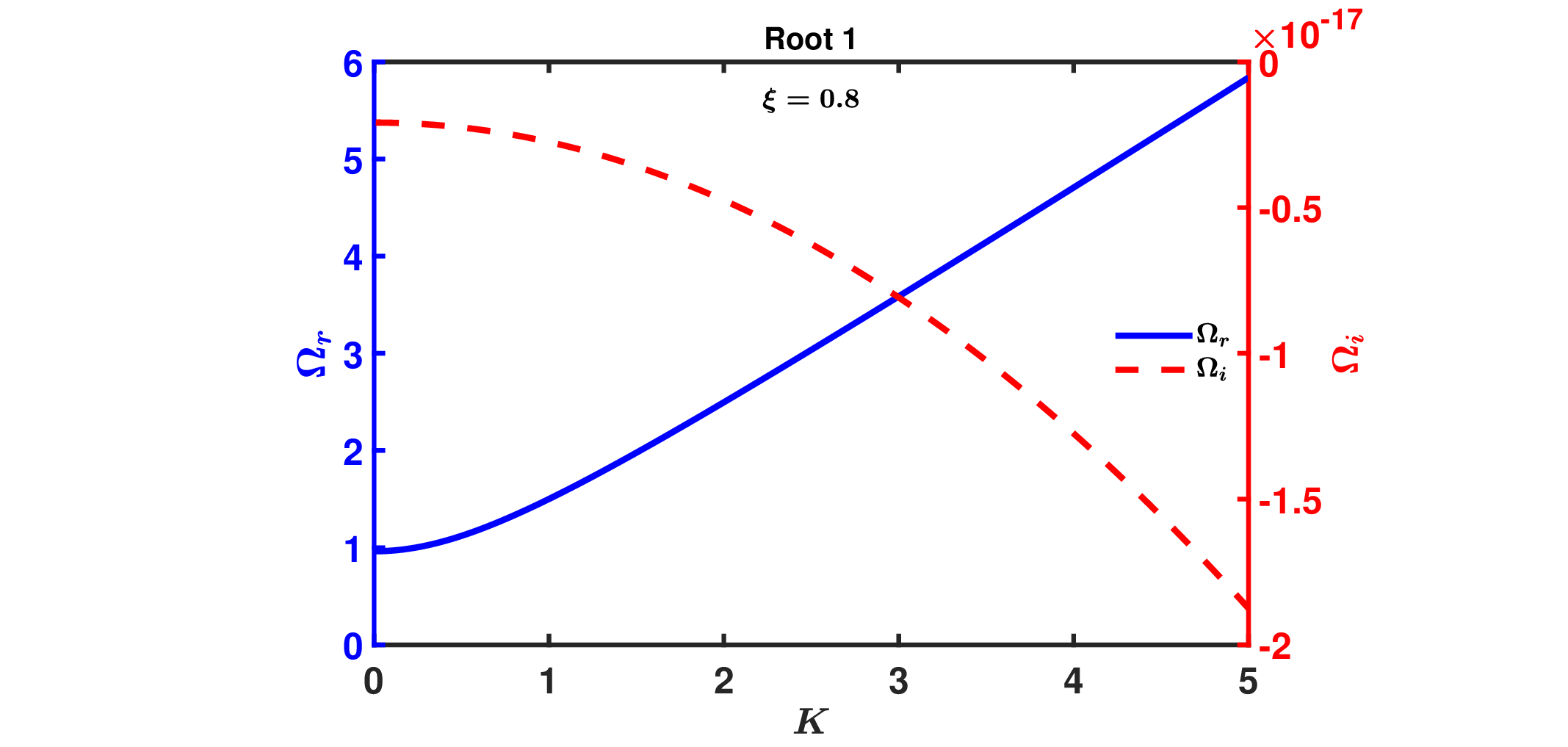}&
		\includegraphics[trim={5cm 0cm 4cm 0cm},clip,width=8cm]{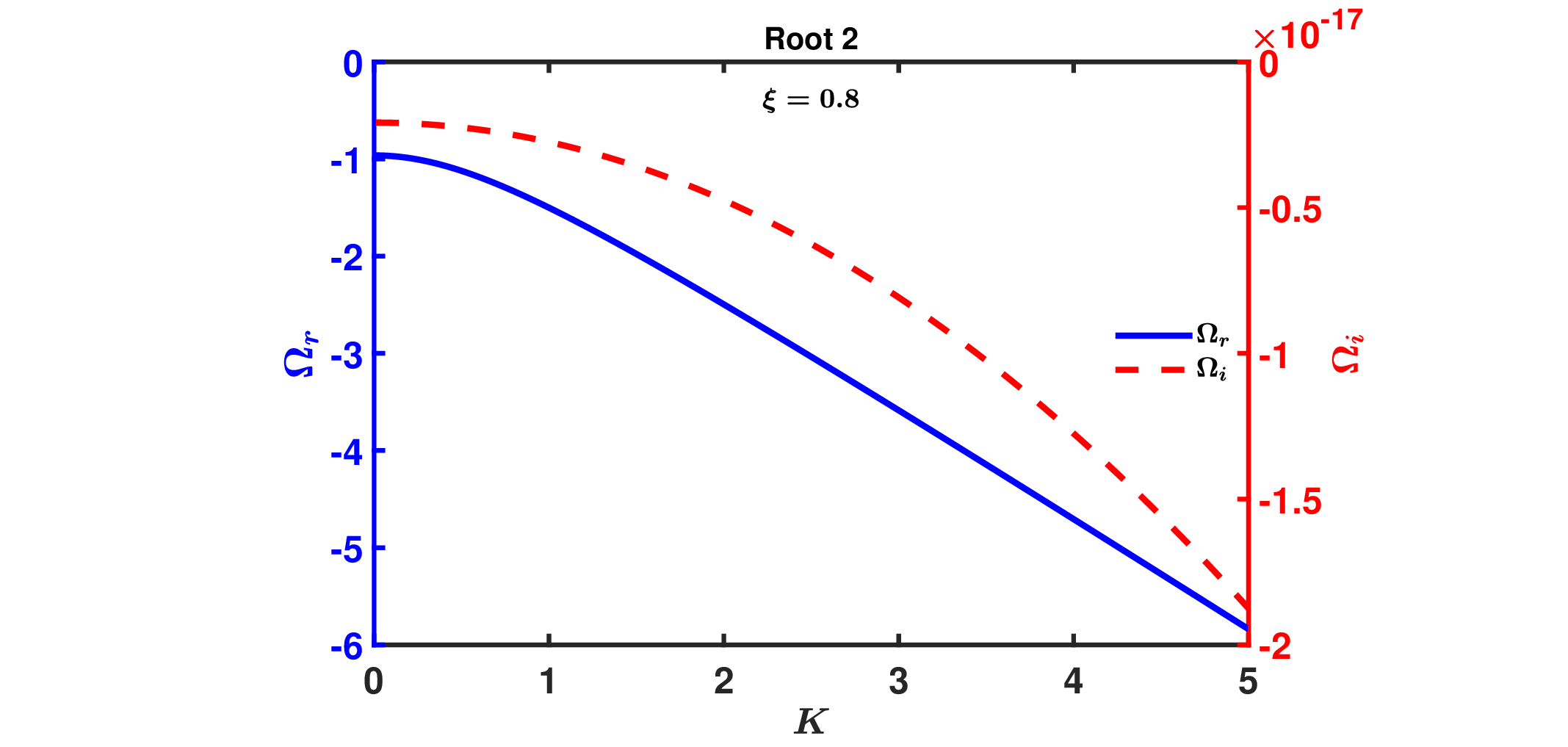}
	\end{tabular}
	\caption{Two roots of the Jeans-normalized composite frequency $(\Omega)$ as a function of the Jeans-normalized wavenumber $(K)$. The EiBI gravity parameter is fixed at $\chi=3\times10^7$ m$^5$\,kg$^{-1}$\,s$^{-2}$, and the relative polytropic sound speed is set to $\beta=\Gamma$.}
	\label{fig:2}
\end{figure*}

Both roots of the quadratic dispersion relation are numerically evaluated and shown in Figure~\ref{fig:2}. The second root (Root 2) yields a negative oscillatory component ($\Omega_r < 0$), which is not physically meaningful for the present solar plasma configuration and therefore discarded. Consequently, only the first root (Root 1) is retained as the physically admissible solution and forms the basis of all subsequent analysis. For numerical evaluation, the normalized electron temperature is fixed at $\epsilon_{T0}^{-1}=10$, and the viscosity is set to $\eta = 0.1$ kg\,m$^{-1}$\,s$^{-1}$, ensuring a consistent baseline damping throughout.

\subsection{EiBI gravity--modified stability}
\begin{figure*}
	\centering
	\begin{tabular}{c c}
		\includegraphics[trim={5cm 0cm 5cm 0cm},clip,width=8cm]{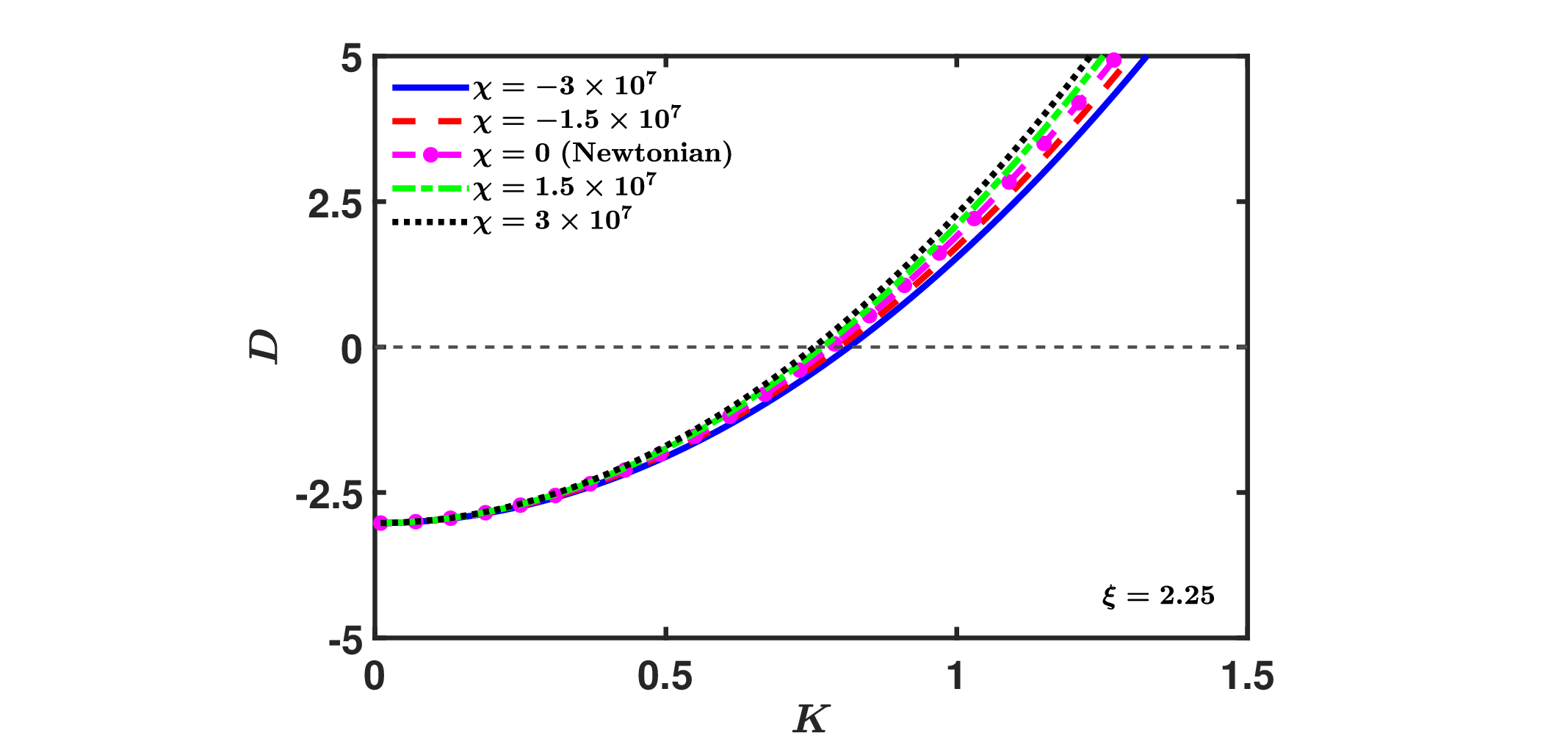}&
		\includegraphics[trim={5cm 0cm 5cm 0cm},clip,width=8cm]{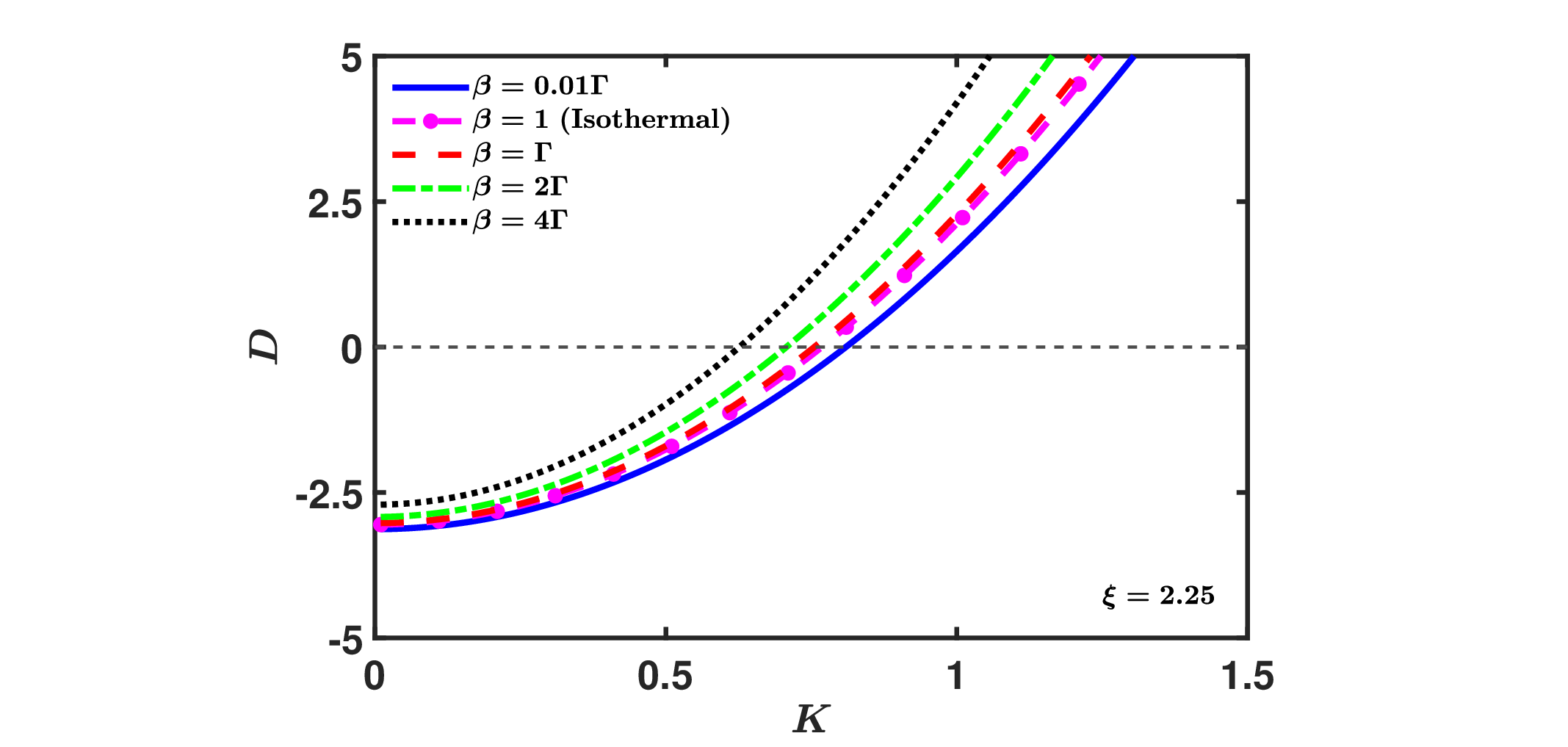}\\
		\includegraphics[trim={5cm 0cm 5cm 0cm},clip,width=8cm]{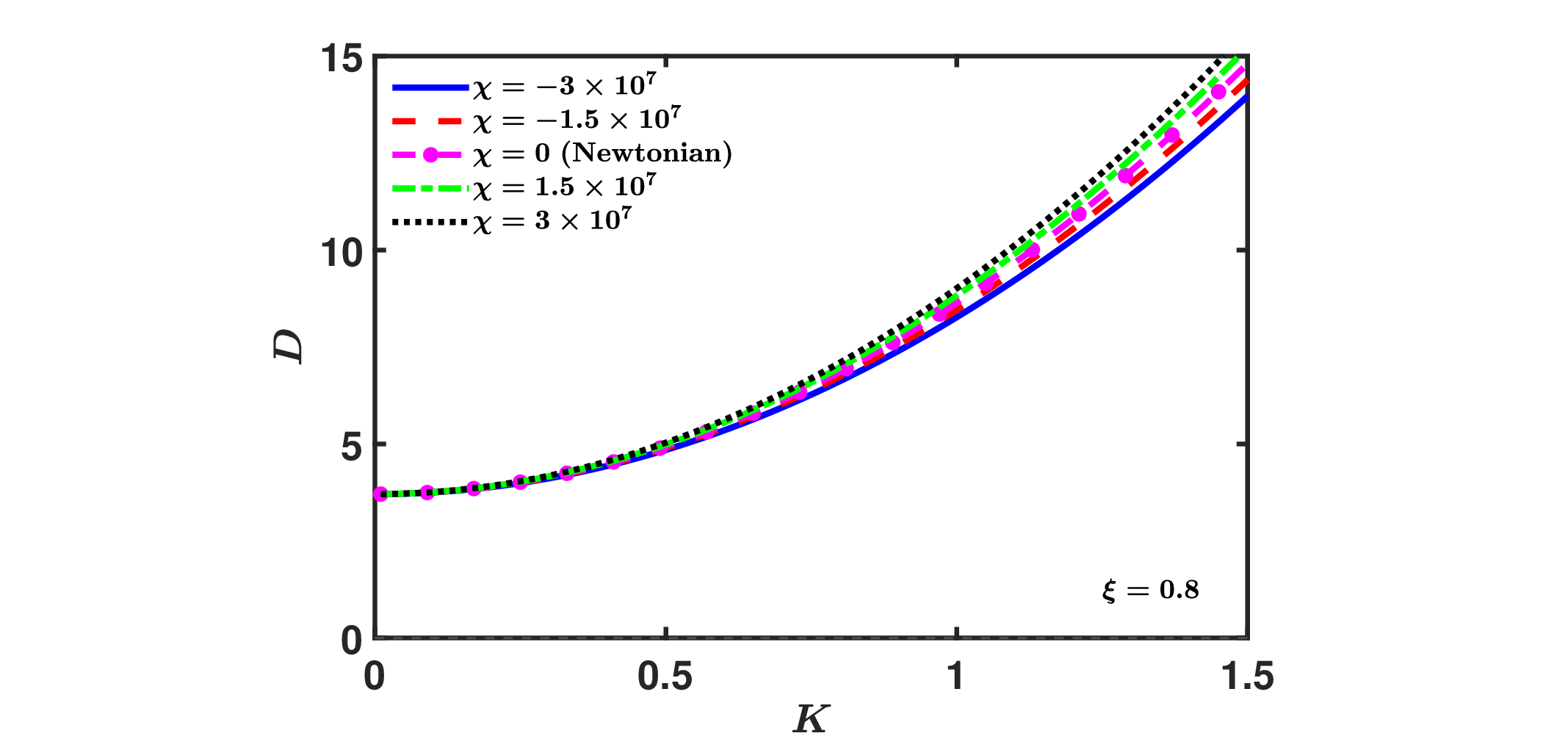}&
		\includegraphics[trim={5cm 0cm 5cm 0cm},clip,width=8cm]{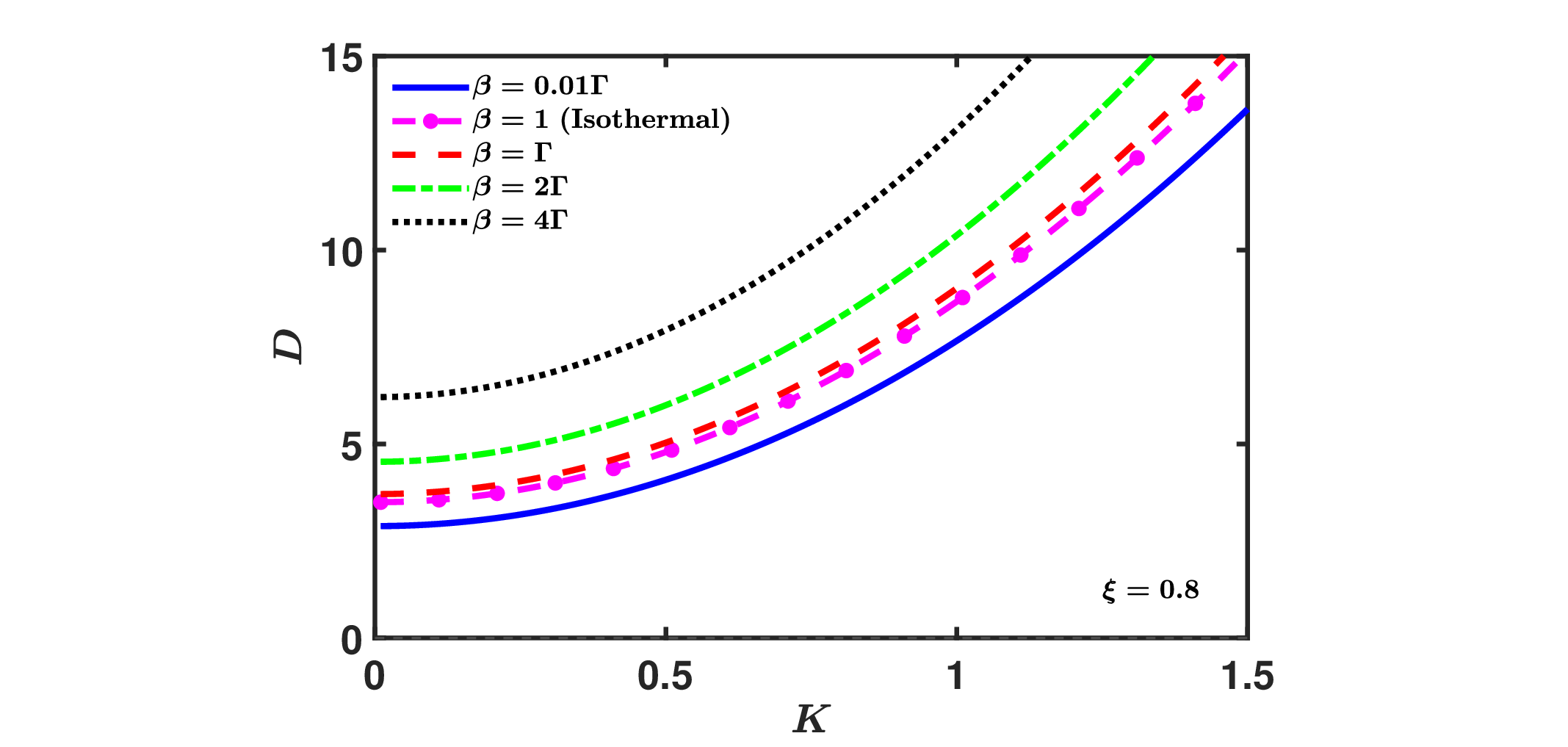}
	\end{tabular}
	\caption{Variation of the discriminant $(D)$ as a function of Jeans-normalized angular wavenumber $(K)$ for different values of the EiBI gravity parameter $(\chi)$ and the relative polytropic sound speed $(\beta)$. Here, $\chi$ is expressed in SI units.}
	\label{fig:3}
\end{figure*}

\begin{figure*}
	\centering
	\begin{tabular}{c c}
		\includegraphics[trim={8cm 0cm 5cm 0cm},clip,width=8cm]{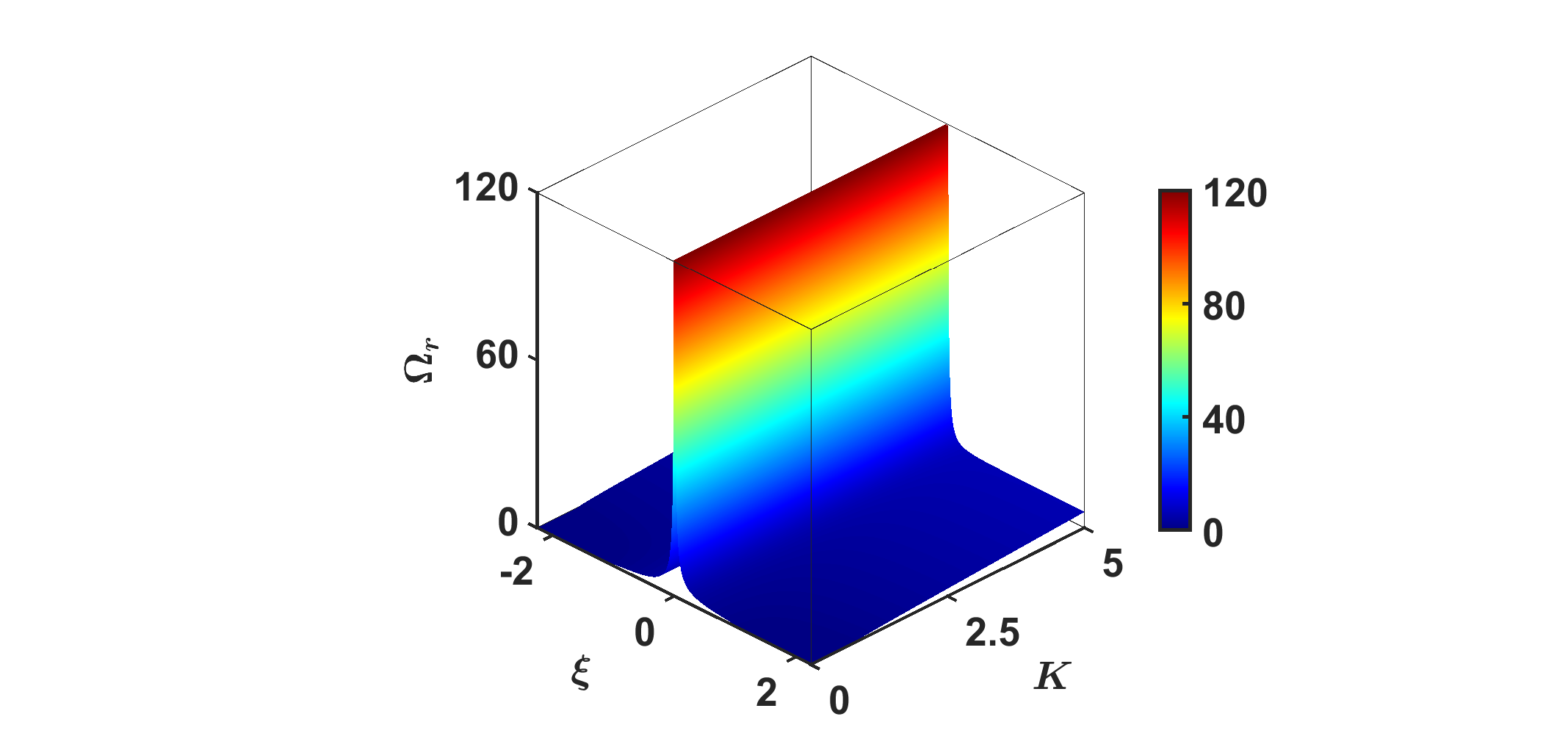}&
		\includegraphics[trim={8cm 0cm 5cm 0cm},clip,width=8cm]{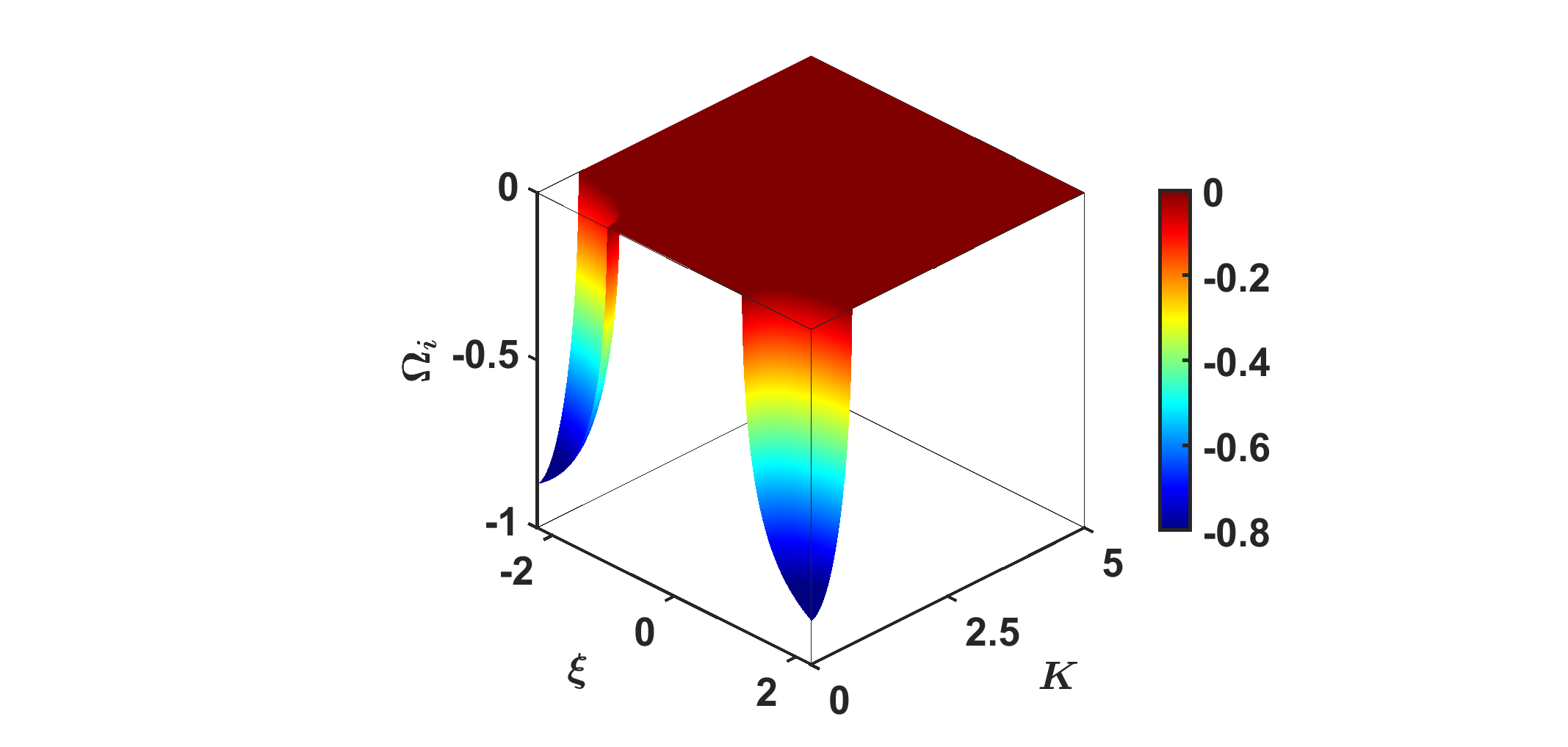}\\
		\includegraphics[trim={8cm 0cm 5cm 0cm},clip,width=8cm]{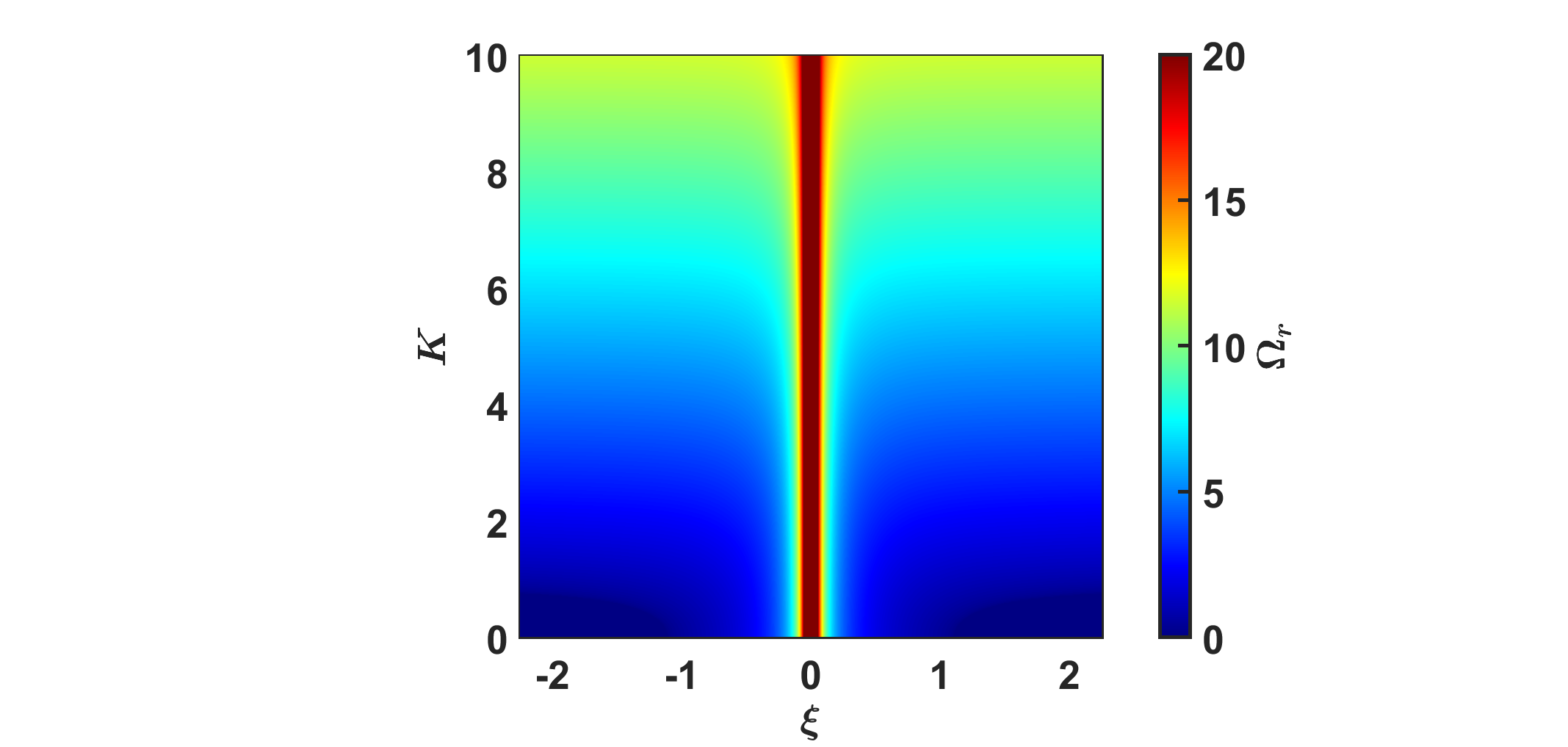}&
		\includegraphics[trim={8cm 0cm 5cm 0cm},clip,width=8cm]{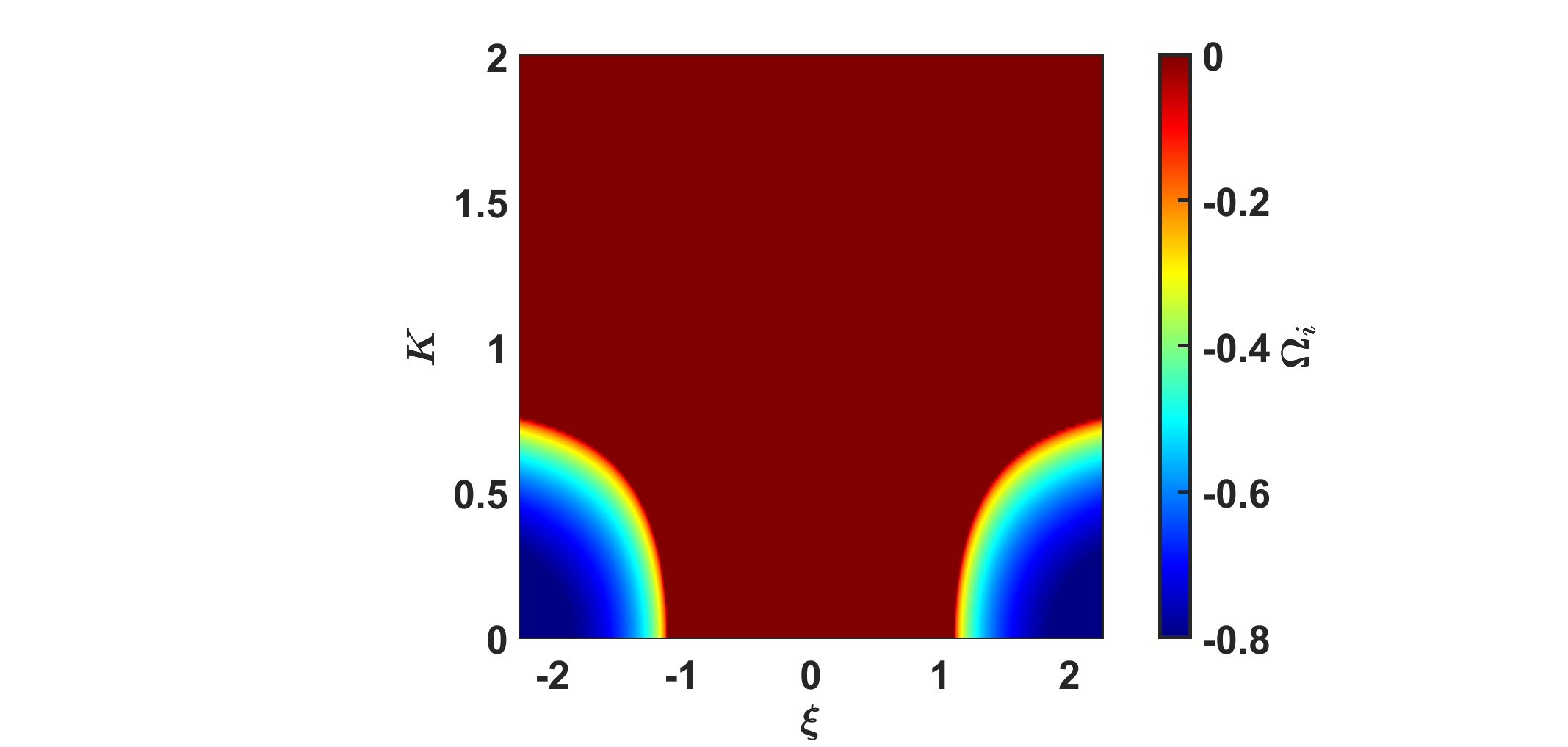}
	\end{tabular}
	\caption{Profiles of the Jeans-normalized oscillation frequency $(\Omega_r)$ and the growth rate $(\Omega_i)$ as functions of the Jeans-normalized angular wavenumber $(K)$ and radial distance $(\xi)$. The EiBI gravity parameter is fixed at $\chi=3\times10^7$ m$^5$\,kg$^{-1}$\,s$^{-2}$, and the relative polytropic sound speed is set to $\beta=\Gamma$.}
	\label{fig:4}
\end{figure*}

\begin{figure*}
	\centering
	\begin{tabular}{c c}
		\includegraphics[trim={5cm 0cm 5cm 0cm},clip,width=8cm]{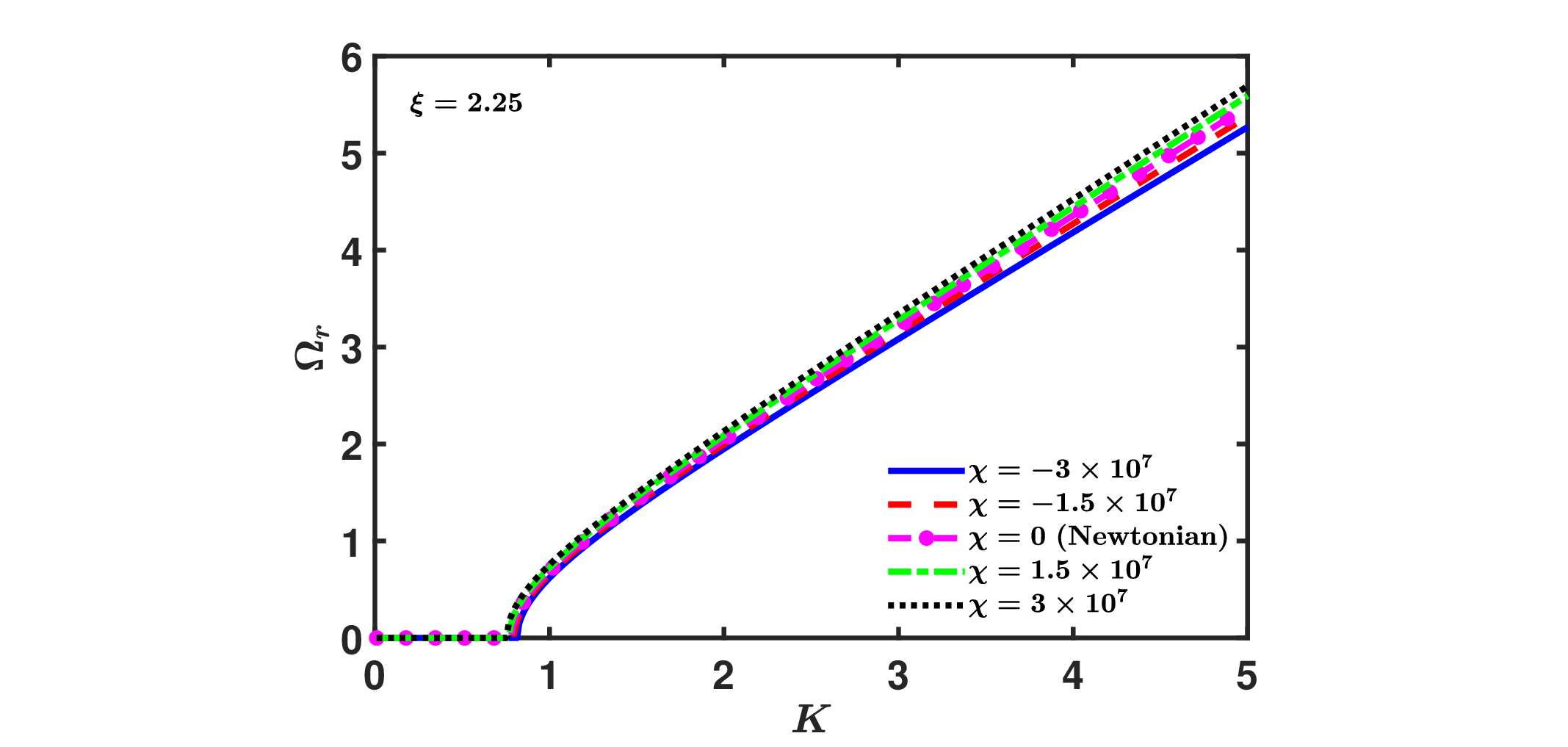}&
		\includegraphics[trim={5cm 0cm 5cm 0cm},clip,width=8cm]{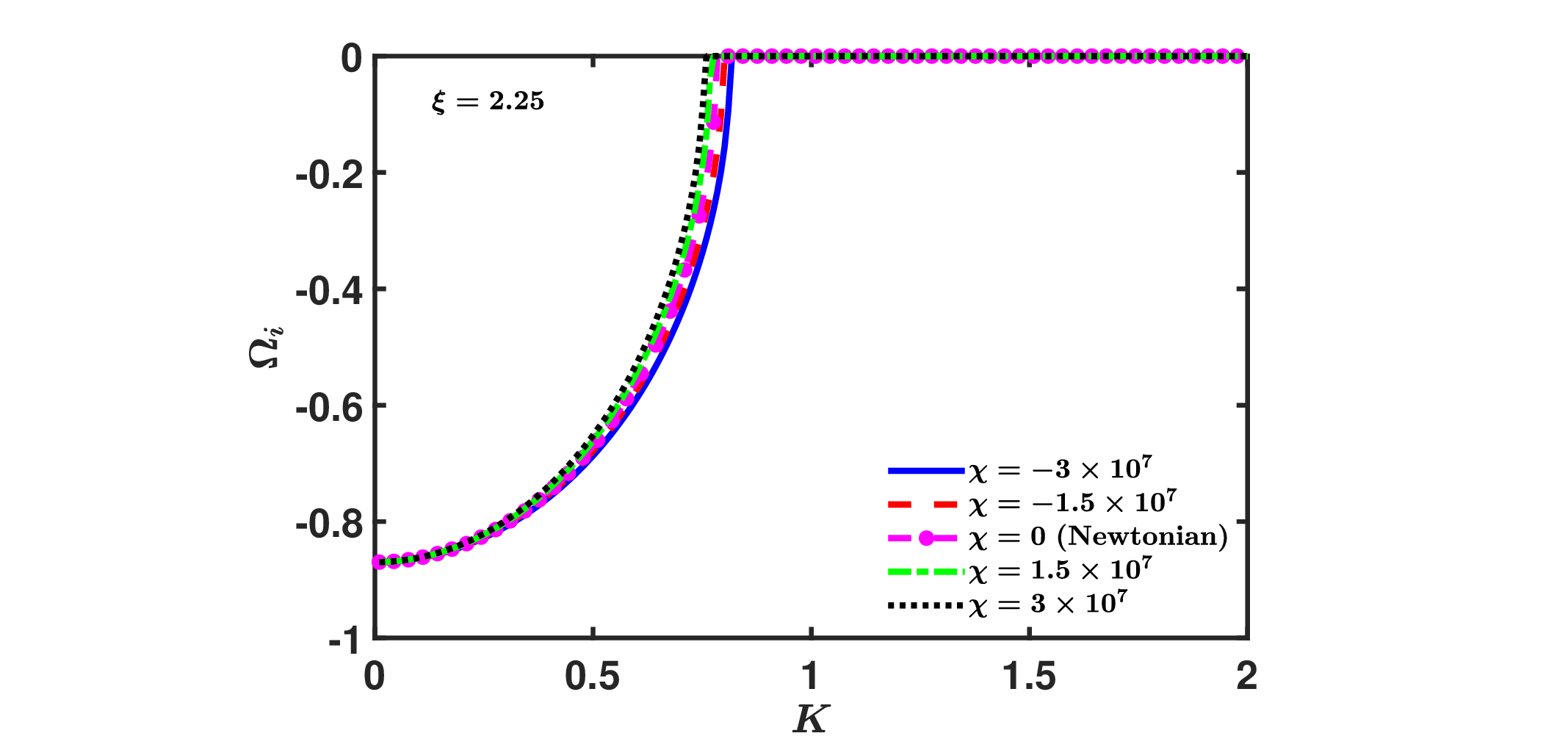}\\
		\includegraphics[trim={5cm 0cm 5cm 0cm},clip,width=8cm]{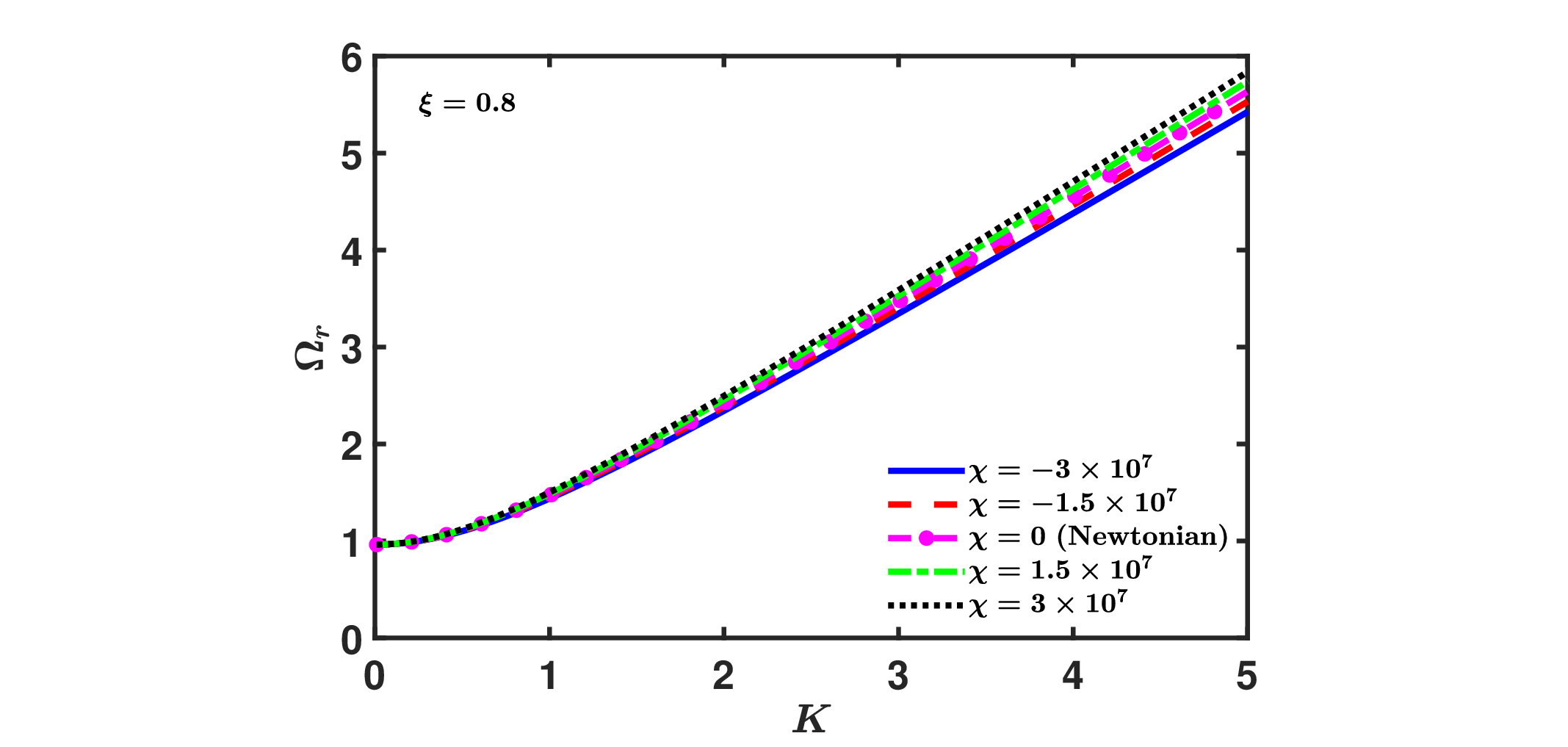}&
		\includegraphics[trim={5cm 0cm 5cm 0cm},clip,width=8cm]{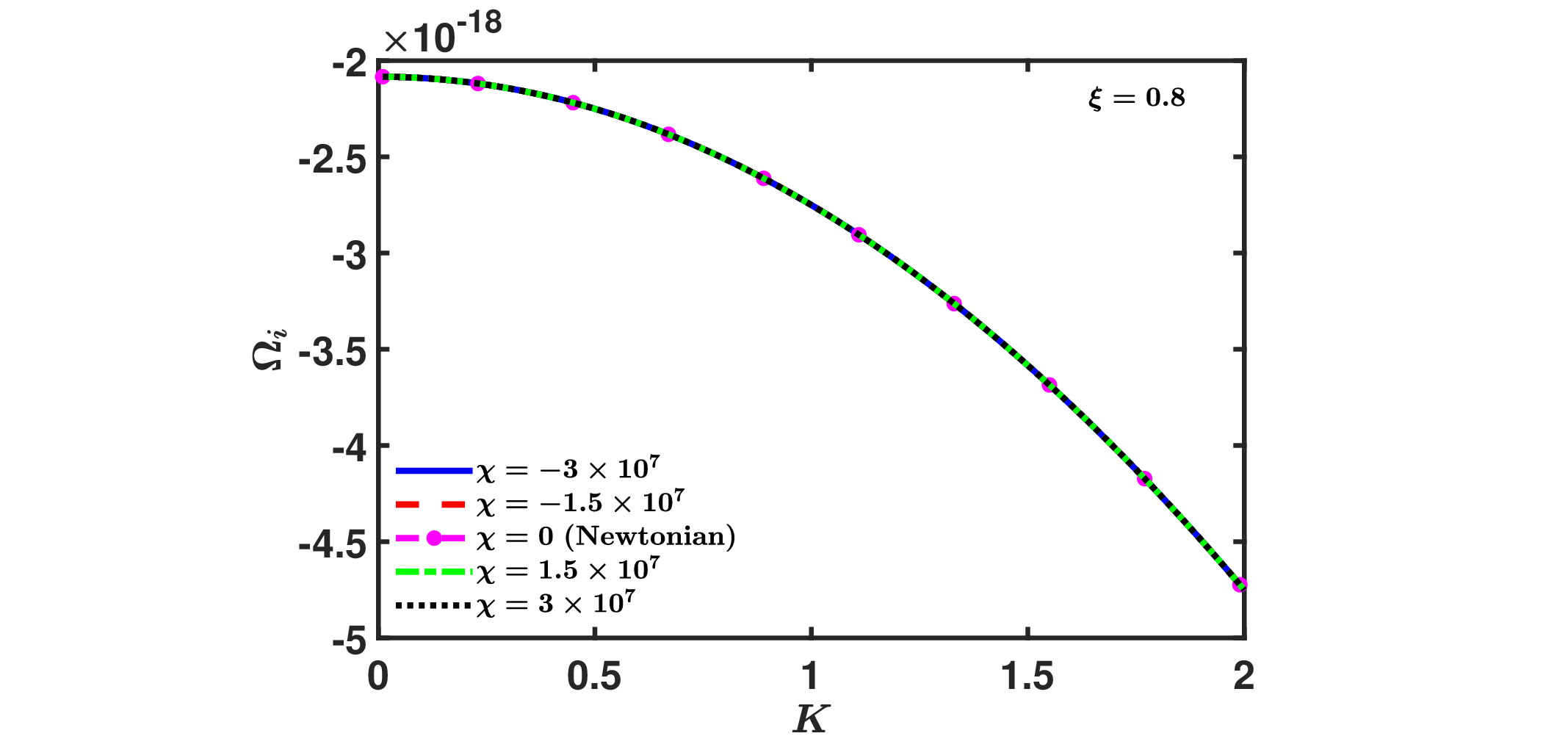}
	\end{tabular}
	\caption{Profiles of the Jeans-normalized oscillation frequency $(\Omega_r)$ and the growth rate $(\Omega_i)$ as functions of the Jeans-normalized angular wavenumber $(K)$ for different representative values of the EiBI gravity parameter $(\chi)$ in SI units. The relative polytropic sound speed is fixed at $\beta=\Gamma$.}
	\label{fig:5}
\end{figure*}

\begin{figure*}
	\centering
	\begin{tabular}{c c}
		\includegraphics[trim={5cm 0cm 5cm 0cm},clip,width=8cm]{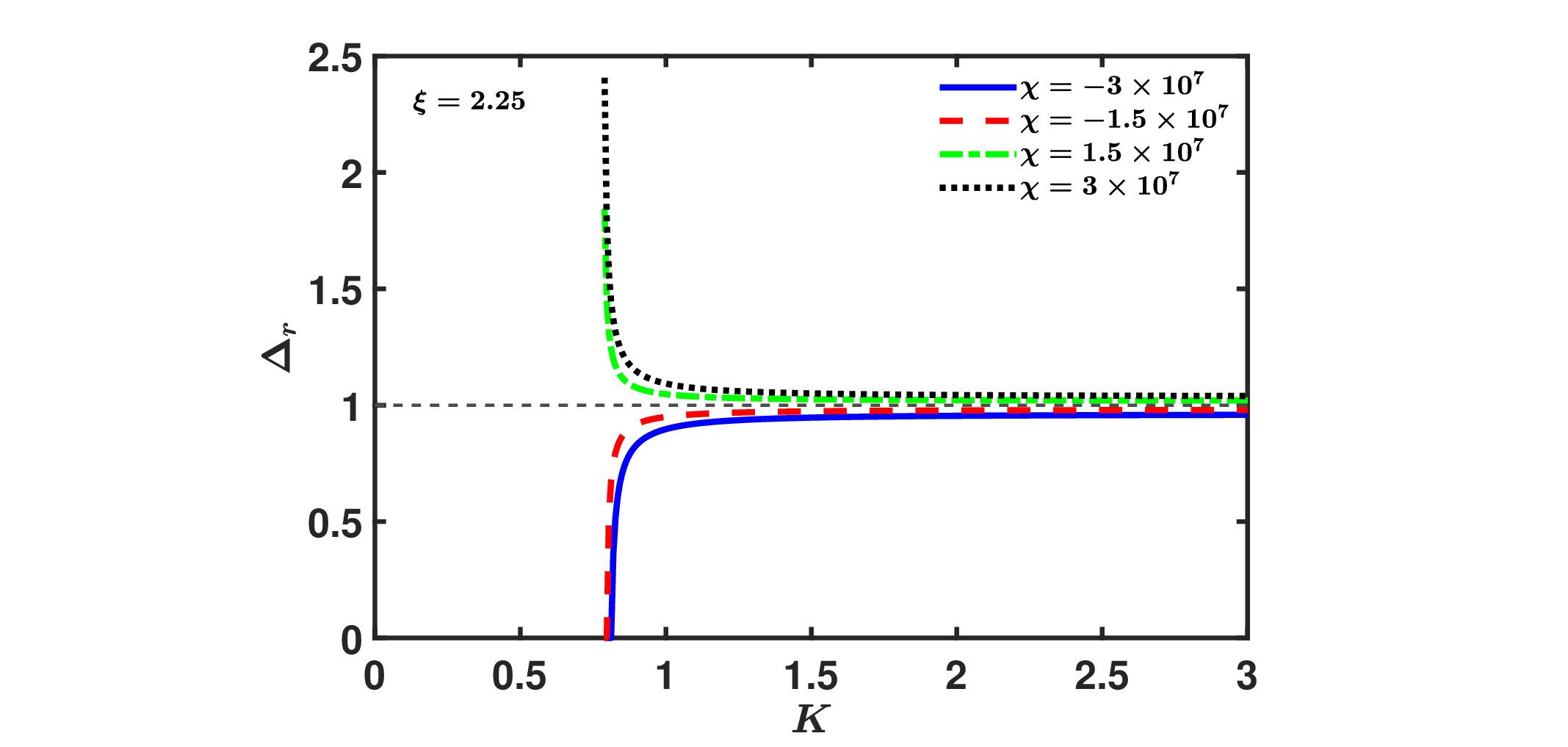}&
		\includegraphics[trim={5cm 0cm 5cm 0cm},clip,width=8cm]{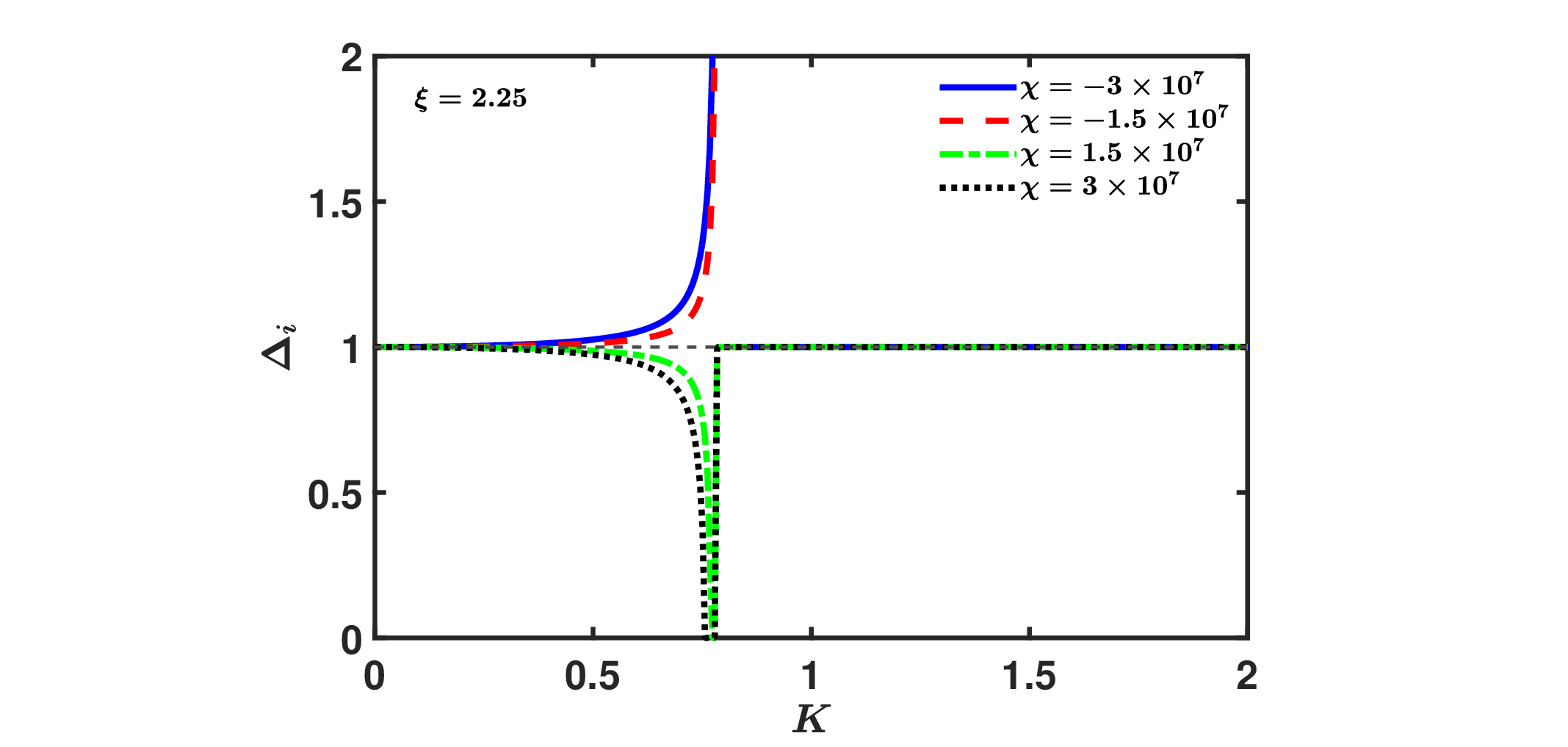}\\
		\includegraphics[trim={5cm 0cm 5cm 0cm},clip,width=8cm]{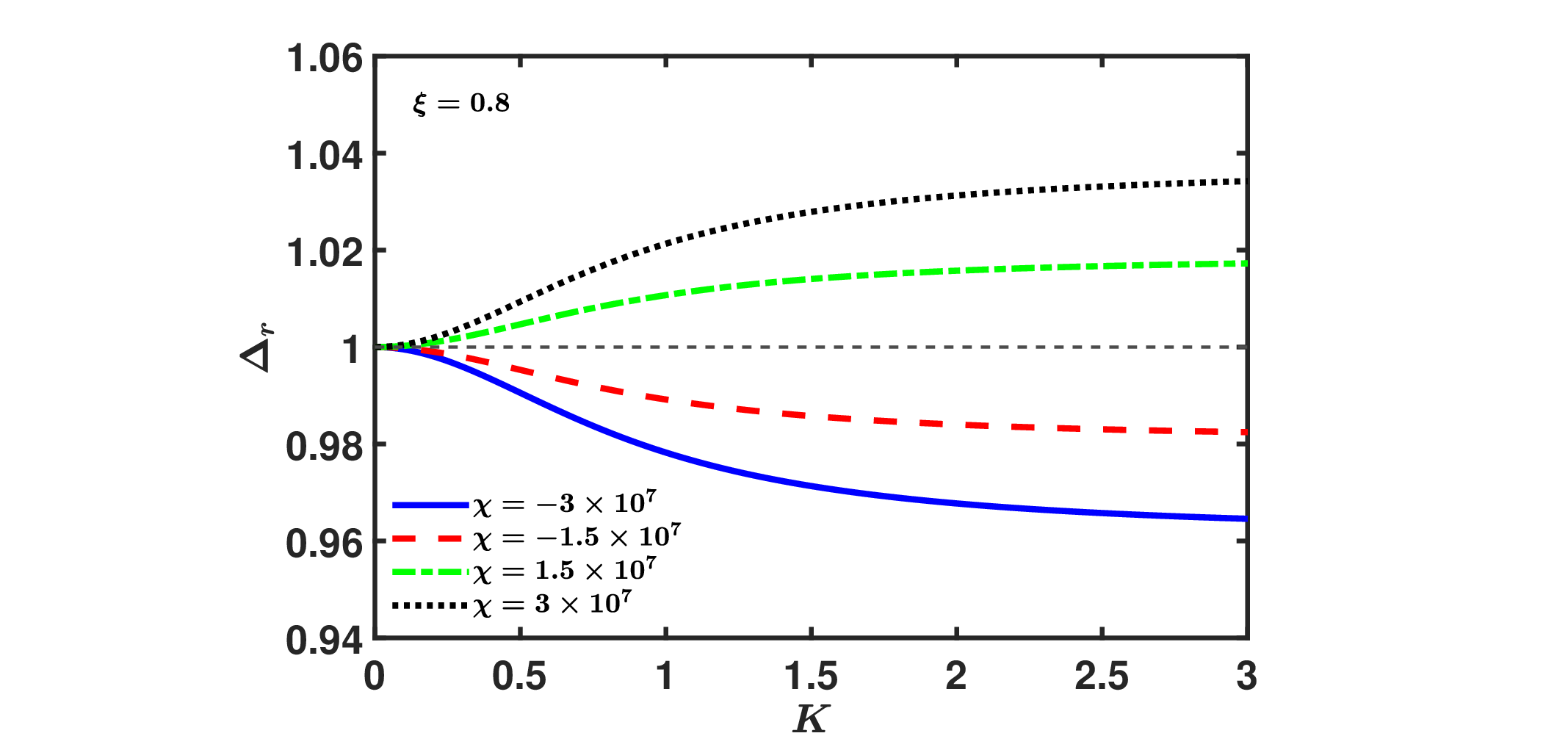}&
		\includegraphics[trim={5cm 0cm 5cm 0cm},clip,width=8cm]{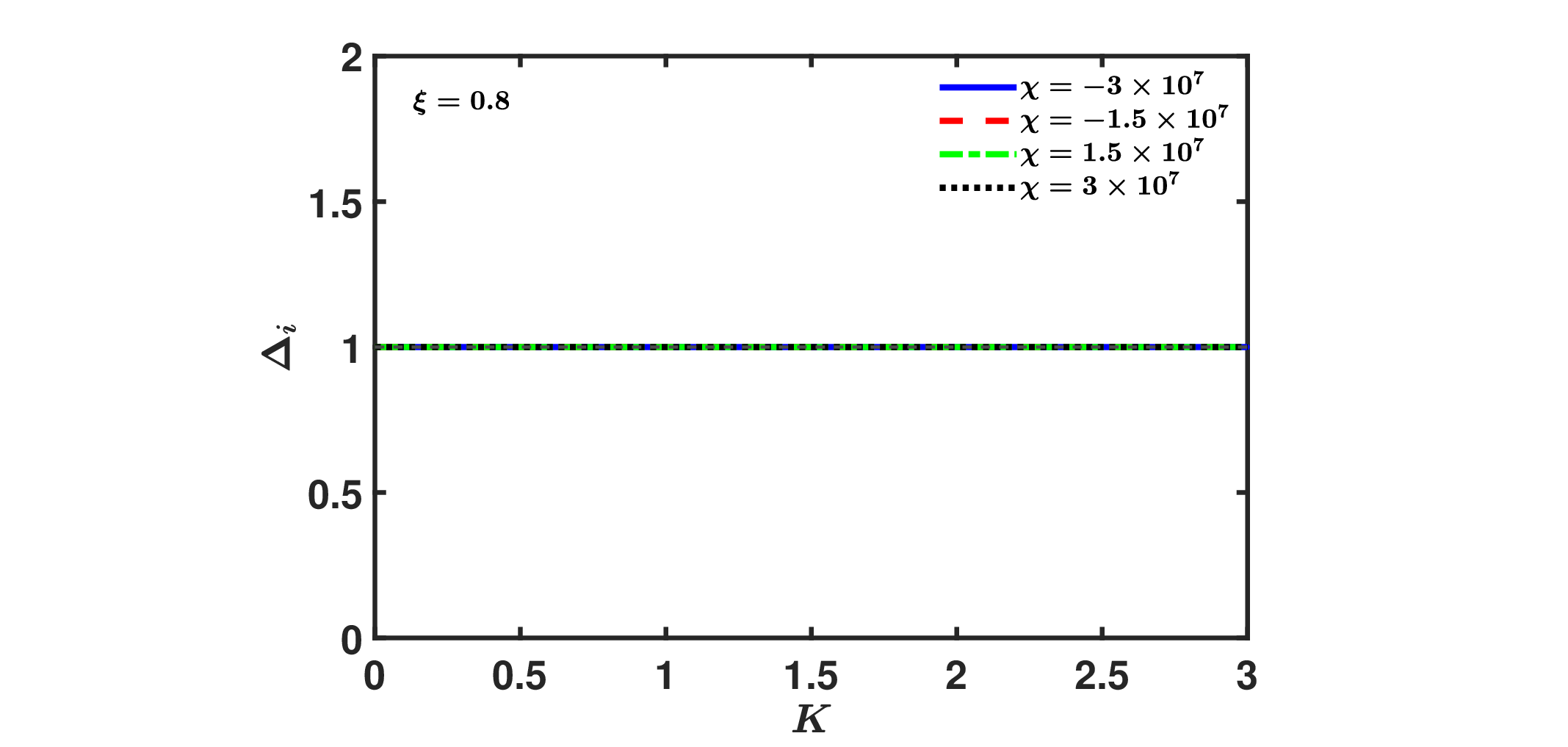}
	\end{tabular}
	\caption{Relative shifts of the Jeans-normalized modal frequency properties in EiBI gravity, expressed as 
	$\Delta_r = \Omega_r^{\mathrm{EiBI}} / \Omega_r^{\mathrm{Newtonian}}$ for the real oscillation frequency (propagation) and $\Delta_i = \Omega_i^{\mathrm{EiBI}} / \Omega_i^{\mathrm{Newtonian}}$ for the imaginary oscillation frequency (growth), plotted as a function of the Jeans-normalized angular wavenumber $(K)$ for different representative values of the EiBI gravity parameter $(\chi)$ in SI units. The dashed horizontal line at $\Delta = 1$ clearly indicates the Newtonian $(\chi = 0)$ baseline.}
	\label{fig:6}
\end{figure*}

\begin{figure*}
	\centering
	\begin{tabular}{c c}
		\includegraphics[trim={5cm 0cm 5cm 0cm},clip,width=8cm]{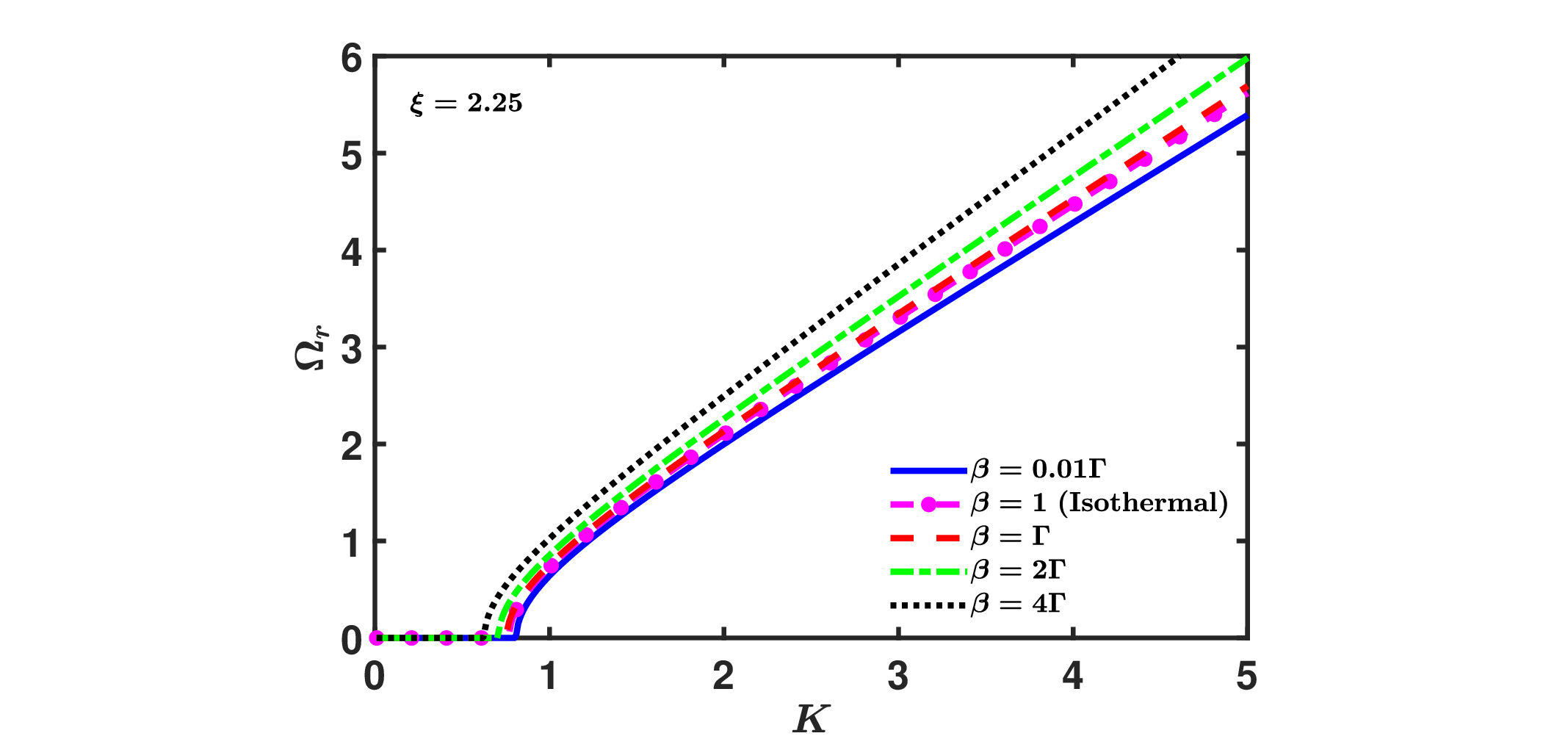}&
		\includegraphics[trim={5cm 0cm 5cm 0cm},clip,width=8cm]{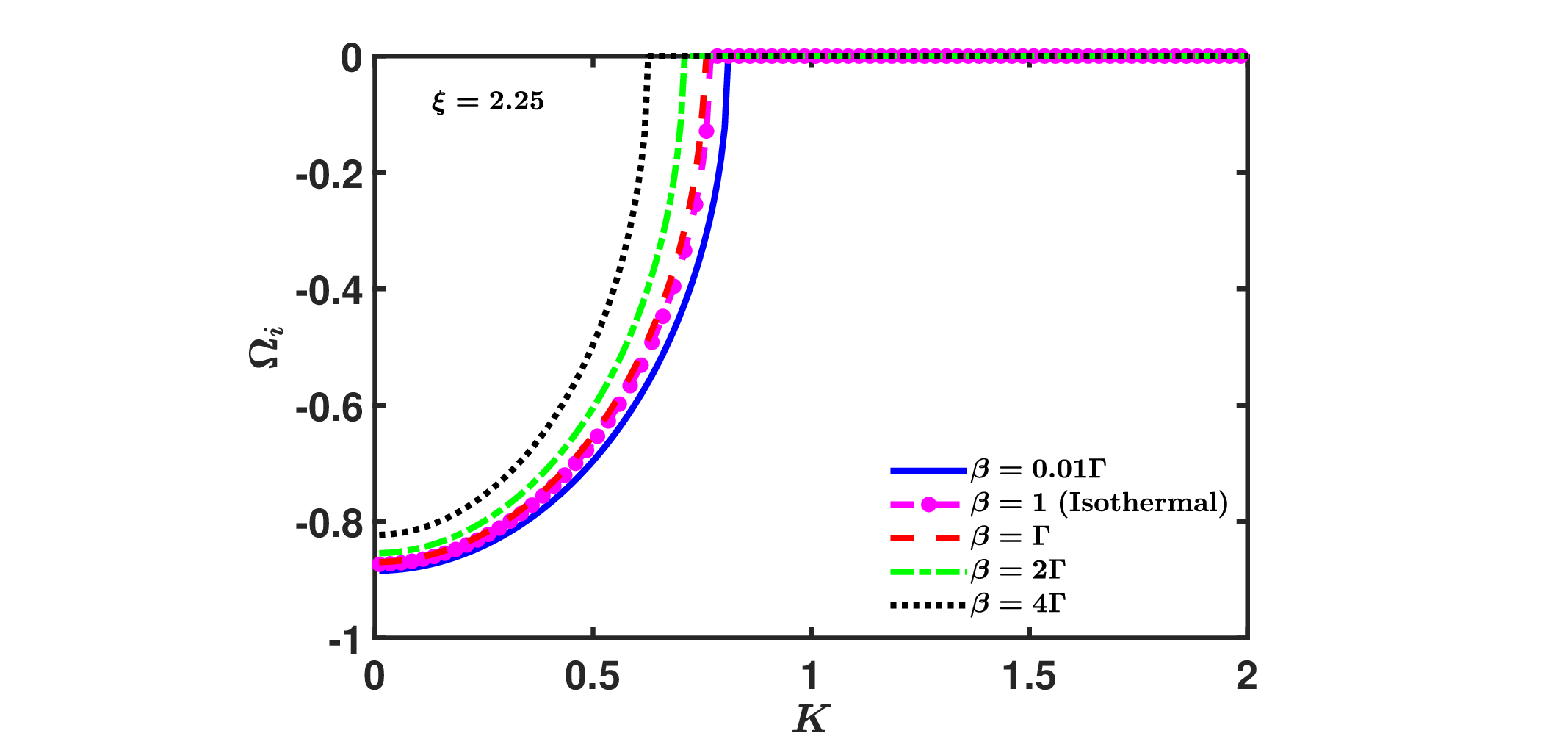}\\
		\includegraphics[trim={5cm 0cm 5cm 0cm},clip,width=8cm]{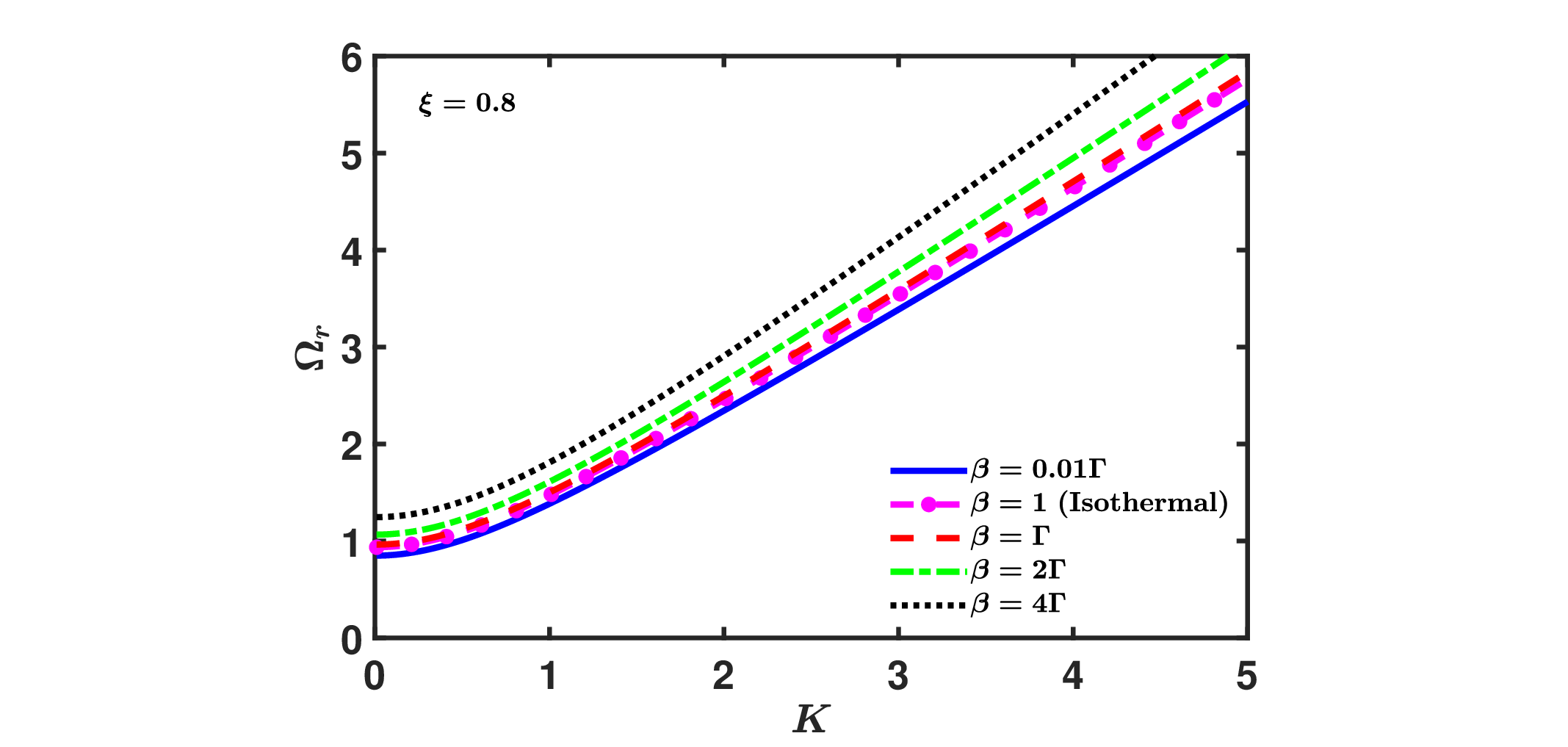}&
		\includegraphics[trim={5cm 0cm 5cm 0cm},clip,width=8cm]{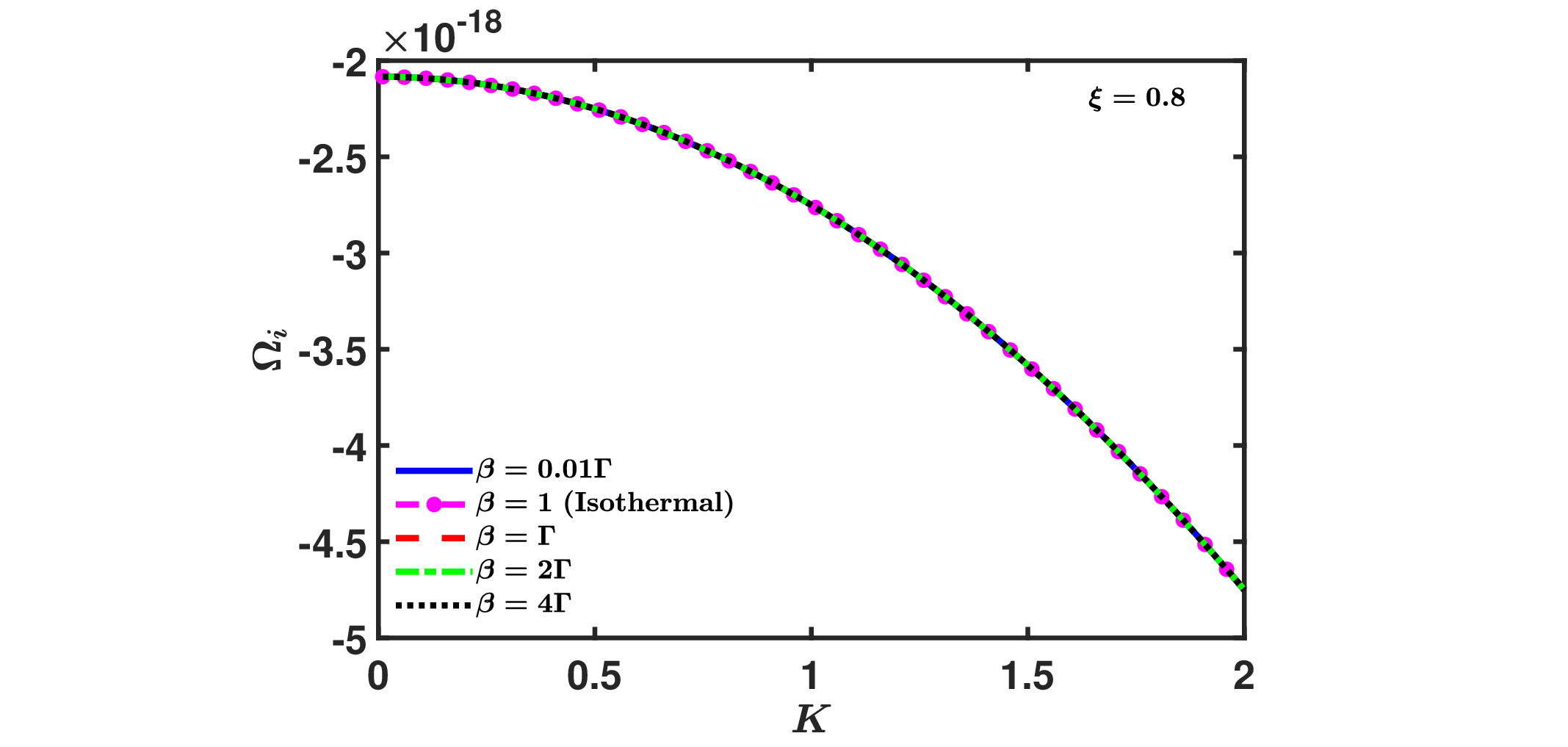}
	\end{tabular}
	\caption{Profiles of the Jeans-normalized oscillation frequency $(\Omega_r)$ and the growth rate $(\Omega_i)$ as functions of the Jeans-normalized angular wavenumber $(K)$ for different indicated values of relative polytropic sound speed $(\beta)$. The EiBI gravity parameter is fixed at $\chi=3\times10^7$ m$^5$\,kg$^{-1}$\,s$^{-2}$.}
	\label{fig:7}
\end{figure*}

\begin{figure*}
	\centering
	\begin{tabular}{c c}
		\includegraphics[trim={5cm 0cm 5cm 0cm},clip,width=8cm]{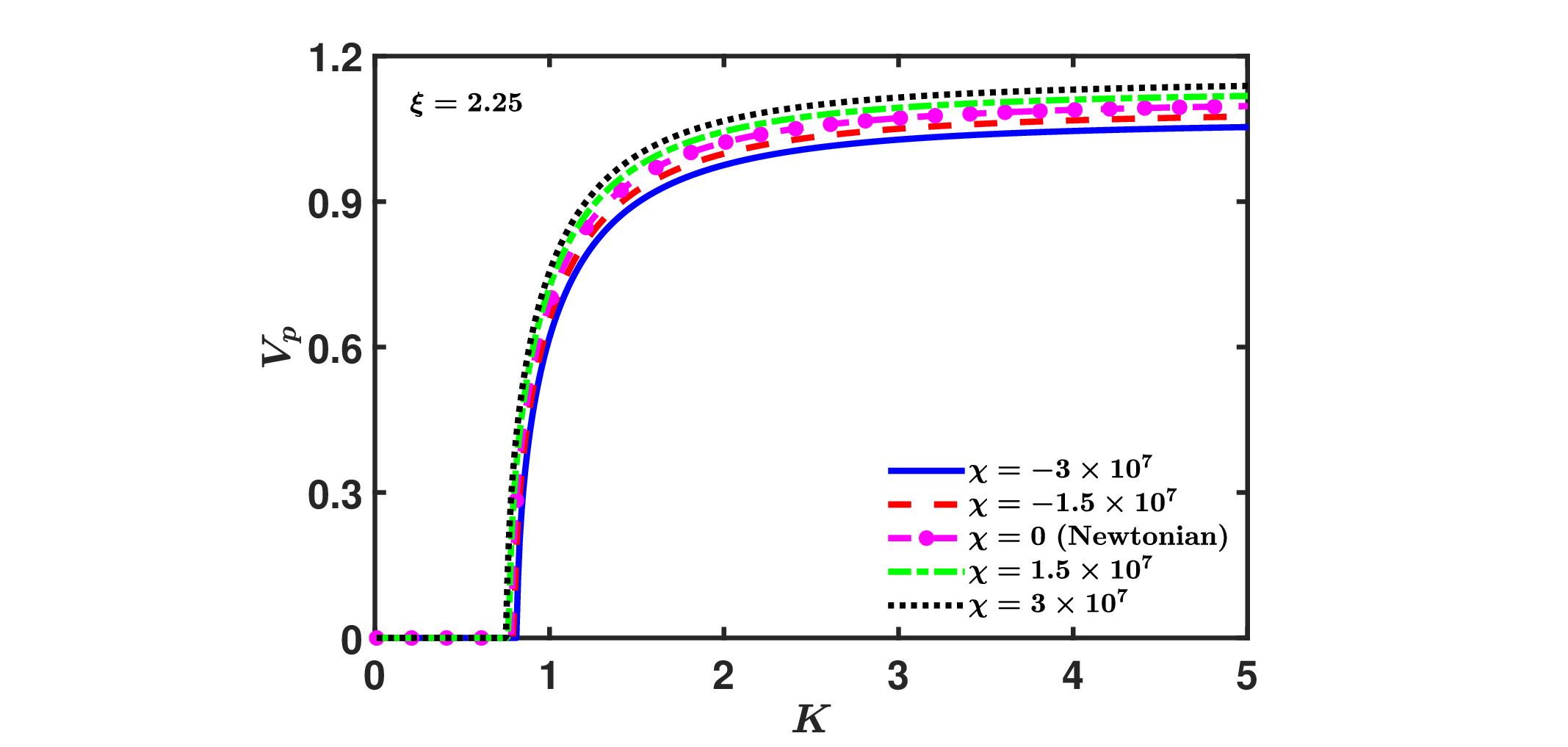}&
		\includegraphics[trim={5cm 0cm 5cm 0cm},clip,width=8cm]{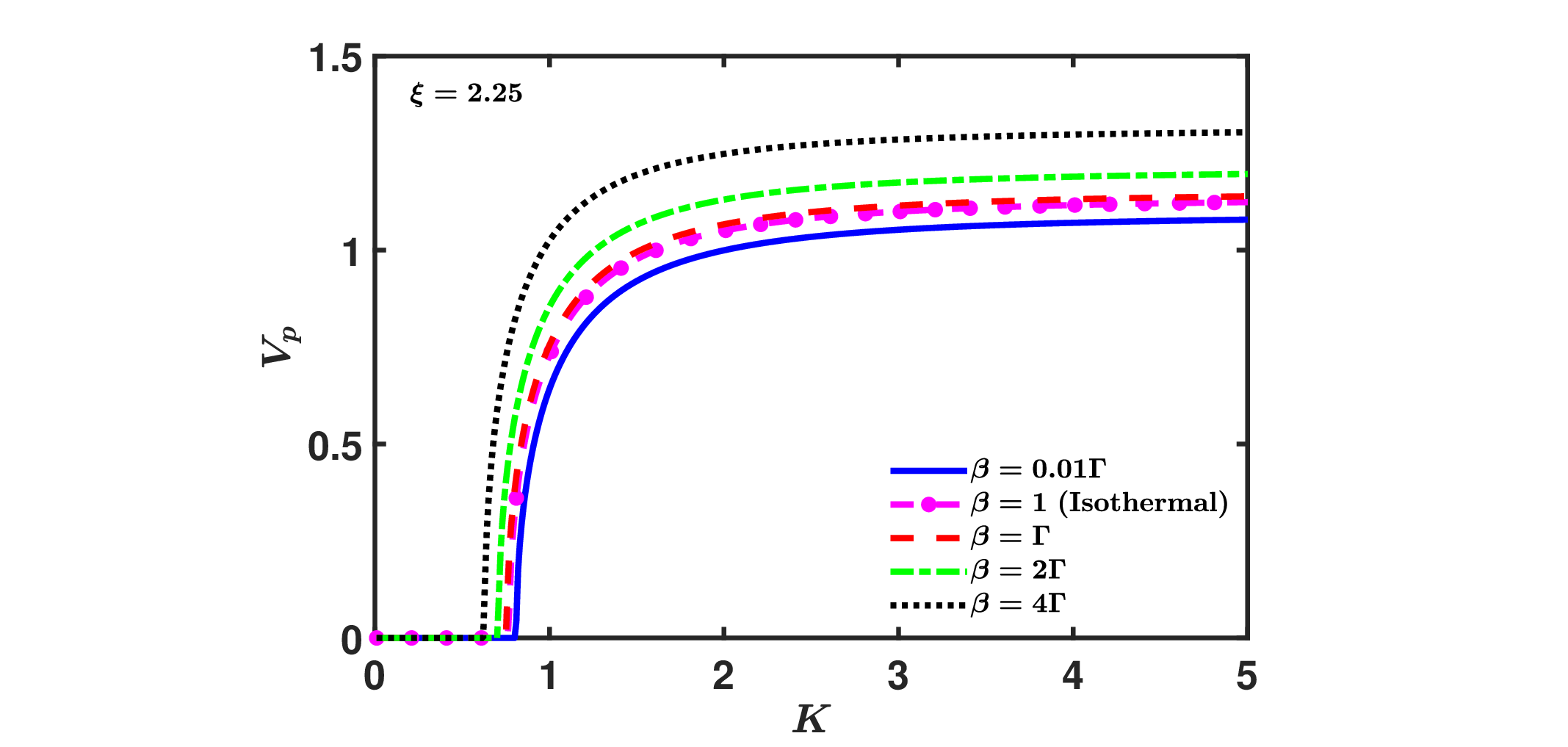}\\
		\includegraphics[trim={5cm 0cm 5cm 0cm},clip,width=8cm]{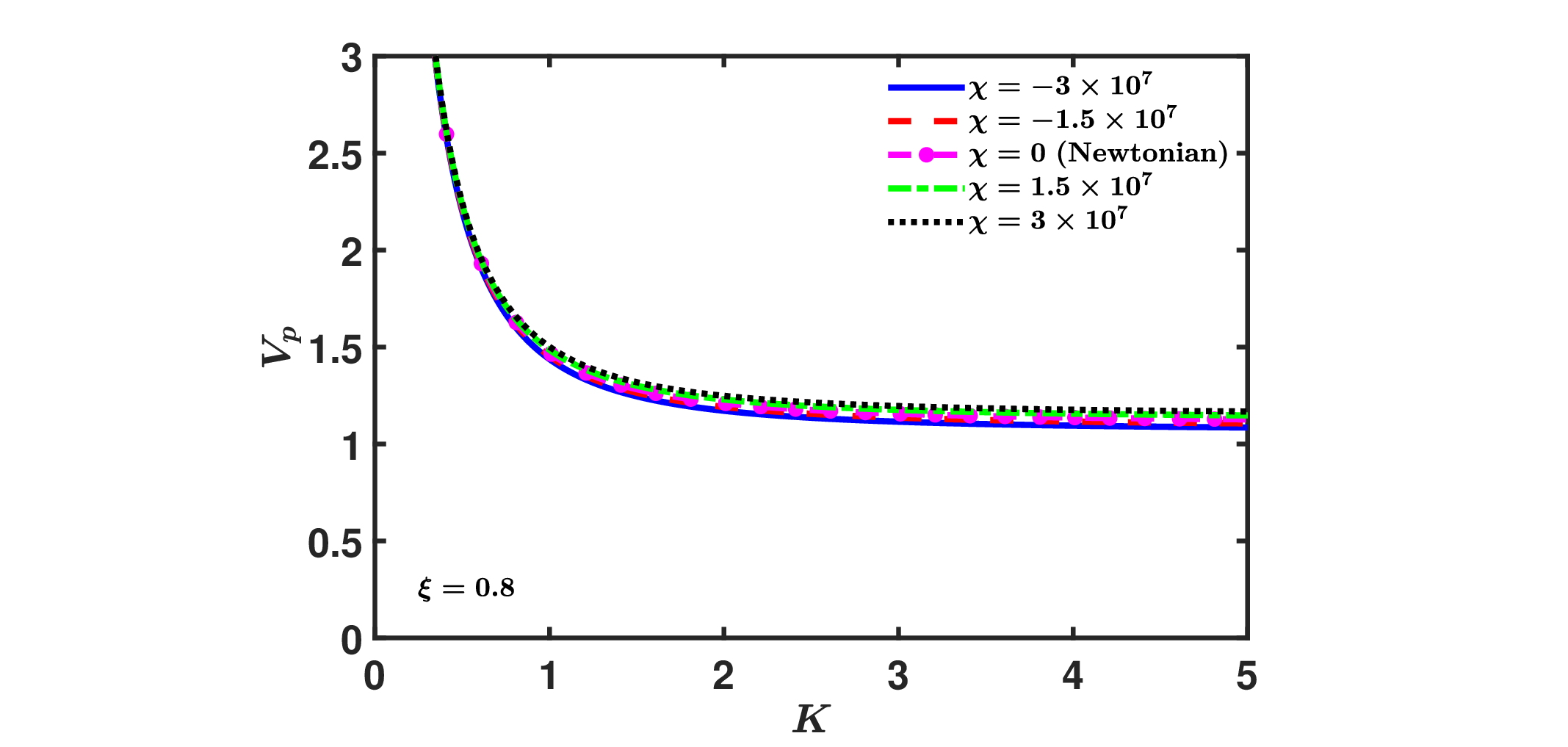}&
		\includegraphics[trim={5cm 0cm 5cm 0cm},clip,width=8cm]{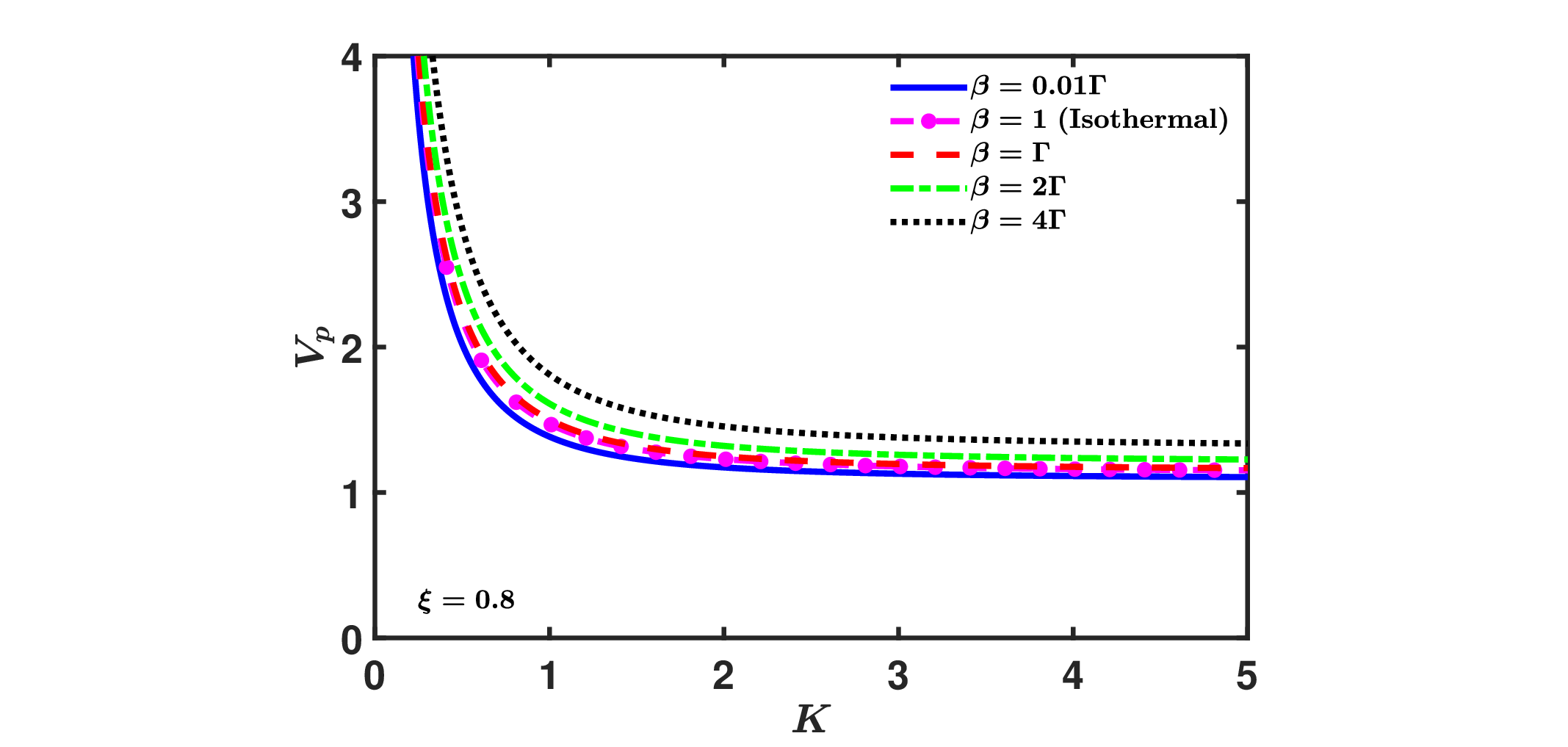}
	\end{tabular}
	\caption{Profiles of the Jeans-normalized phase speed $(V_p)$ as a function of the Jeans-normalized angular wavenumber $(K)$ for different indicated values of EiBI gravity parameter $(\chi)$ and relative polytropic sound speed $(\beta)$. Here, $\chi$ is expressed in SI units.}
	\label{fig:8}
\end{figure*}

\begin{figure*}
	\centering
	\begin{tabular}{c c}
		\includegraphics[trim={9cm 0cm 4cm 0cm},clip,width=8cm]{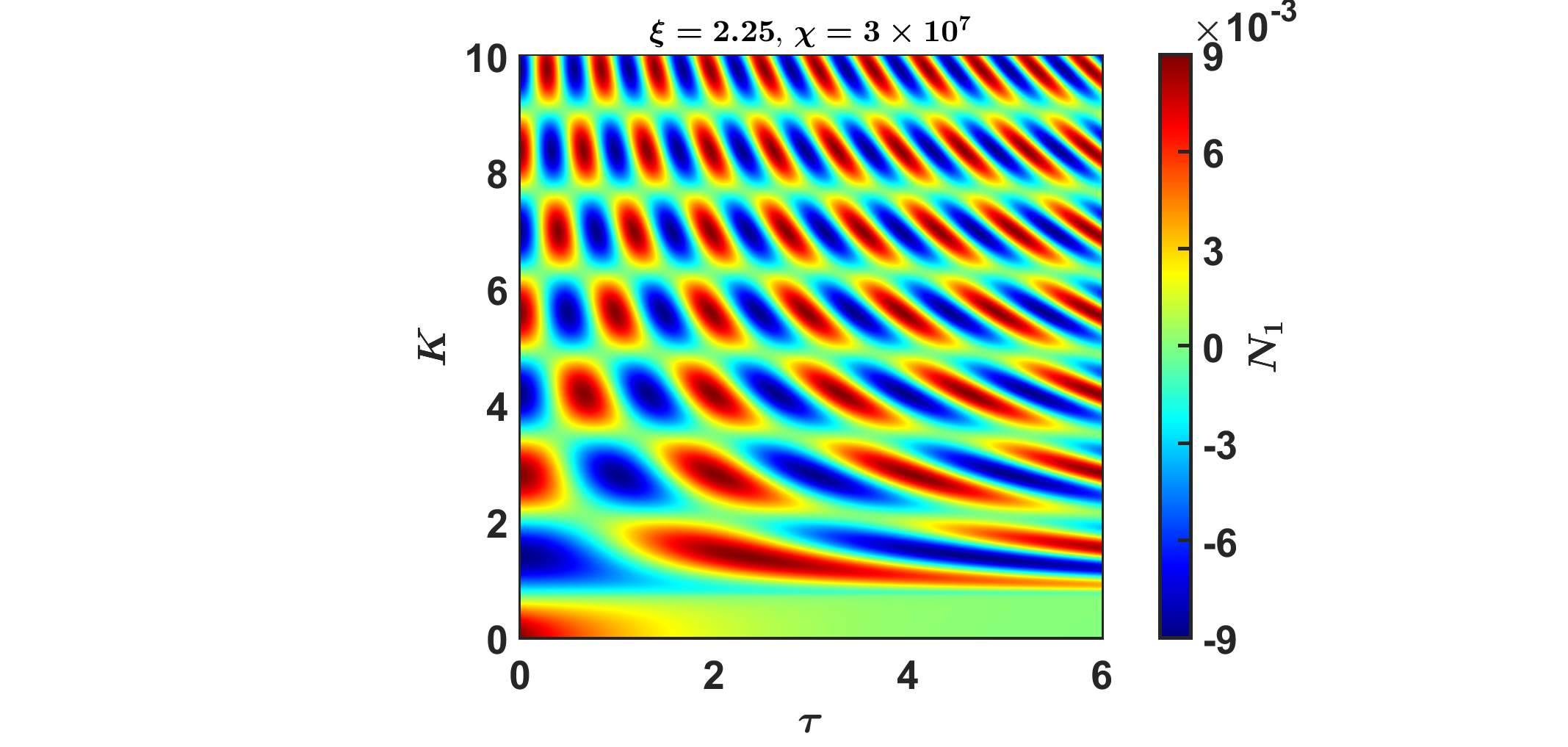}&
		\includegraphics[trim={9cm 0cm 4cm 0cm},clip,width=8cm]{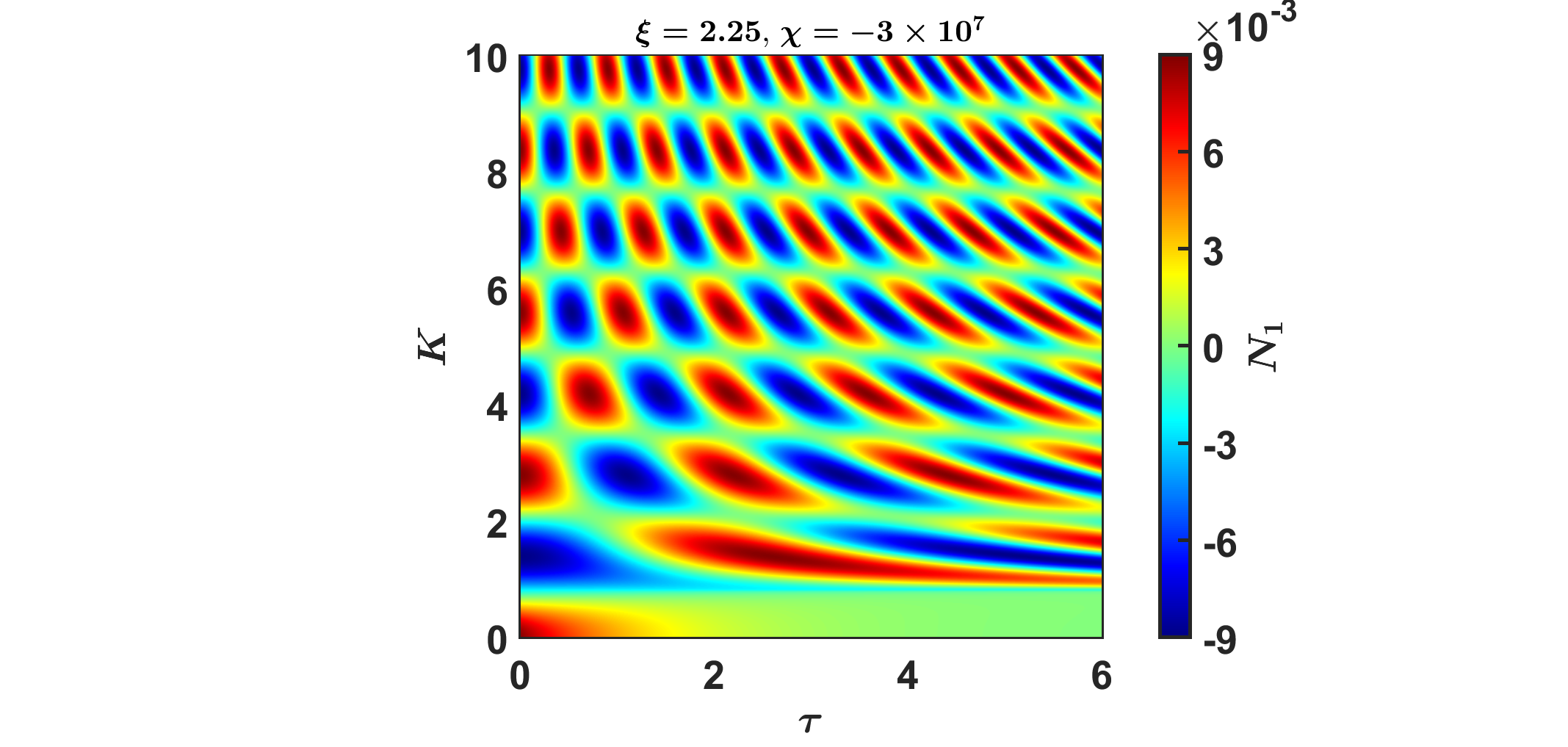}\\
		\includegraphics[trim={9cm 0cm 4cm 0cm},clip,width=8cm]{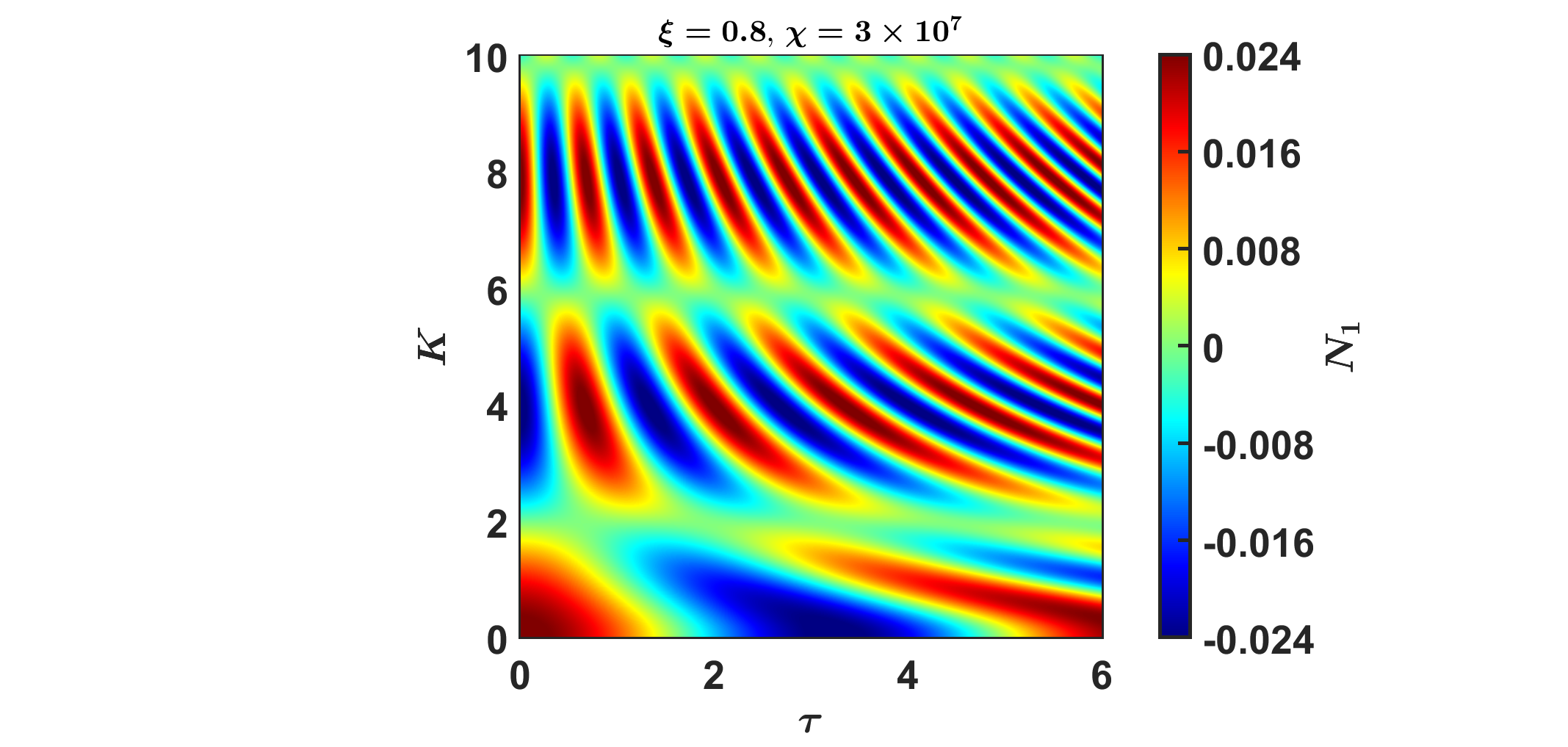}&
		\includegraphics[trim={9cm 0cm 4cm 0cm},clip,width=8cm]{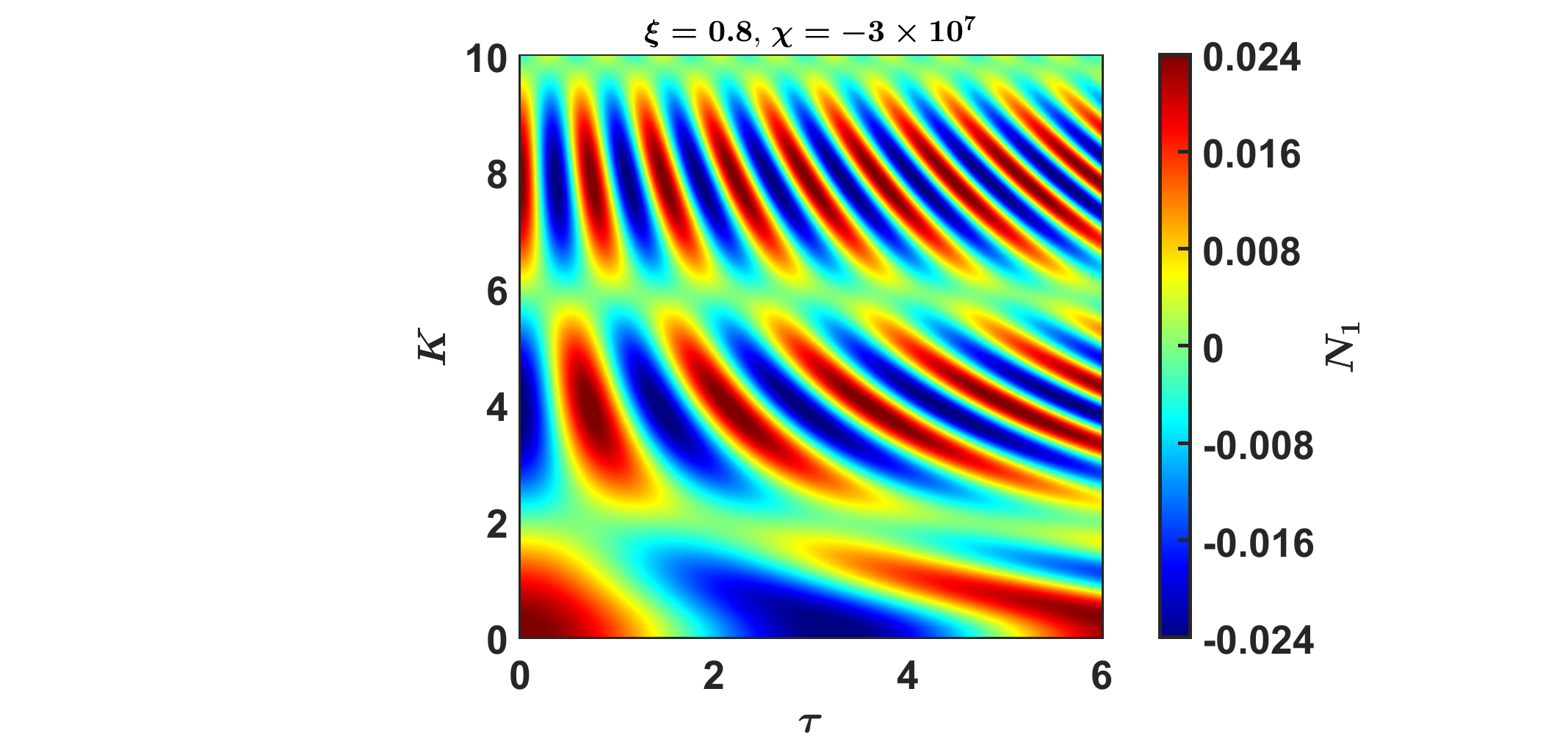}
	\end{tabular}
	\caption{Profiles of the normalized perturbed density $(N_1)$ as a function of the Jeans-normalized angular wavenumber $(K)$ and time $(\tau)$ for different representative values of the EiBI gravity parameter $(\chi)$ in SI units.}
	\label{fig:9}
\end{figure*}

\begin{figure*}
	\centering
	\begin{tabular}{c c c}
		\includegraphics[trim={9cm 0cm 10cm 0cm},clip,width=5cm]{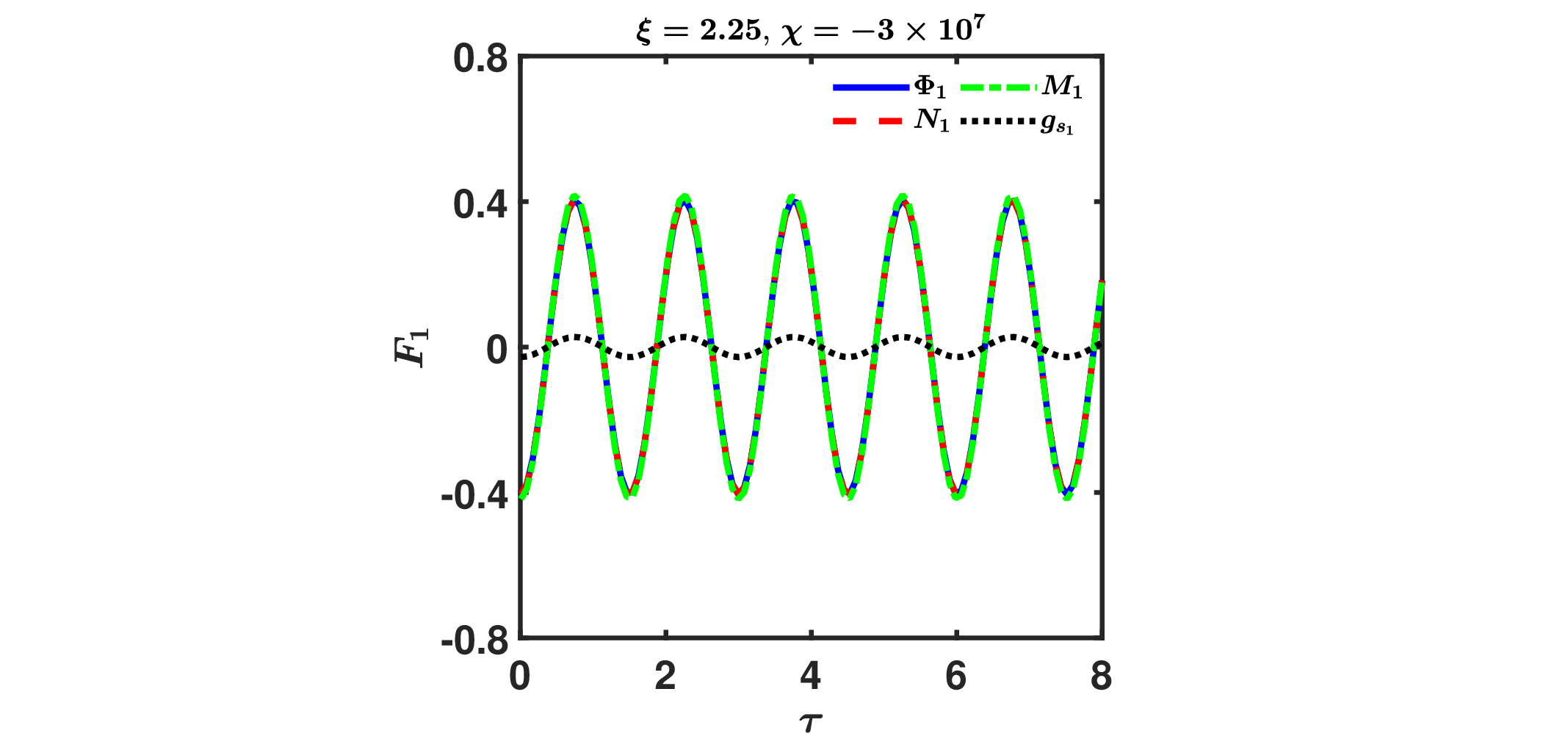}&
		\includegraphics[trim={9cm 0cm 10cm 0cm},clip,width=5cm]{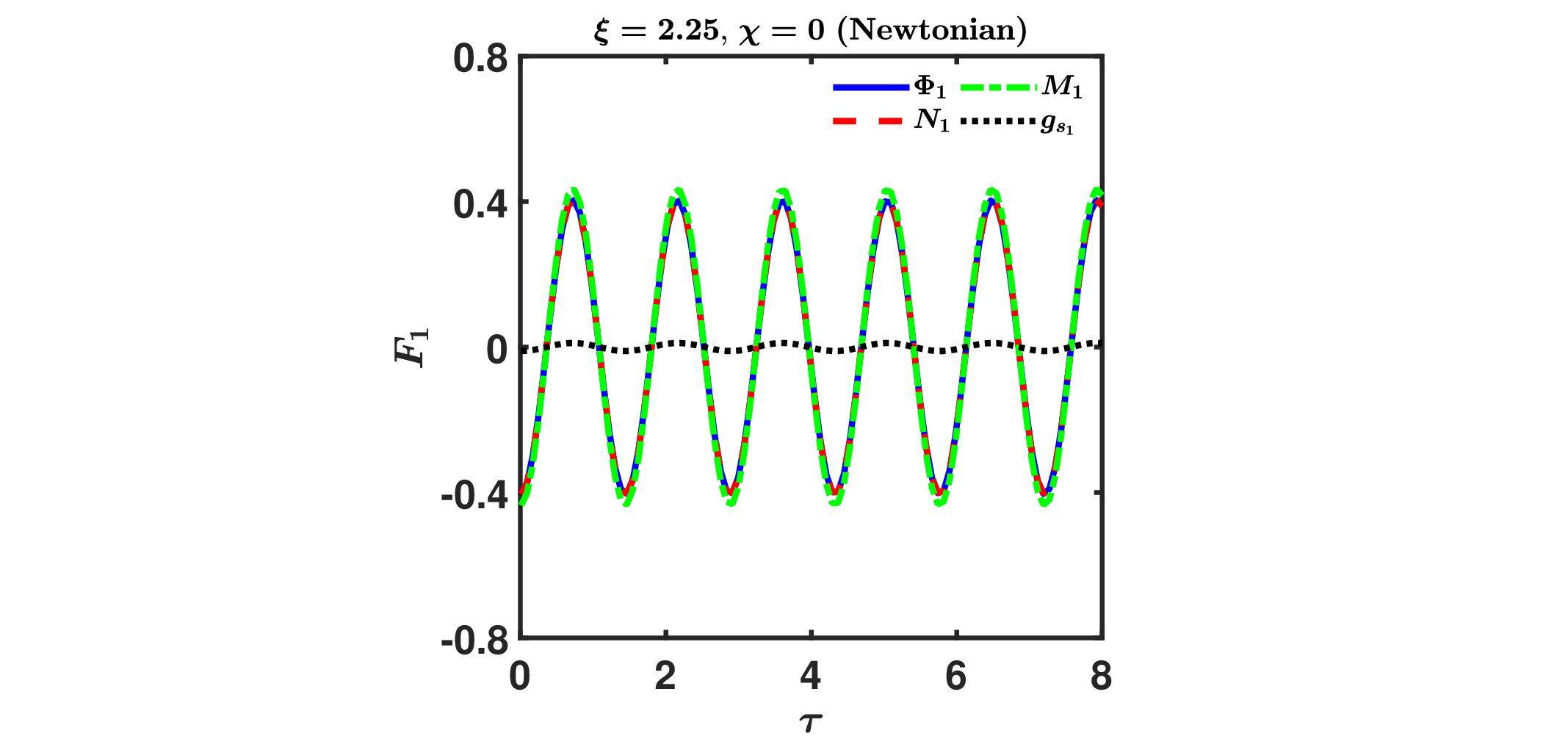}&
		\includegraphics[trim={9cm 0cm 10cm 0cm},clip,width=5cm]{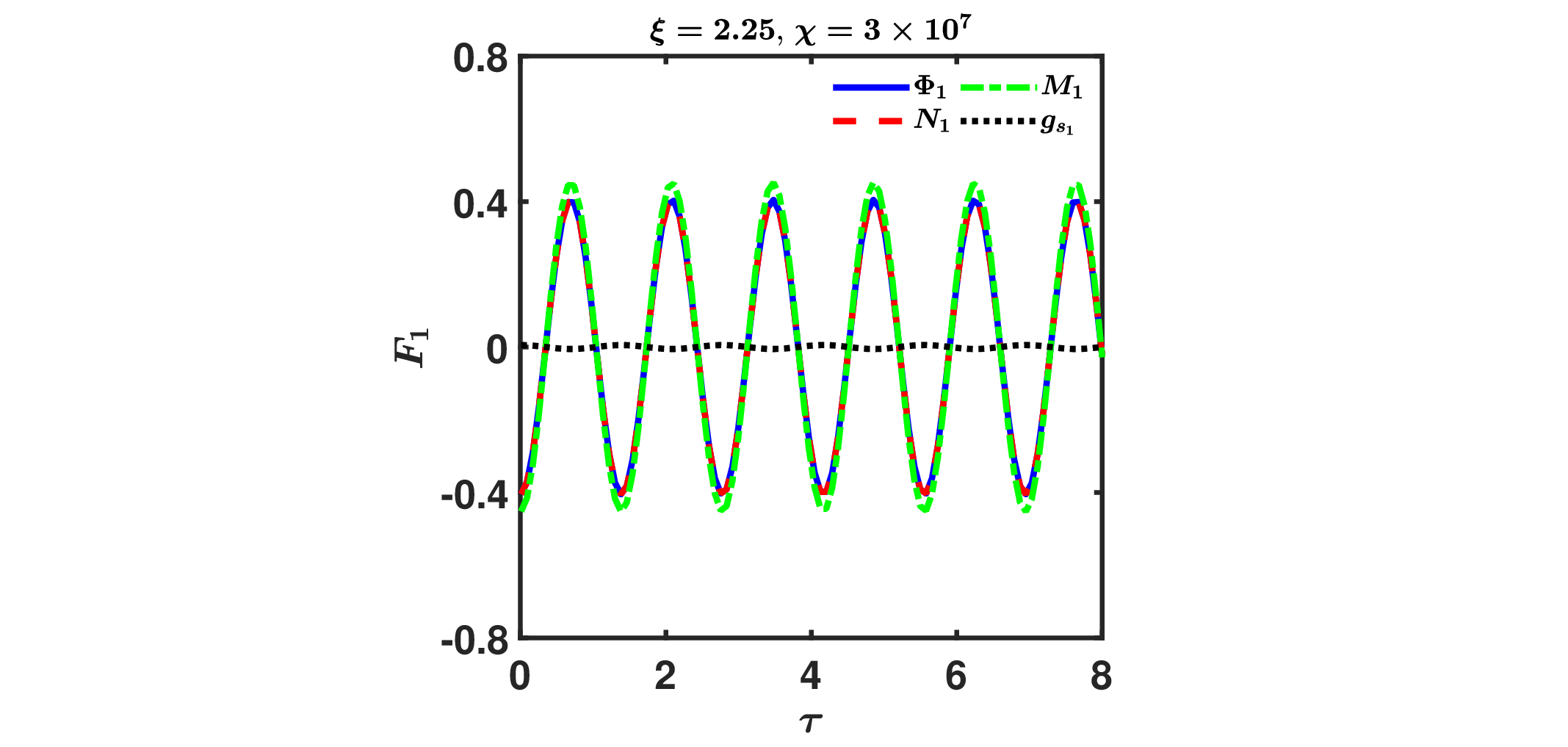}\\
		\includegraphics[trim={9cm 0cm 10cm 0cm},clip,width=5cm]{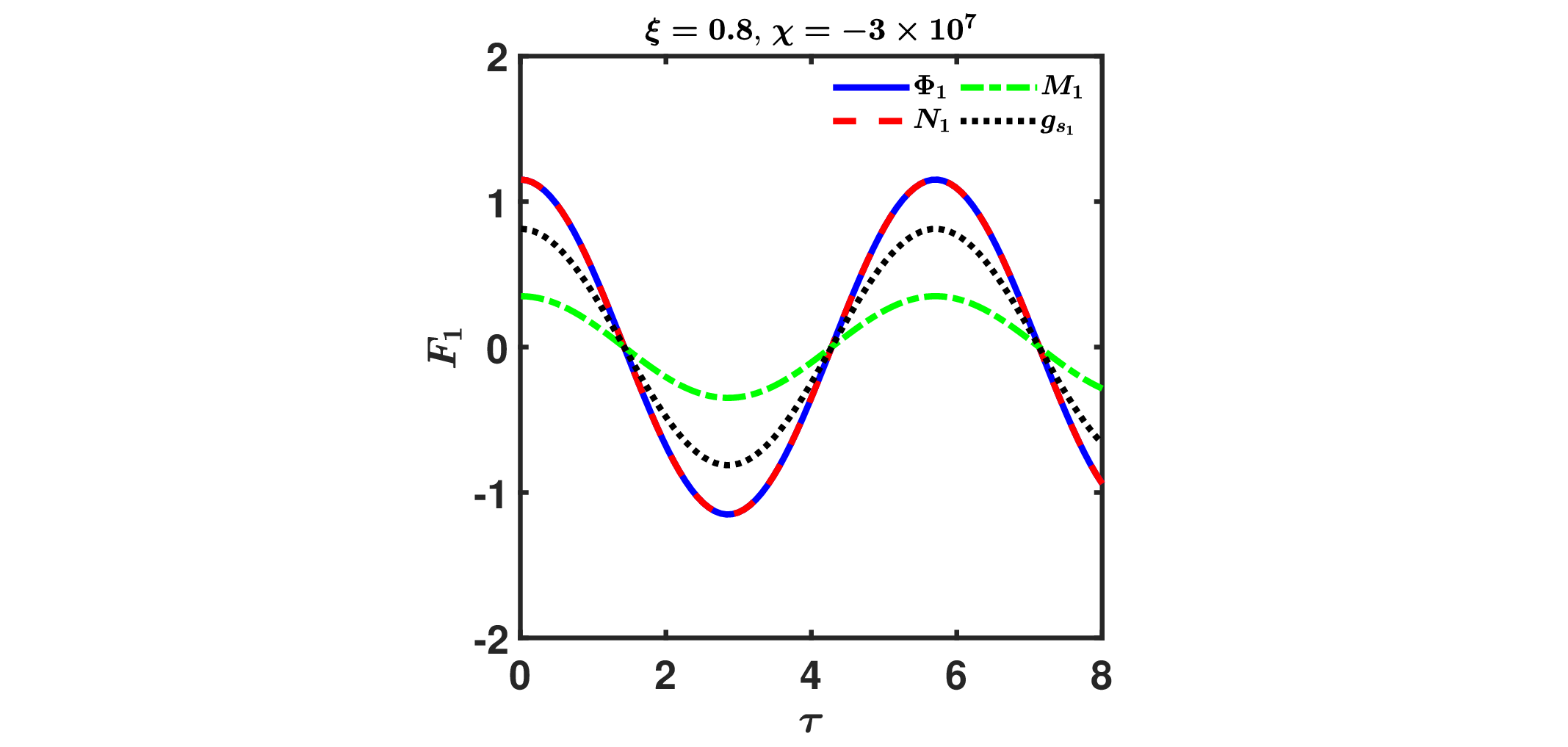}&
		\includegraphics[trim={9cm 0cm 10cm 0cm},clip,width=5cm]{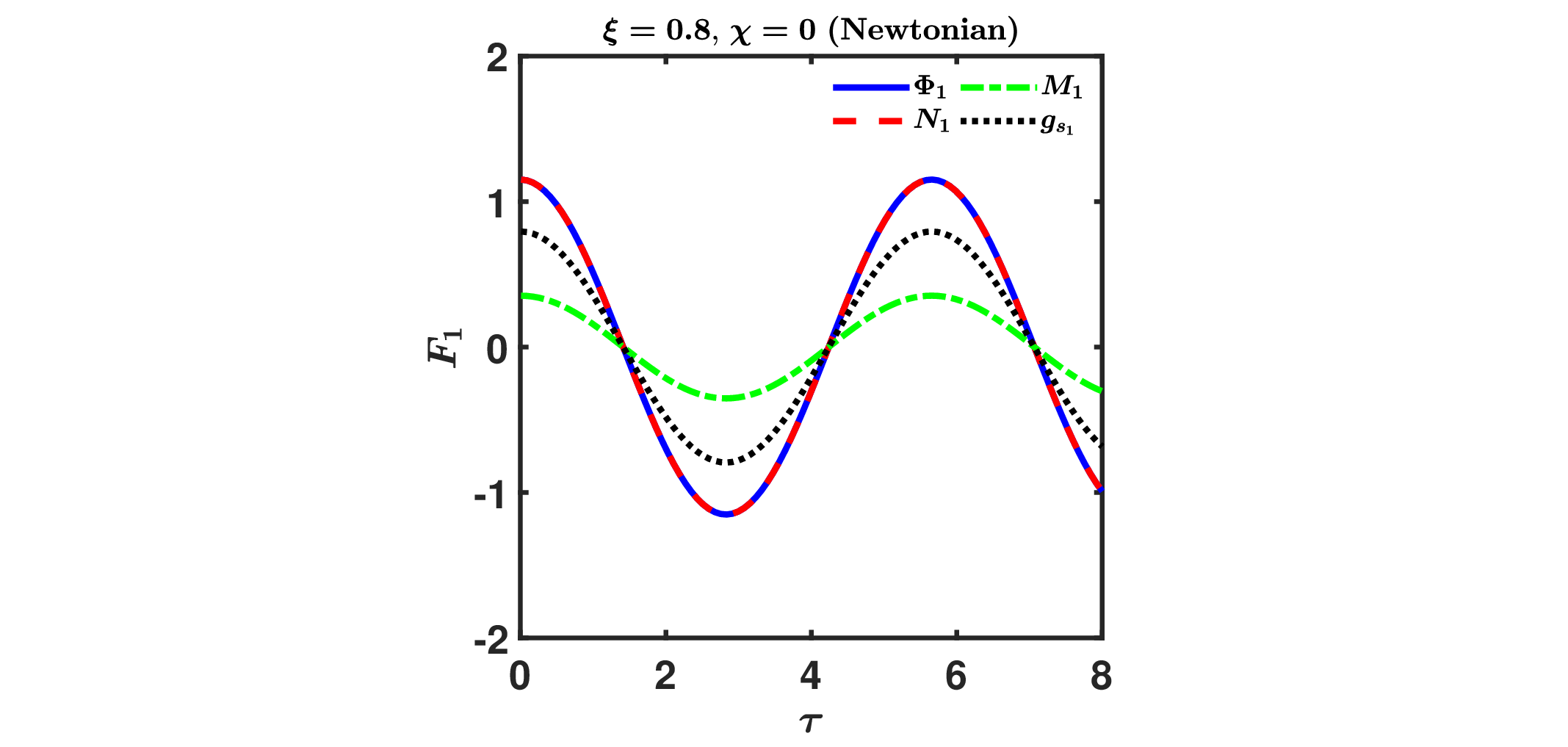}&
		\includegraphics[trim={9cm 0cm 10cm 0cm},clip,width=5cm]{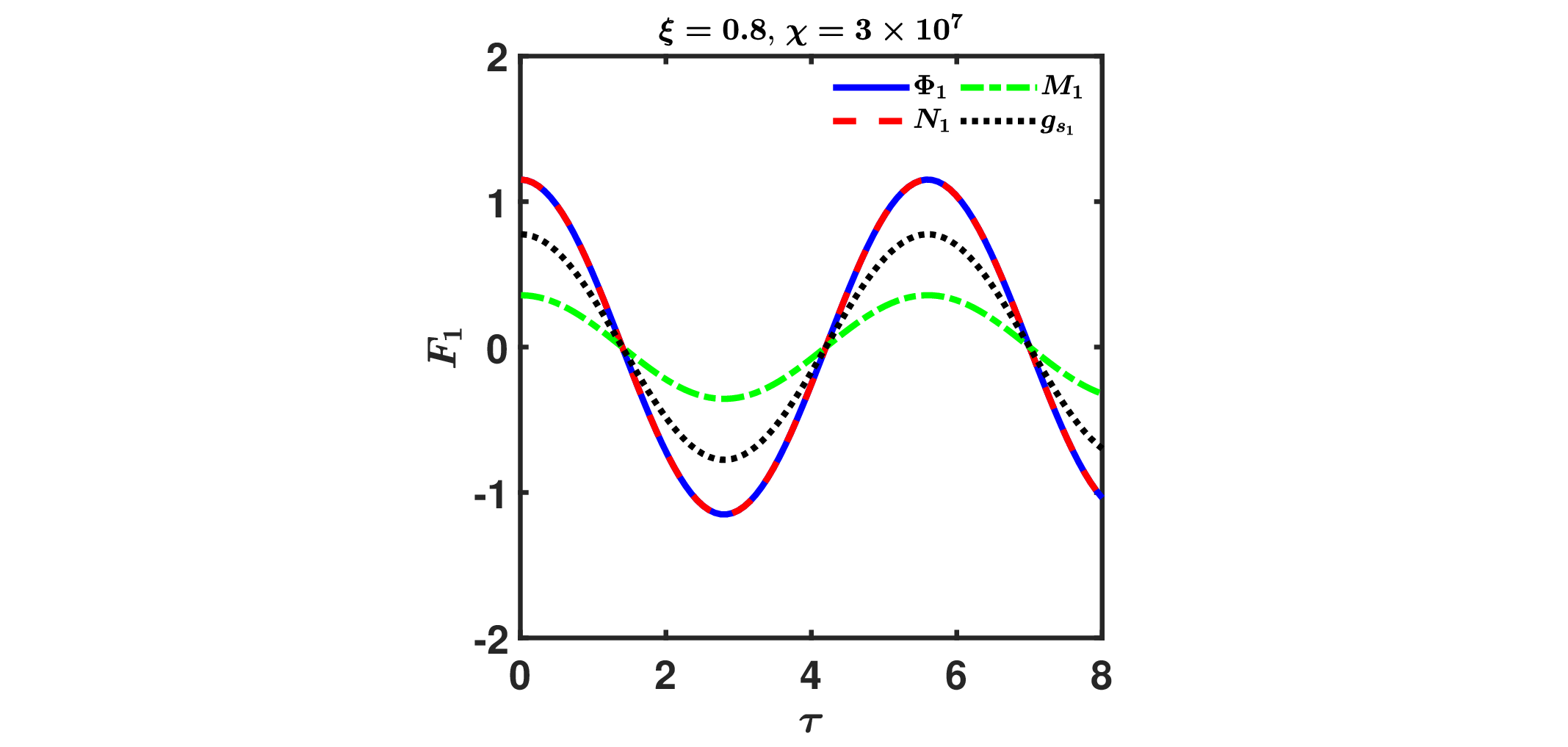}
	\end{tabular}
	\caption{Profiles of the temporal evolution of normalized solar fluctuation variables ($\phi_1$, $M_1$, $N_1$, and ${g_s}_1$) as functions of the Jeans-normalized time $(\tau)$ for different representative values of the EiBI gravity parameter $(\chi)$ in SI units. The Jeans-normalized angular wavenumber is fixed at $K=4$ in the surface region $(\xi=2.25)$ and at $K=0.5$ in the deep interior $(\xi=0.8)$.}
	\label{fig:10}
\end{figure*}

\begin{figure*}
	\centering
	\begin{tabular}{c c}
		\includegraphics[trim={5cm 0cm 5cm 0cm},clip,width=8cm]{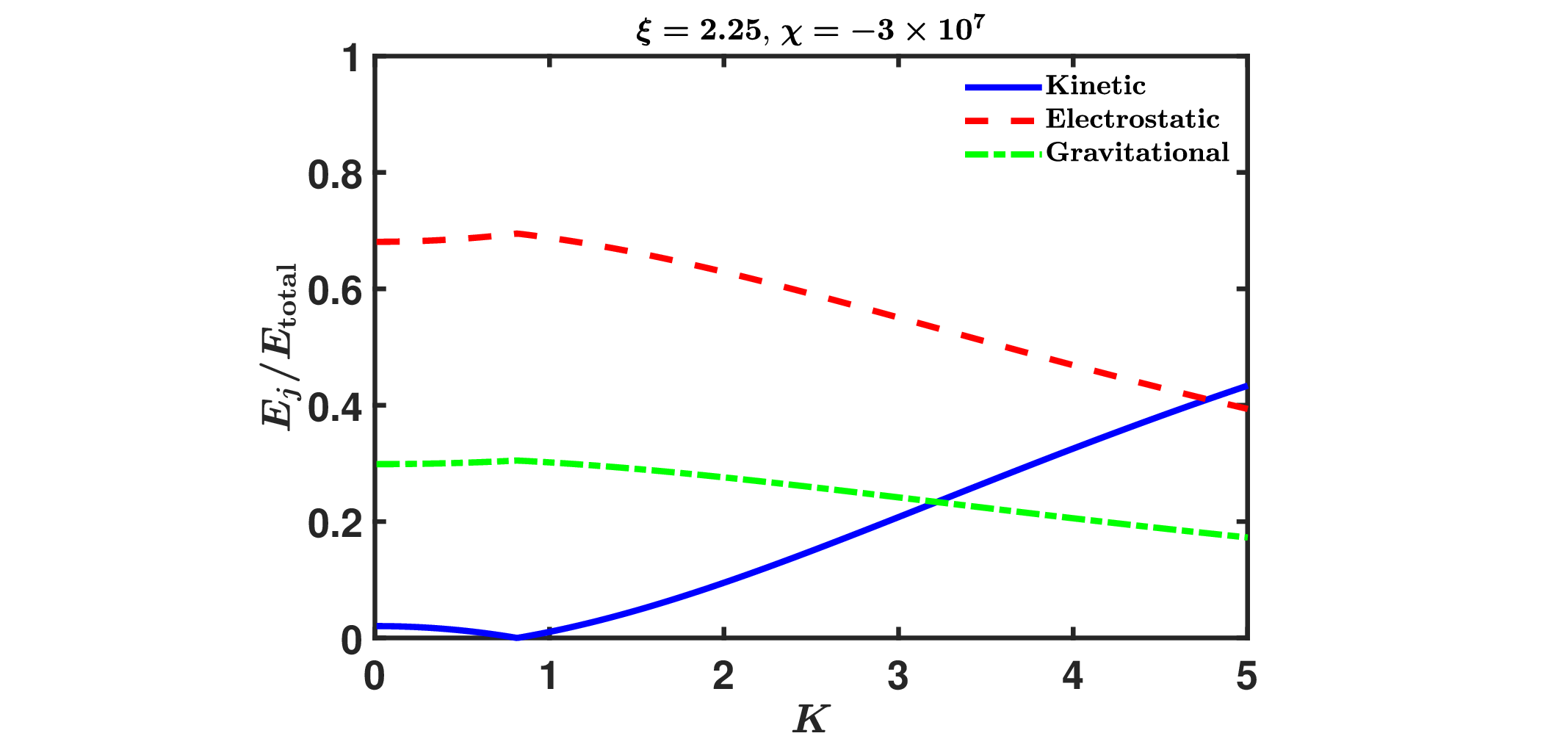}&
		\includegraphics[trim={5cm 0cm 5cm 0cm},clip,width=8cm]{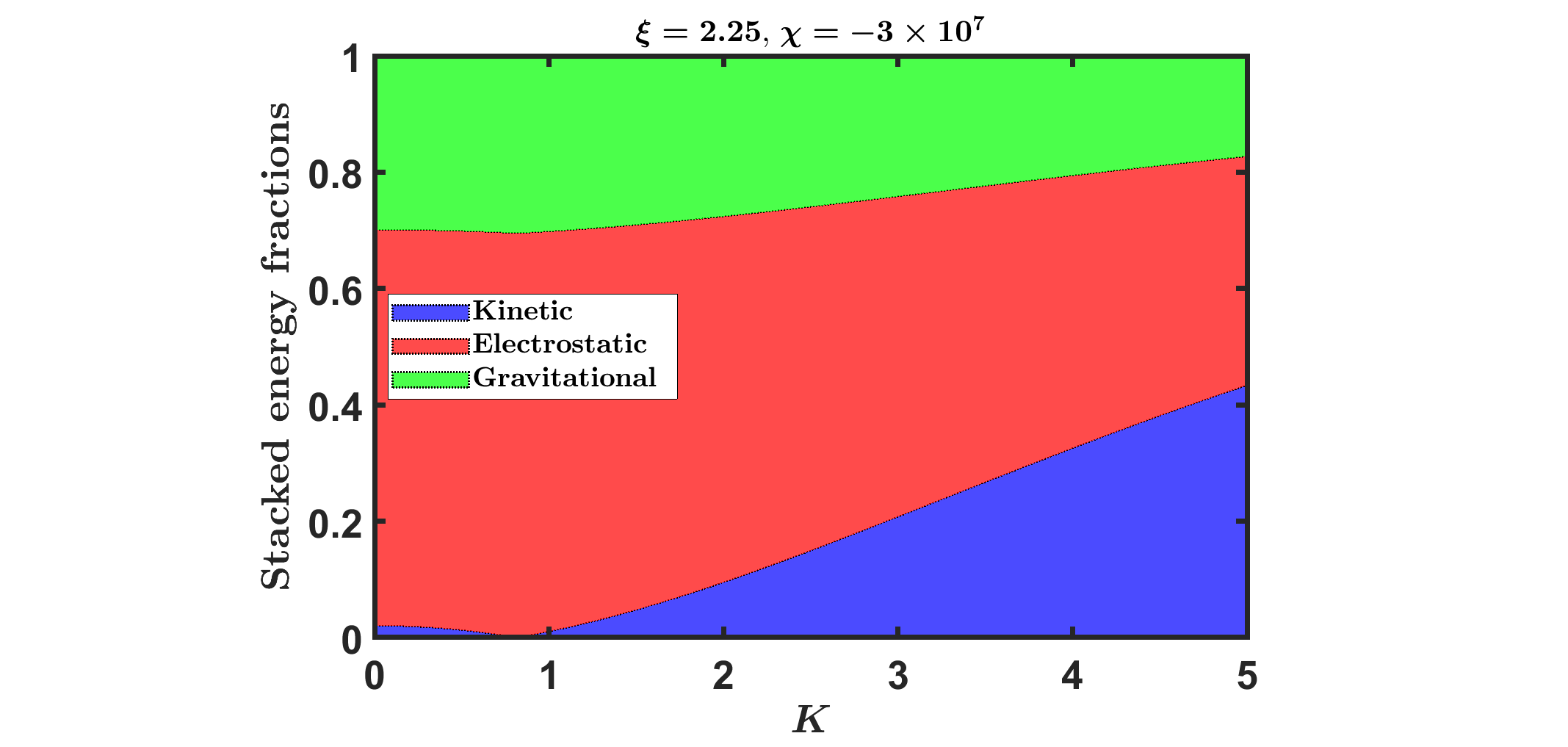}\\
		\includegraphics[trim={5cm 0cm 5cm 0cm},clip,width=8cm]{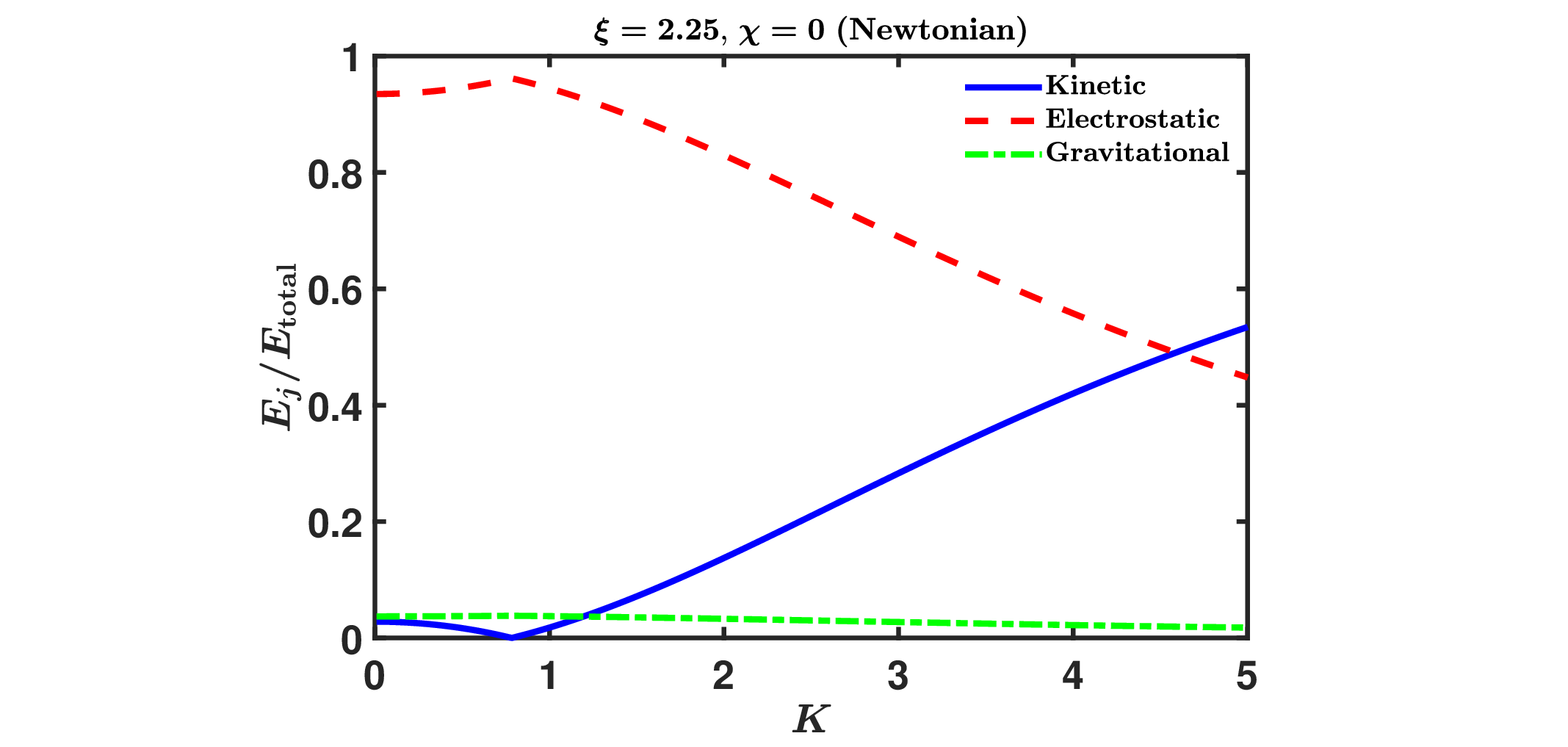}&
		\includegraphics[trim={5cm 0cm 5cm 0cm},clip,width=8cm]{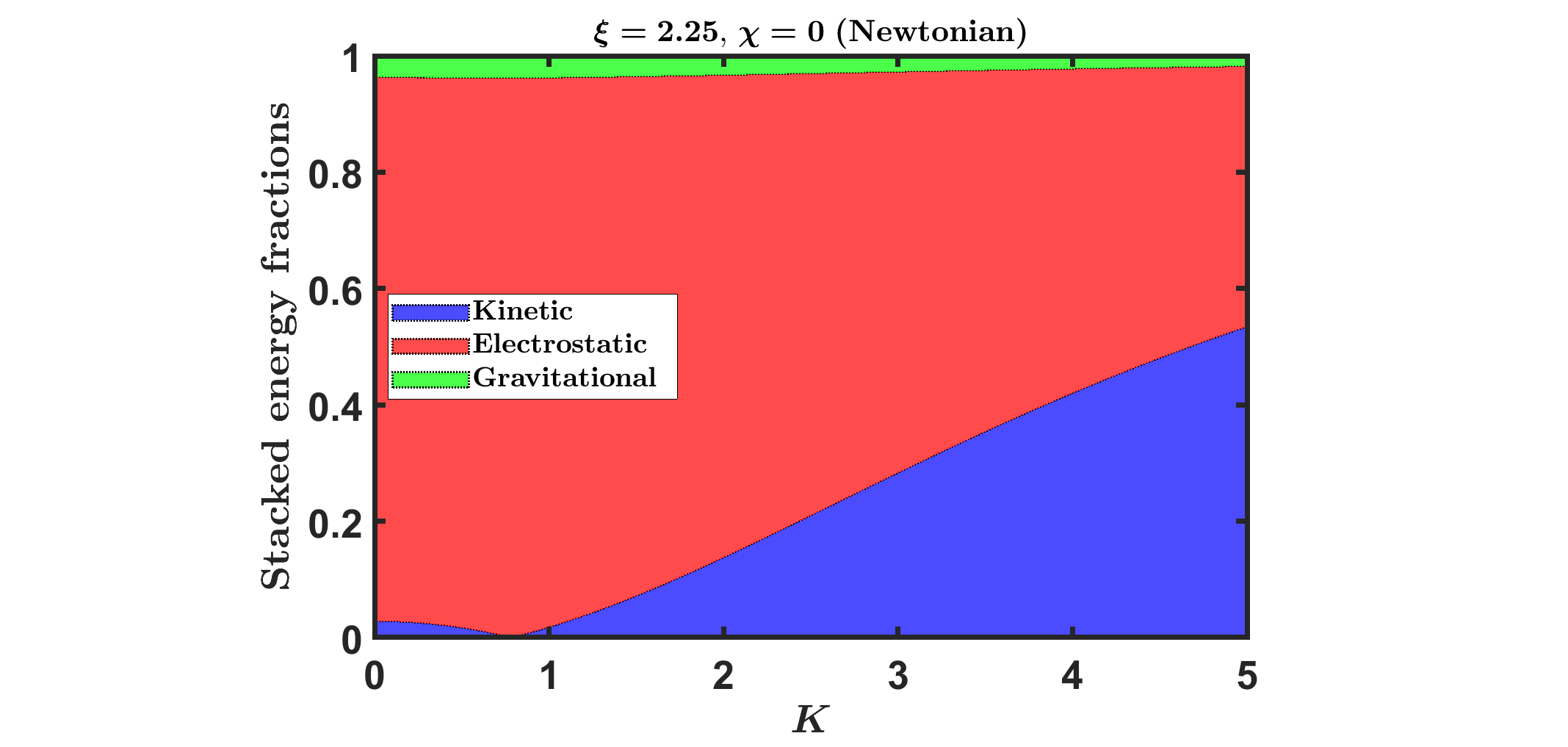}\\
		\includegraphics[trim={5cm 0cm 5cm 0cm},clip,width=8cm]{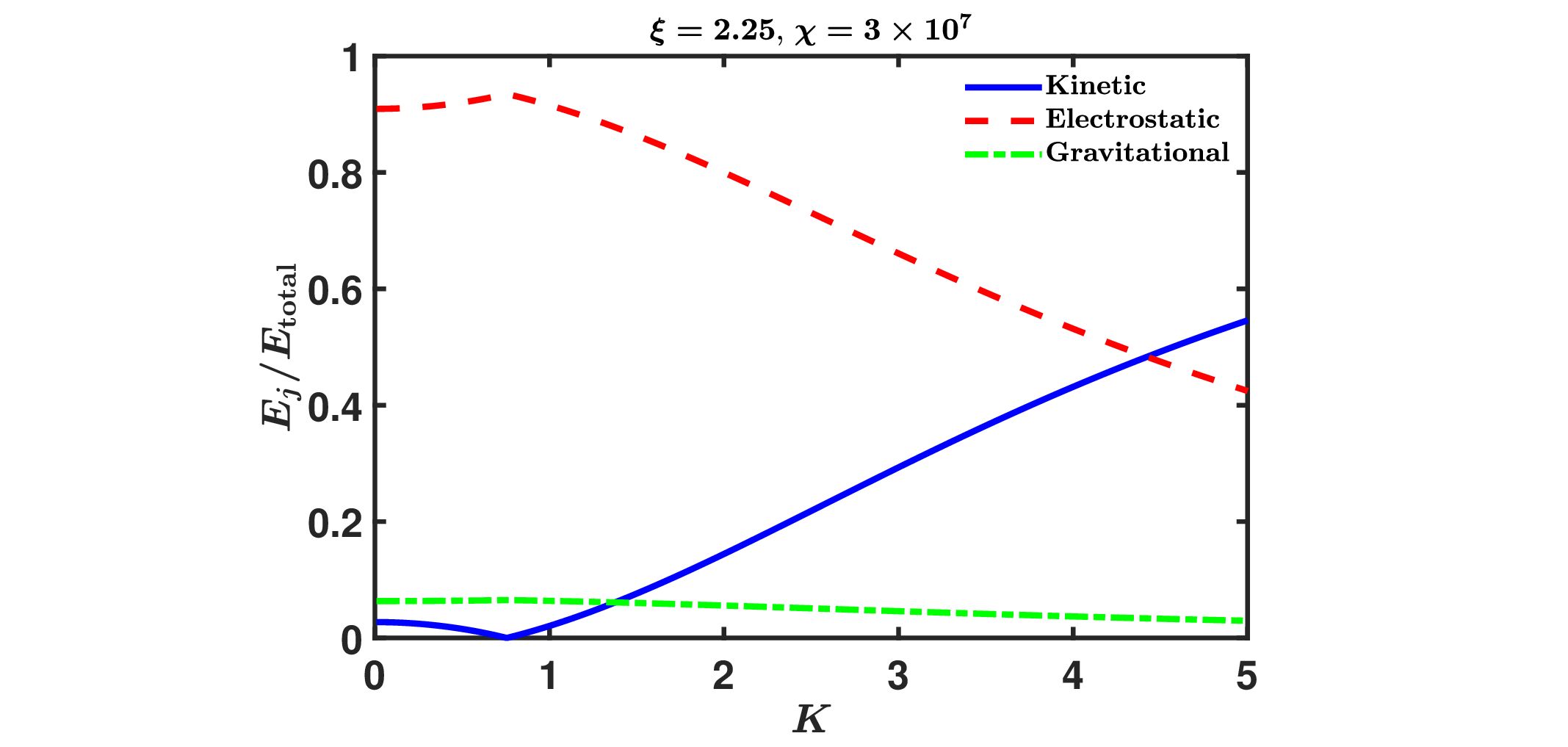}&
		\includegraphics[trim={5cm 0cm 5cm 0cm},clip,width=8cm]{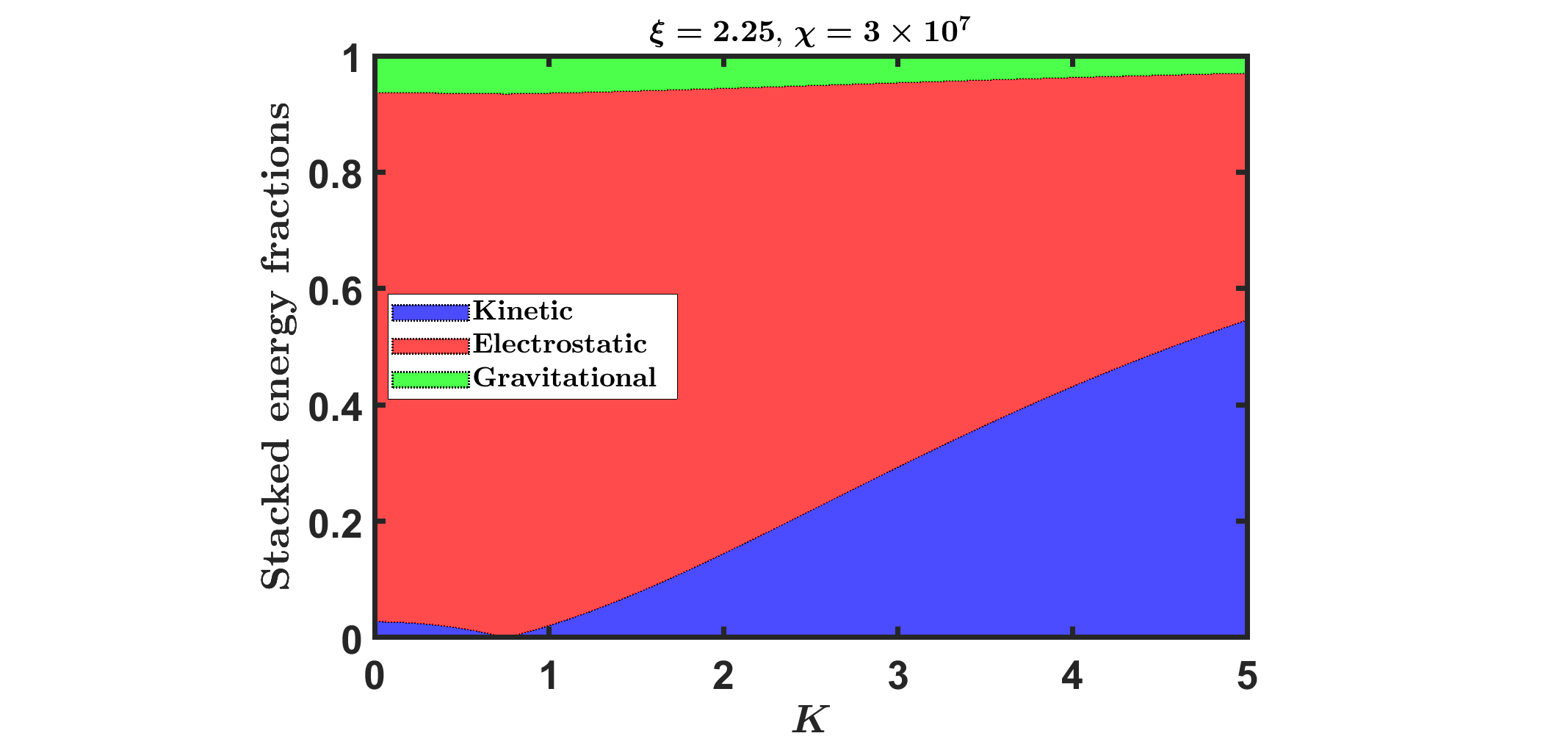}
	\end{tabular}
	\caption{Profiles of the normalized energy fractions --- kinetic, electrostatic, and gravitational --- as functions of the Jeans-normalized angular wavenumber $(K)$ at the solar surface $(\xi=2.25)$ for different representative values of the EiBI gravity parameter $(\chi)$ in SI units. The left column shows individual contributions, while the right column presents stacked energy fractions for direct comparison.}
	\label{fig:11}
\end{figure*}

We numerically analyse the dispersion characteristics to graphically explore the EiBI gravity--modified solar plasma stability dynamics. A sequence of colour-spectral profiles (Figures~\ref{fig:3}--\ref{fig:11}) illustrate diversified features of waves and oscillations, and the diverse wave and oscillatory behaviours arising from the system. To highlight the role of spatial scales, the parametric variations are examined in two representative solar regimes: the radiative interior ($\xi=0.8,\, r=0.36R_{\odot}$) and the solar surface ($\xi=2.25,\, r=R_{\odot}=6.96 \times 10^8\,\mathrm{m}$).\par
Figure~\ref{fig:3} displays the discriminant $D=A_1^2-4A_0$ of the quadratic dispersion relation as a function of Jeans-normalized angular wavenumber $K$, systematically illustrating how variations in both the EiBI gravity parameter $\chi$ and the relative polytropic sound speed $\beta$ modulate the stability behaviour of solar plasma oscillations. Each family of curves, plotted for either different values of $\chi$ (at $\beta=\Gamma$) or different values of $\beta$ (at $\chi=3\times10^7$ m$^5$\,kg$^{-1}$\,s$^{-2}$), is shown for typical surface ($\xi=2.25$) and deep interior ($\xi=0.8$) regions of the Sun. The discriminant $D$ formally classifies the mathematical regimes as: $D>0$ (purely stable oscillatory), $D<0$ (growing/damped perturbation), and $D=0$ (transition or critical point). However, within the $D<0$ regime, only the decaying (damped) branch is physically realized for the admissible Root 1 (see Figure~\ref{fig:2}). Increasing $\beta$ or adopting a more positive $\chi$ shifts the discriminant upward, expanding the stable domain and moving the transition threshold to lower $K$. Conversely, negative $\chi$ or reduced $\beta$ shrinks the stability domain by 18--23\%, demonstrating the bidirectional capacity for both destabilization and stabilization within the EiBI framework. Such sensitivity is most pronounced in the interior, underscoring the crucial roles of non-Newtonian gravity and thermodynamics. These behaviours collectively demonstrate how both the gravity modifications (through $\chi$) and enhanced pressure support (through $\beta$) are crucial in shaping the stability boundaries of solar plasma oscillations, with the discriminant plots serving as a clear graphical diagnostic for predicting and interpreting stable and unstable wave behaviour in the Sun.\par

In Figure~\ref{fig:4}, we show how the Jeans-normalized oscillation frequency ($\Omega_r$) and growth rate ($\Omega_i$) vary as functions of the Jeans-normalized angular wavenumber ($K$) and radial coordinate ($\xi$), with fixed values of $\chi$, $\beta$, and $\eta$. The frequency peaks near the solar core ($\xi=0$) and rises sharply with $K$. Both short-wavelength (\textit{p}-mode) and long-wavelength (\textit{g}-mode) oscillations are strong here. However, near the surface, long-wavelength modes show almost no propagation. This reflects the trapping of \textit{g}-modes near the radiative zone. The growth rate approaches zero near the core and becomes more negative outward, indicating damping. Hence, the system shows an overall stability. This is because of the viscosity effect considered here. Thus, together, these profiles reveal that EiBI gravity accentuates frequency localization and energetics in the interior, while the combined dependencies on $K$ and $\xi$ regulate both the oscillatory and stability properties of solar plasma modes.\par

Figure~\ref{fig:5} demonstrates the variation of the Jeans-normalized oscillation frequency ($\Omega_r$, left panels) and growth rate ($\Omega_i$, right panels) with the Jeans-normalized angular wavenumber ($K$) for different representative values of the EiBI gravity parameter ($\chi$). At both the surface ($\xi=2.25$, top row) and in the interior ($\xi=0.8$, bottom row), positive $\chi$ elevates oscillation frequencies by approximately 4--10\%, reflecting plasma stiffening; negative $\chi$ suppresses frequencies by up to 4--10\% relative to Newtonian gravity. In the \textit{g}-mode regime, positive $\chi$ decreases damping rate by 65\%; while damping rate is increased by 30--38\% with negative $\chi$, confirming robust stability. Damping is strongest for low-$K$ \textit{g}-modes near the surface, but vanishes deep in the interior. The plots elaborate that at the solar surface, particularly the high-$K$ \textit{p}-modes show pure stable oscillatory features and the low-$K$ \textit{g}-modes show only damping propensity without any propagatory nature. Conversely, in the deep interior, both the short and long waves show highly oscillatory nature with very negligible damping rate. The EiBI gravity directly regulates both the energetic and stability properties of solar plasma waves, enhancing oscillation frequencies without introducing instability under the explored conditions.\par

Figure~\ref{fig:6} delineates the relative modification of the Jeans-normalized oscillation properties introduced by the EiBI gravity. The relative oscillation frequency $(\Delta_r=\Omega_r^{\mathrm{EiBI}}/\Omega_r^{\mathrm{Newtonian}})$ and relative growth rate $(\Delta_i=\Omega_i^{\mathrm{EiBI}}/\Omega_i^{\mathrm{Newtonian}})$ are plotted as functions of the Jeans-normalized angular wavenumber $(K)$ for several representative values of the EiBI gravity parameter $(\chi)$. The profiles clearly demonstrate that EiBI corrections produce systematic and measurable deviations from the Newtonian values, with both the magnitude and sign of the shift depending sensitively on $\chi$ and $K$. The dashed line at $\Delta=1$ marks the Newtonian $(\chi=0)$ baseline, highlighting the relative enhancement or suppression of the oscillation frequency and growth rate across the full wavenumber spectrum.

In Figure~\ref{fig:7}, we present the Jeans-normalized oscillation frequency ($\Omega_r$, left panels) and growth rate ($\Omega_i$, right panels) as functions of the Jeans-normalized angular wavenumber ($K$) for different values of the relative polytropic sound speed ($\beta$), with the EiBI gravity parameter held fixed. At both the surface ($\xi=2.25$, top row) and interior ($\xi=0.8$, bottom row) regions, increasing $\beta$ raises frequency by about 20--55\% at fixed $K$, reflecting a stiffer and more responsive plasma due to enhanced pressure support. Larger $\beta$ reduces surface damping by up to 70\% for low-$K$ \textit{g}-modes, but shows no influence on $\Omega_i$ in the interior. These results reveal that higher polytropic sound speed not only strengthens wave propagation but also maintains stability by increasing oscillation frequency without causing instability, particularly in the deep solar interior. The collective behaviour underscores the stabilizing and frequency-amplifying influence of thermal and polytropic effects under the EiBI gravity in both surface and core regions.\par

Figure~\ref{fig:8} illustrates the variation of Jeans-normalized phase speed $V_p=\Omega_r/K$ of plasma oscillations as a function of Jeans-normalized angular wavenumber ($K$), systematically comparing the influence of EiBI gravity parameter ($\chi$, left panels) and relative polytropic sound speed ($\beta$, right panels). At the solar surface ($\xi=2.25$, top row), the low-$K$ \textit{g}-modes cannot propagate ($V_p=0$) up to a threshold point beyond which $V_p$ rises sharply and then saturates with $K$. Meanwhile, in the deep interior ($\xi=0.8$, bottom row), $V_p$ starts at a higher value and decreases monotonically with $K$, reflecting the fact that the \textit{g}-modes can only propagate in the solar interior. Positive $\chi$ enhances the asymptotic phase speed up to 10\%, whereas negative $\chi$ reduces it by 4--10\% relative to the Newtonian case. Likewise, larger $\beta$ systematically increases $V_p$ by up to 55\%, consistent with stronger thermal support. These results demonstrate how nonlinear gravity corrections and stiffer plasma equations of state amplify collective wave velocities, shaping both surface and interior dynamics of the Sun.\par

In Figure~\ref{fig:9}, we elucidate the temporal evolution of the normalized perturbed density ($N_1$) as a function of the Jeans-normalized angular wavenumber ($K$) and time ($\tau$) for representative positive and negative values of the EiBI gravity parameter ($\chi$), at both the surface ($\xi=2.25$, top row) and deep interior ($\xi=0.8$, bottom row). At the surface, the perturbations remain weak, with amplitudes of order $10^{-3}$, and exhibit periodic oscillatory patterns in both $K$ and $\tau$, reflecting predominantly wave-like behaviour modulated by the modified gravity contribution. The effect of $\chi$-sign is visible through subtle phase shifts in the oscillation bands, with positive $\chi$ slightly amplifying the density fluctuations compared to negative $\chi$. In contrast, within the deep interior, the perturbations grow substantially stronger (about an order of magnitude higher), and the oscillatory bands become sharper and more densely packed, signifying enhanced wave activity and stronger mode coupling under high gravitational confinement. The symmetry between positive and negative $\chi$ remains, but the interior dynamics show markedly greater sensitivity to the modified gravity corrections, indicating that EiBI effects act more prominently in regions of higher plasma density. Overall, these density fluctuation profiles reveal how nonlinear gravity corrections, combined with radial stratification, govern the amplitude, periodicity, and localization of collective oscillations in solar plasmas.\par

Figure~\ref{fig:10} exhibits the time evolution of normalized fluctuations in key solar variables---electrostatic potential ($\Phi_1$), velocity perturbation ($M_1$), density perturbation ($N_1$), and self-gravitational acceleration (${g_s}_1$)---as functions of Jeans-normalized time ($\tau$), for typical surface ($\xi=2.25$, top row; $K=4$) and interior ($\xi=0.8$, bottom row; $K=0.5$) regions, across various values of the EiBI gravity parameter ($\chi$) in SI units. At the surface, all variables oscillate nearly in phase and exhibit similar amplitudes (except ${g_s}_1$). The influence of $\chi$ is manifested primarily in subtle phase shifts and amplitude modulation, where positive $\chi$ enhances the oscillatory strength while negative $\chi$ slightly suppresses it relative to the Newtonian case. In contrast, in the deep interior, the oscillations become slower and of significantly larger amplitude, consistent with the excitation of gravity-dominated \textit{g}-modes. Here, the EiBI corrections introduce marked differences in the amplitude of density and gravitational fluctuations, indicating a strong sensitivity of core dynamics to nonlinear gravitational effects. The consistent phase-locking between $\Phi_1$, $M_1$, and $N_1$, along with the lagging response of ${g_s}_1$, highlights the tight coupling of plasma and self-gravity perturbations across both regions. These results confirm that the EiBI gravity systematically modulates the collective oscillatory behaviour of solar plasmas, with surface modes remaining robust and interior modes exhibiting amplified gravity-sensitive fluctuations.\par

\begin{figure*}
	\centering
	\begin{tabular}{c c}
		\includegraphics[trim={5cm 0cm 5cm 0cm},clip,width=8cm]{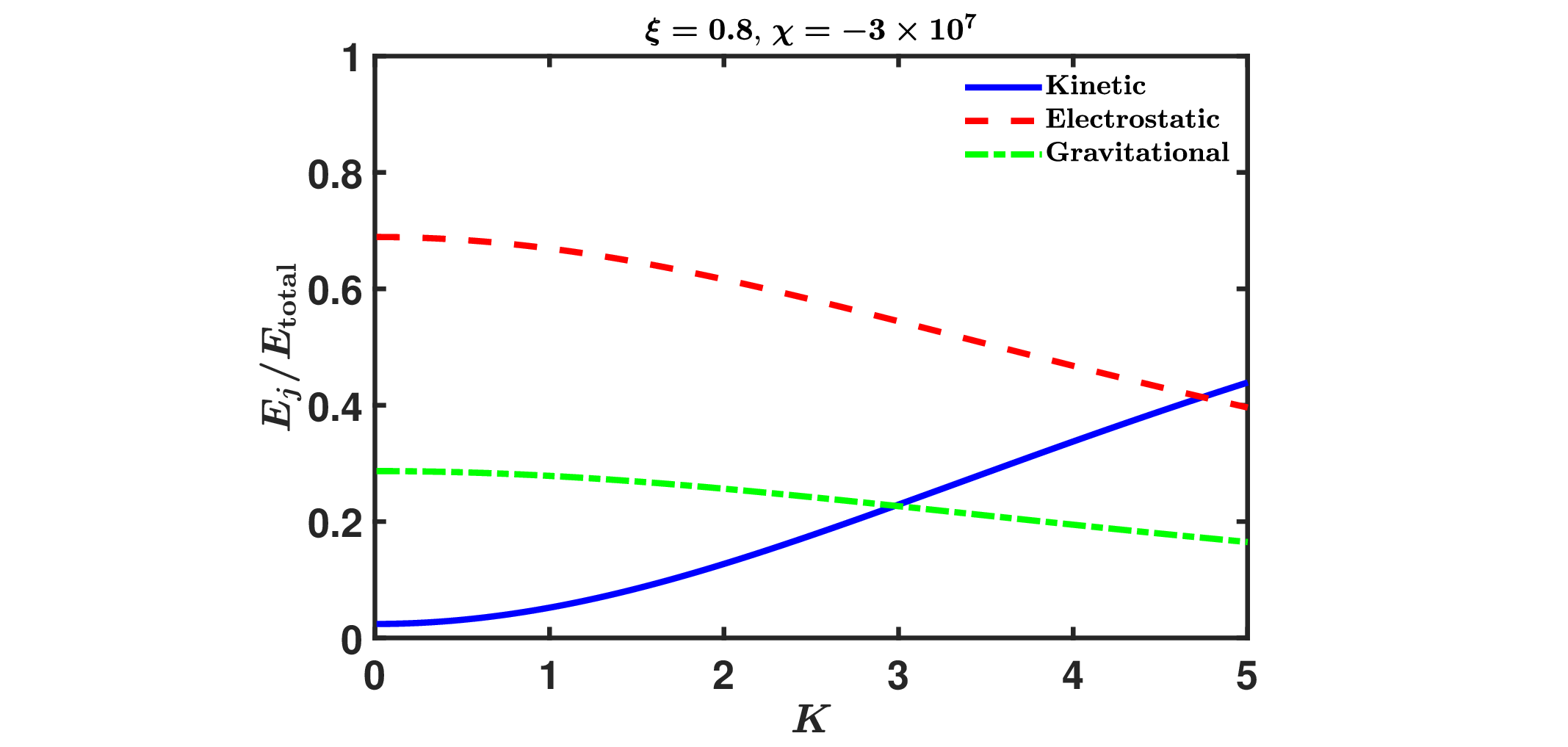}&
		\includegraphics[trim={5cm 0cm 5cm 0cm},clip,width=8cm]{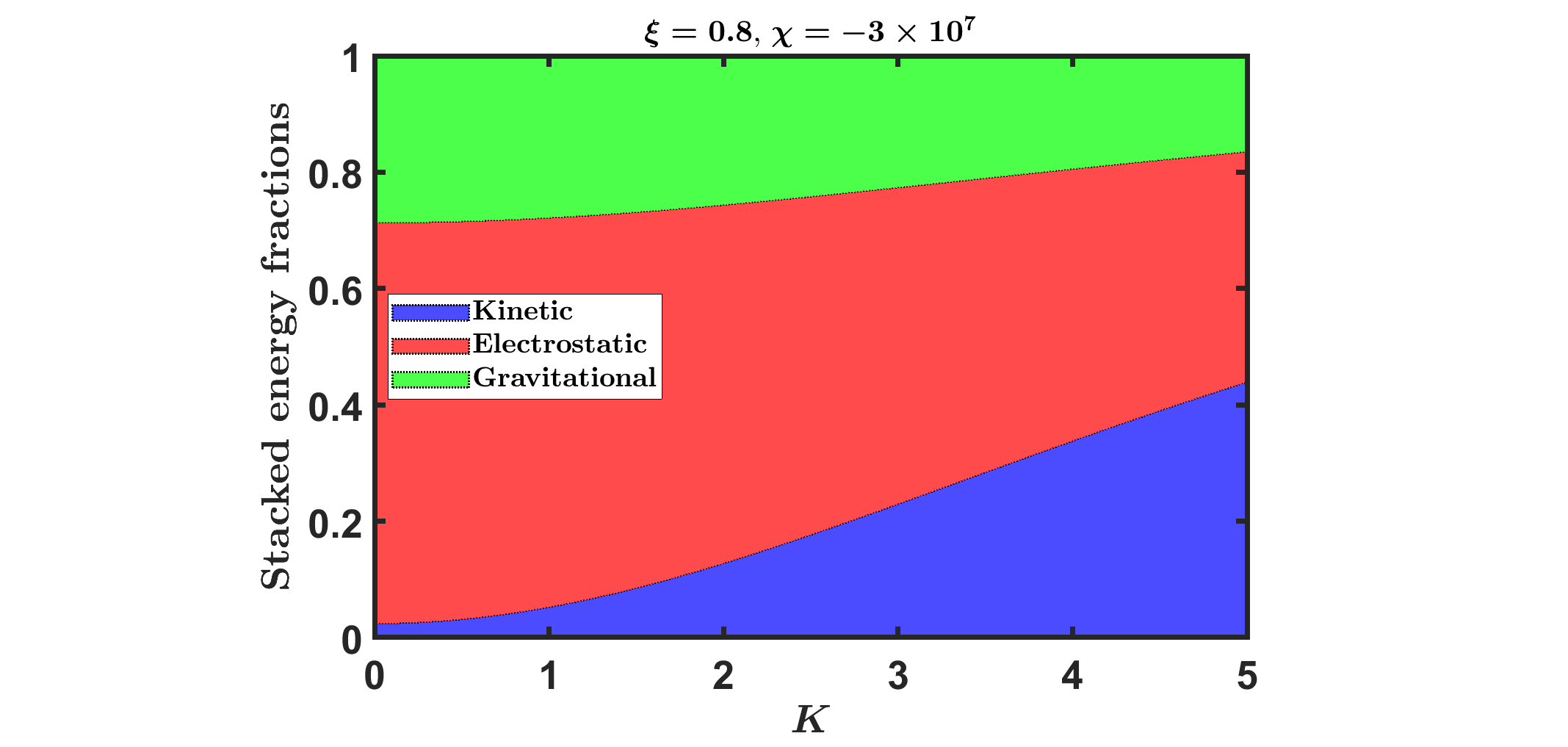}\\
		\includegraphics[trim={5cm 0cm 5cm 0cm},clip,width=8cm]{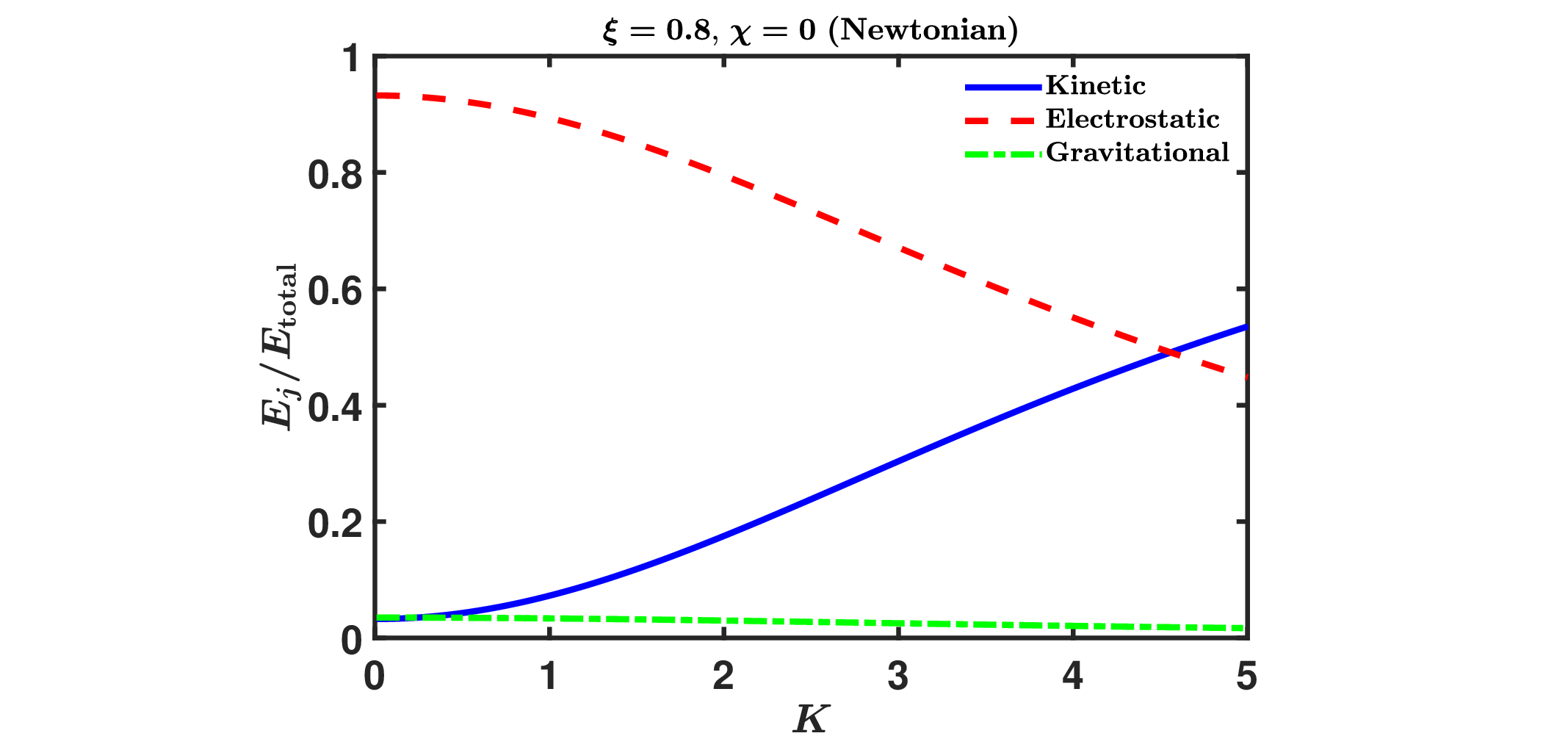}&
		\includegraphics[trim={5cm 0cm 5cm 0cm},clip,width=8cm]{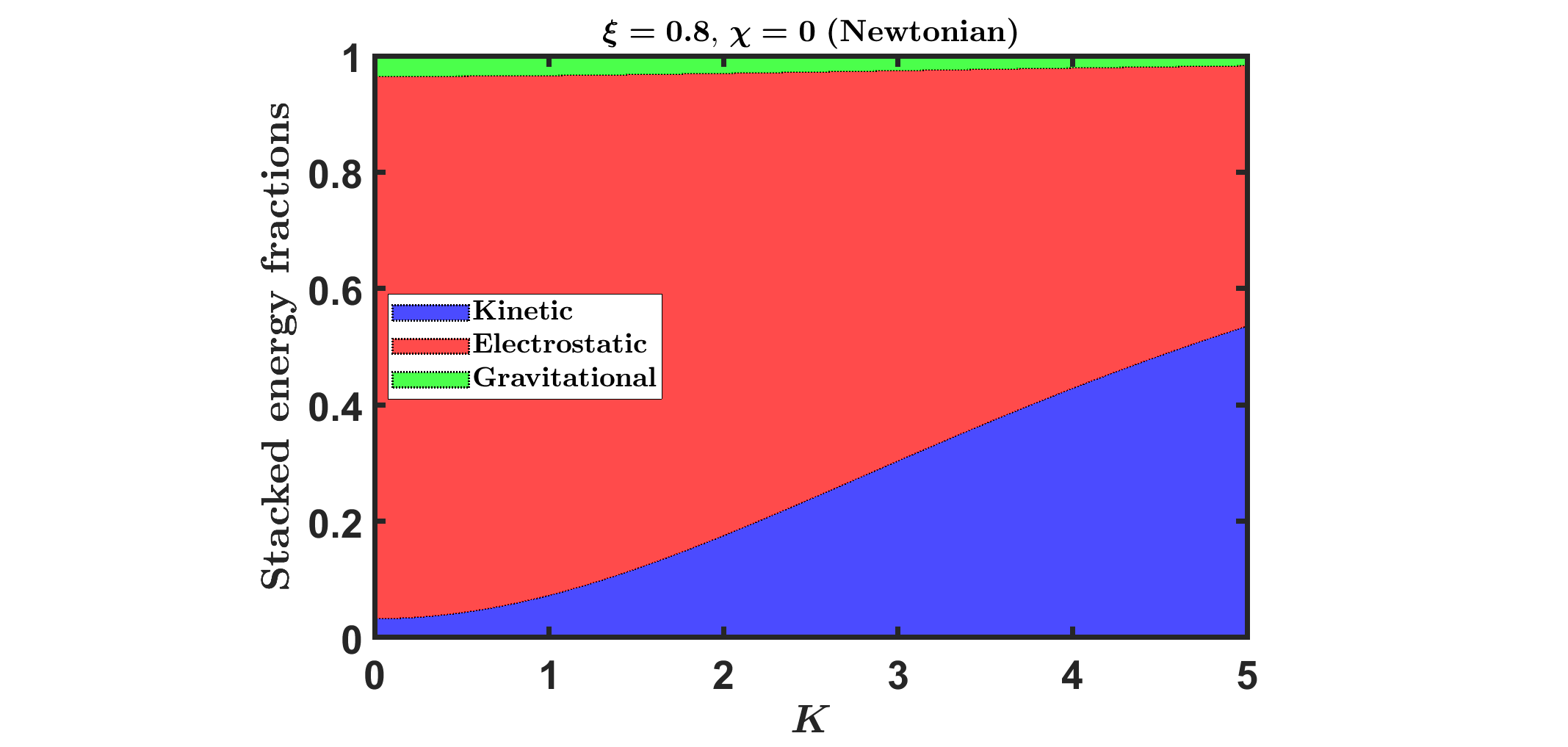}\\
		\includegraphics[trim={5cm 0cm 5cm 0cm},clip,width=8cm]{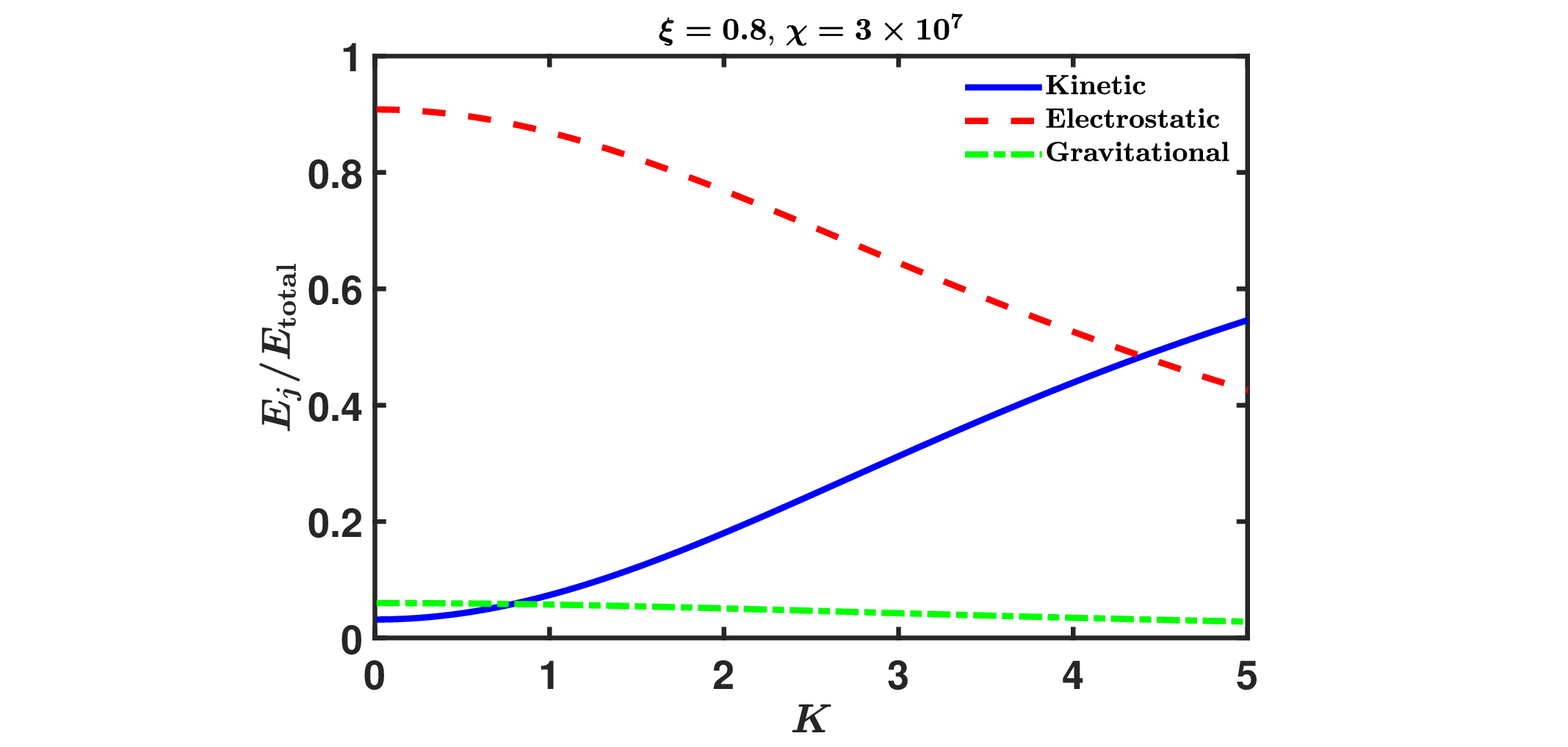}&
		\includegraphics[trim={5cm 0cm 5cm 0cm},clip,width=8cm]{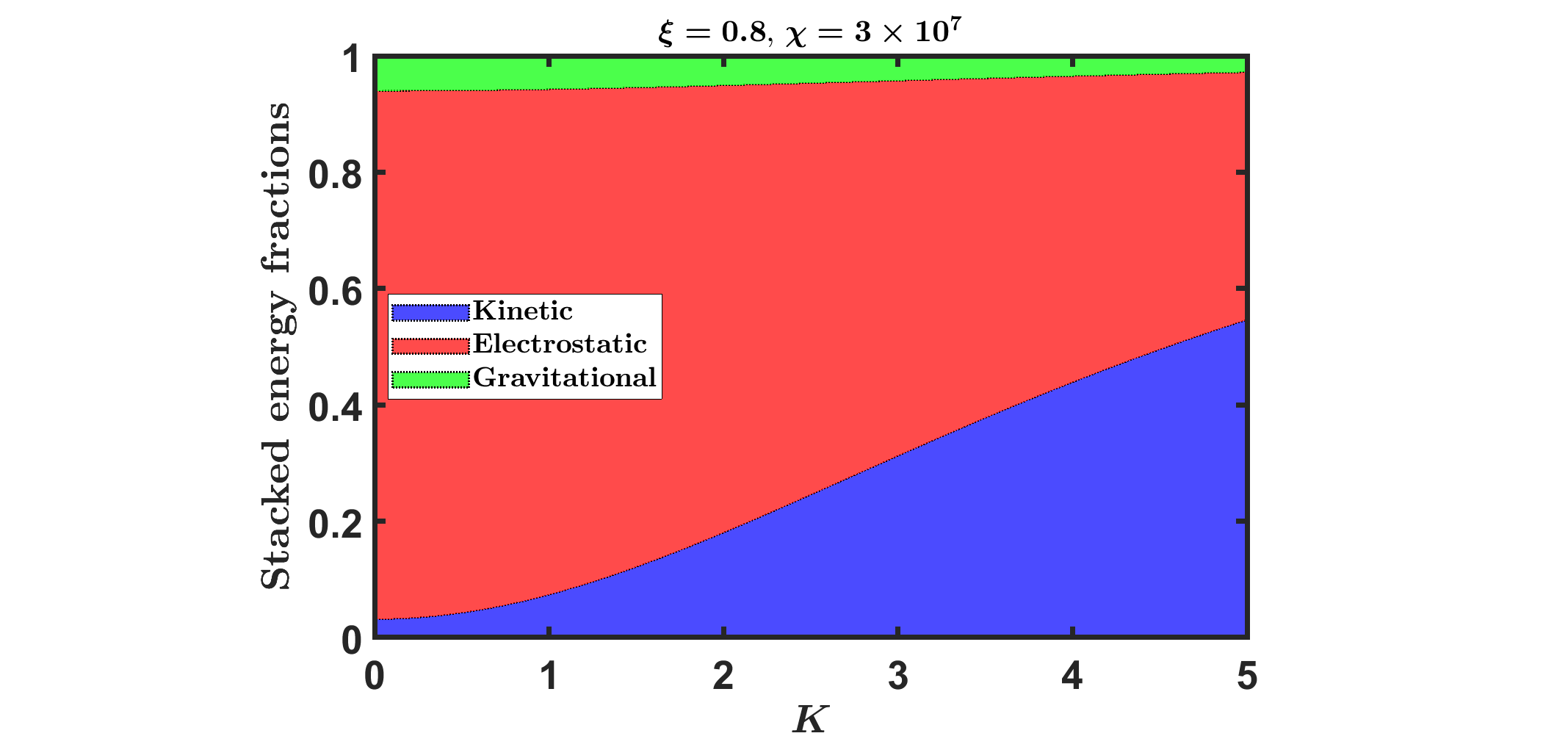}
	\end{tabular}
	\caption{Profiles of the normalized energy fractions --- kinetic, electrostatic, and gravitational --- as functions of the Jeans-normalized angular wavenumber $(K)$ at the solar interior $(\xi=0.8)$ for different representative values of the EiBI gravity parameter $(\chi)$ in SI units. The left column shows individual contributions, while the right column presents stacked energy fractions for direct comparison.}
	\label{fig:12}
\end{figure*}

Figures~\ref{fig:11}--\ref{fig:12} depict the fractional contributions of kinetic, electrostatic, and gravitational energies to the total energy budget as functions of the Jeans-normalized angular wavenumber $(K)$, for different representative values of the EiBI gravity parameter $(\chi)$ at the solar surface ($\xi=2.25$, Figure~\ref{fig:11}) and in the interior ($\xi=0.8$, Figure~\ref{fig:12}). The perturbation energies are defined as
\[
E_M \sim \tfrac{1}{2} \left|M_1\right|^2, \quad
E_\Phi \sim \tfrac{1}{2} \left|\Phi_1\right|^2, \quad
E_\Psi \sim \tfrac{1}{2} \left|{g_s}_1\right|^2,
\]
with
\[E_{\mathrm{total}} = E_M + E_\Phi + E_\Psi.\]
In the Newtonian case ($\chi=0$), the energy budget is dominated by electrostatic contributions at small $K$, which gradually transfer into kinetic energy as $K$ increases, while the gravitational energy fraction remains negligible ($\sim 4\%$). For non-zero $\chi$, however, the gravitational fraction becomes significant and competes directly with the electrostatic contribution across a wide range of $K$. Importantly, for negative $\chi$, the gravitational energy fraction is markedly enhanced with peaks at 28--32\%, while positive $\chi$ lowers the peak gravitational fraction to 7\%. The gravitational energy fraction decreases with $K$ and the released energy is progressively channelled into kinetic energy, reflecting the increasing role of plasma inertia at shorter scales. The stacked representations emphasize that EiBI corrections fundamentally reshape the partitioning of oscillation energy, enhancing gravitational influence and suppressing electrostatic dominance relative to the Newtonian scenario. This demonstrates that nonlinear gravity corrections not only alter dispersion properties but also significantly regulate the energy transport channels of solar plasma oscillations.

\subsection{Modal outward energy flux}
The outward acoustic energy flux density of the collective helioseismic modes at the solar surface can be expressed, using standard notations, as (in W\,m$^{-2}$) \cite{DePontieu2007, Shoda2018}
\begin{equation}
	{f_r}_s = \tfrac{1}{2}\rho_s {v_p}_s {v_r}_s^2.
	\label{eq:27}
\end{equation}
Here, $\rho_s$ denotes the surface fluid density. The quantities ${v_p}_s=\omega_r/k$ and ${v_r}_s$ represent the surface phase velocity and the velocity amplitude of radially outward propagating acoustic waves, respectively.\par
Accordingly, the normalized surface radiant energy flux density can be estimated as
\begin{equation}
	{F_r}_s = \frac{{f_r}_s}{\rho_0 C_s^3} = \tfrac{1}{2} \rho^* {V_p}_s {V_r}_s^2,
	\label{eq:28}
\end{equation}
where $\rho_0$ refers to the mean solar interior density, $\rho^*=\rho/\rho_0$ is the normalized surface fluid density, ${V_p}_s=V_p(\xi=2.25)$ denotes the Jeans-normalized surface phase velocity, and ${V_r}_s={v_r}_s/C_s$ is the Jeans-normalized velocity amplitude. The typical value of the energy flux normalizing parameter is $\rho_0 C_s^3 = 5\times10^{19}$ W m$^{-2}$.\par
\begin{figure*}
	\centering
	\begin{tabular}{c c}
		\includegraphics[trim={5cm 0cm 5cm 0cm},clip,width=8cm]{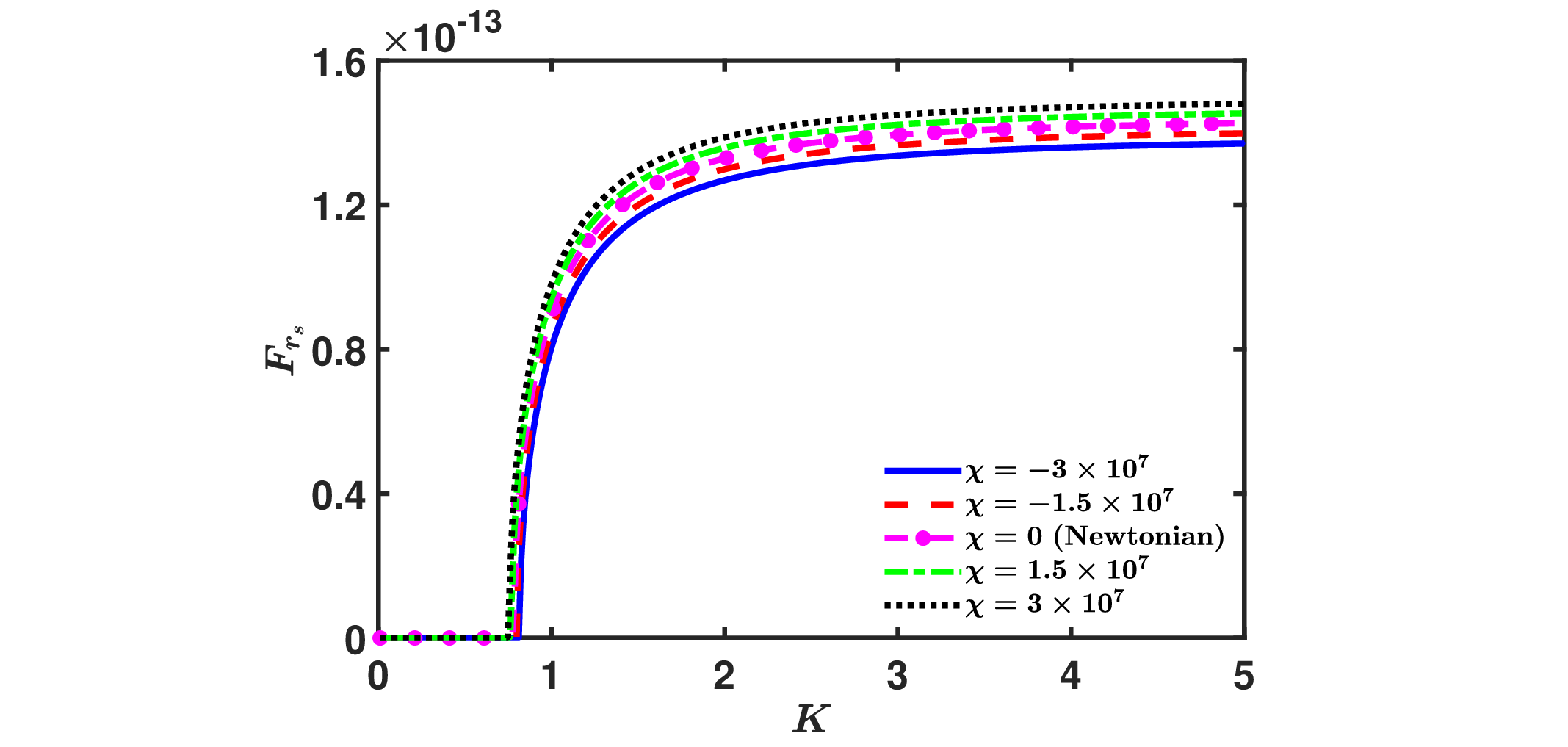}&
		\includegraphics[trim={5cm 0cm 5cm 0cm},clip,width=8cm]{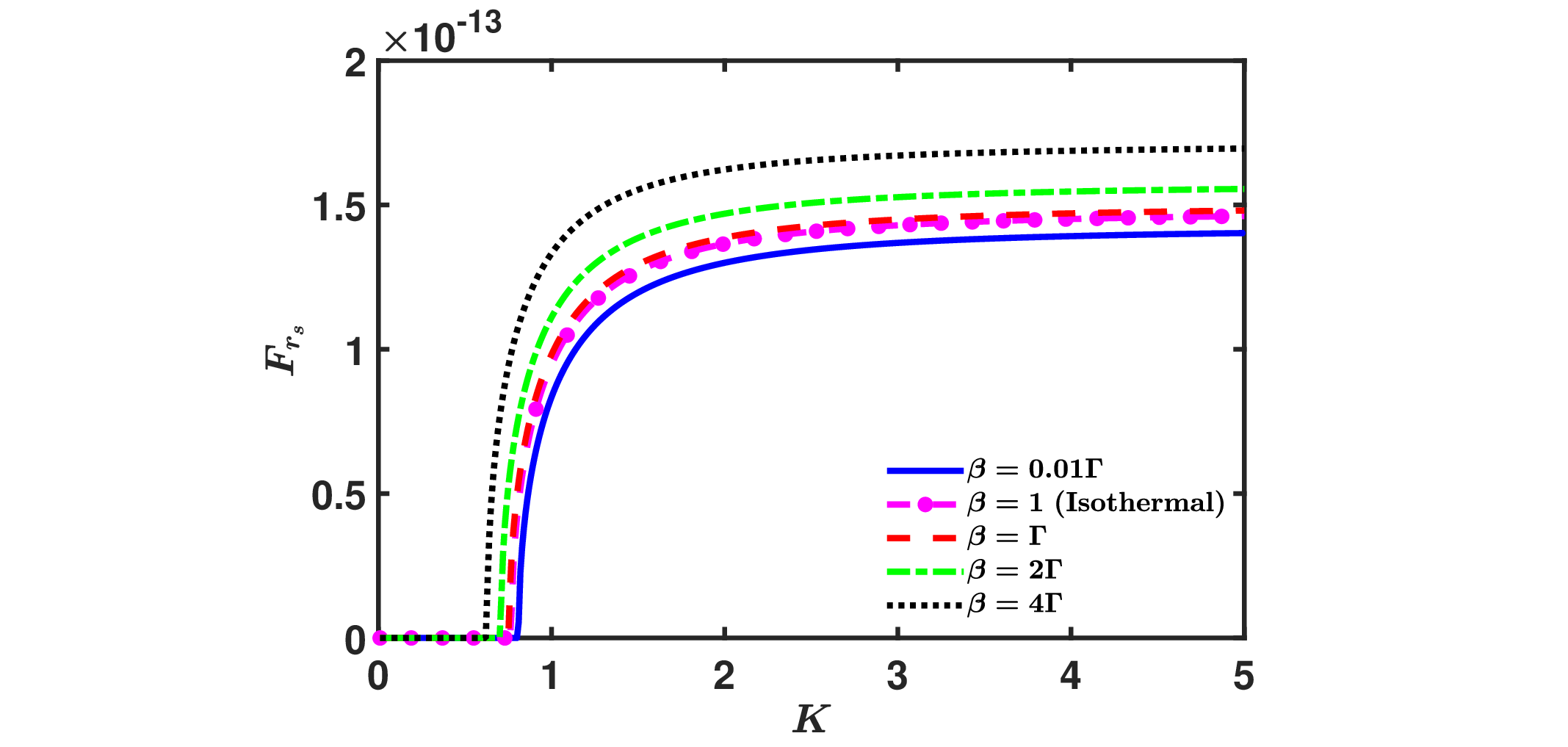}
	\end{tabular}
	\caption{Profiles of the normalized modal surface energy flux density $(F_{r_s})$ as a function of the Jeans-normalized angular wavenumber $(K)$ for different indicated values of EiBI gravity parameter $(\chi)$ and relative polytropic sound speed $(\beta)$. Here, $\chi$ is expressed in SI units.}
	\label{fig:13}
\end{figure*}

Studying the influence of the EiBI gravity parameter on the modal outward acoustic energy flux is crucial for understanding how gravitational modifications regulate energy transport at the solar surface. In Figure~\ref{fig:13}, we expound the normalized modal surface energy flux density ${F_r}_s$ as a function of the Jeans-normalized angular wavenumber $K$, comparing the effects of various EiBI gravity parameters $\chi$ (left panel) and relative polytropic sound speeds $\beta$ (right panel). The results reveal that the flux is negligible for low-$K$ \textit{g}-modes but rises steeply in the \textit{p}-mode regime. It signifies that only \textit{p}-modes contribute effectively to the outward acoustic flux. Positive $\chi$ enhances flux by up to $\sim 10\%$, while negative $\chi$ suppresses it by 4--10\% relative to the Newtonian case. Similarly, larger $\beta$ values systematically increase the flux by up to $\sim 55\%$, consistent with stronger thermal support amplifying acoustic propagation. These findings emphasize that both nonlinear gravitational corrections and polytropic thermodynamic conditions fundamentally regulate the efficiency of outward acoustic energy transport, thereby shaping the excitation and damping of collective solar plasma oscillation modes.\par

To study the outward acoustic energy redistribution, the efficiency of excitation alone does not guarantee effective energy delivery to the outer solar atmosphere. Most of the mechanical energy propagating upward through the solar photosphere is absorbed in the chromosphere due to leakage and mode conversion processes. This leads to a significant reduction of the upward-propagating \textit{p}-mode energy flux with increasing height $z$ above the photospheric base. Fontenla et al. \cite{Fontenla1993} have previously suggested that this decline follows an exponential profile. To model these decay features, we adopt an exponential attenuation characteristic of the outward \textit{p}-mode energy flux density $f_r(z)$ as follows
\begin{equation}
	f_r(z) = \tfrac{1}{2} \rho(z) v_p(z) v_r(z)^2,
	\label{eq:29}
\end{equation}
where the height-dependent parameters are given by,
\[\rho(z) = \rho_s e^{-4z/(3z_0)},\]
\[v_p(z) = {v_p}_s e^{-z/z_0},\]
\[v_r(z) = {v_r}_s e^{-z/z_0}.\]
Here, $\rho_s$, ${v_p}_s$, and ${v_r}_s$ denote the density, phase velocity, and velocity amplitude at the base of the photosphere ($z=0$), respectively. The reference height $z_0 \approx 2$\,Mm serves as a scaling parameter that controls the curvature of the decay profile. It defines the characteristic height over which the energy flux undergoes significant variation, thereby determining how rapidly the flux changes with height. Accordingly, using Equation~\eqref{eq:27}, the resulting energy flux profile as a function of $z$ reads as
\begin{equation}
	f_r(z) = {f_r}_s e^{-13z/(3z_0)}.
	\label{eq:30}
\end{equation}
Equation~\eqref{eq:30} successfully captures the rapid exponential damping of the radially outward \textit{p}-mode acoustic energy flux, representing wave energy attenuation through the lower solar atmosphere.\par
\begin{figure*}
	\centering
	\begin{tabular}{c c c c}
		\includegraphics[trim={5cm 0cm 5cm 0cm},clip,width=8cm]{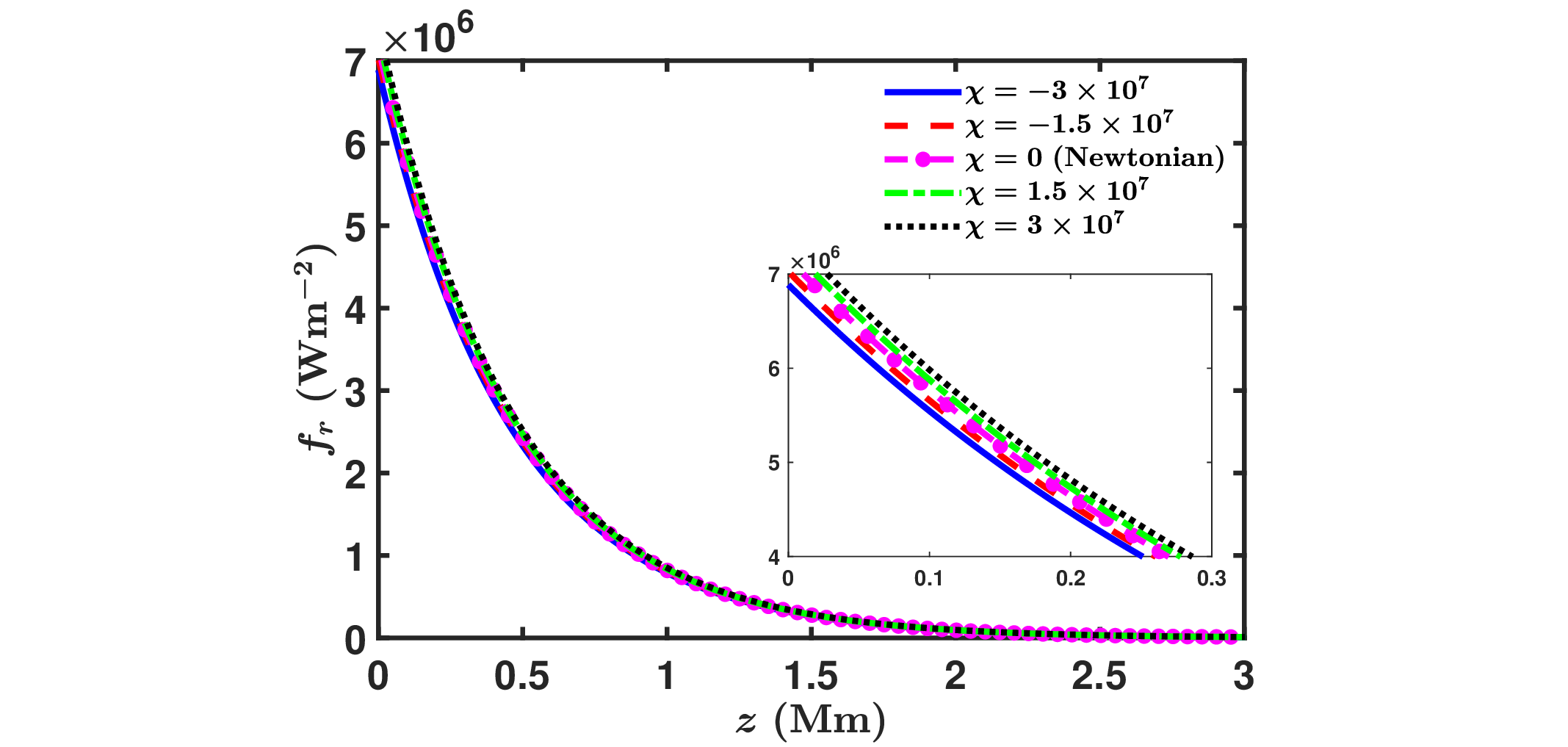}&
		\includegraphics[trim={5cm 0cm 5cm 0cm},clip,width=8cm]{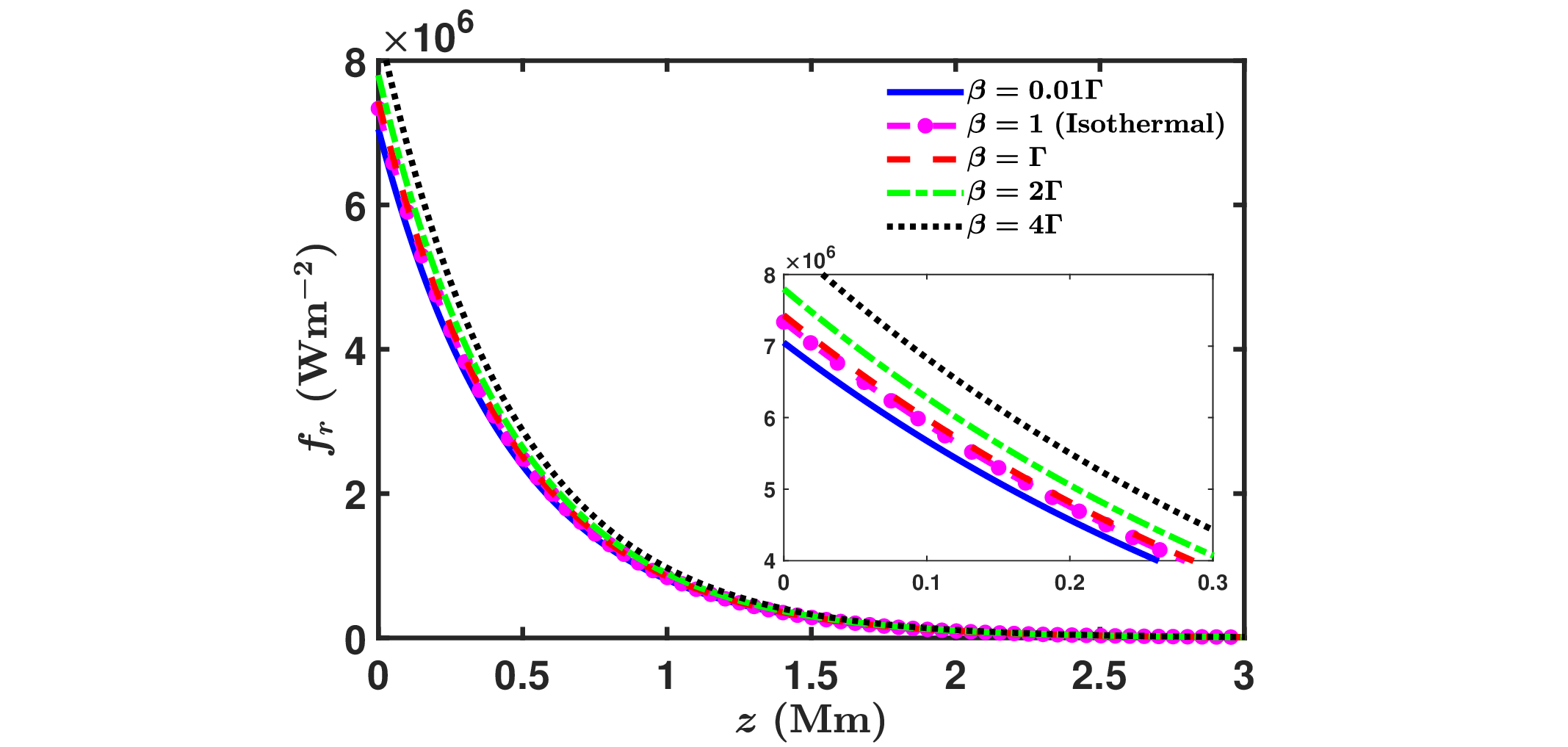}
	\end{tabular}
	\caption{Profiles of the \textit{p}-mode--driven energy flux density $(f_r)$ as a function of height $(z)$ above the base of the photosphere for different indicated values of EiBI gravity parameter $(\chi)$ and relative polytropic sound speed $(\beta)$. Here, $\chi$ is expressed in SI units. The peak velocity amplitude is set to ${v_r}_s=500$ m\,s$^{-1}$, based on observational constraints.}
	\label{fig:14}
\end{figure*}

Figure~\ref{fig:14} elaborates the \textit{p}-mode--driven energy flux density $f_r$ as a function of height $z$ above the base of the photosphere, for various values of the EiBI gravity parameter ($\chi$, left panel) and relative polytropic sound speed ($\beta$, right panel). The calculations adopt a peak photospheric velocity amplitude of ${v_r}_s = 500$ m\,s$^{-1}$, consistent with high-resolution Doppler observations from instruments such as Sunrise/IMaX and GONG. In both parameter sets, $f_r$ exhibits a steep exponential-like decay with height, highlighting the strong attenuation of acoustic flux due to radiative damping, leakage, and mode conversion as waves propagate outward through the stratified photospheric layers. Elevated values of $\chi$ and higher $\beta$ systematically enhance the energy flux across all altitudes, indicating that both nonlinear gravity modifications and enhanced polytropic pressure support increase the efficiency of upward acoustic energy transport. The insets provide a zoomed view near the base ($z \leq 0.3$ Mm), where the separation among curves is most prominent. These findings establish that EiBI gravity and polytropic effects not only modulate the asymptotic flux levels but also crucially shape the near-surface attenuation profile, thereby regulating the fraction of \textit{p}-mode energy available for chromospheric and coronal heating.

\subsection{Observations and reliability analysis}
We employ version 7.0.0 of the \texttt{SunPy} open-source software package \cite{Community2015,Barnes2020} to analyze line-of-sight Doppler velocity time series from Solar Dynamics Observatory/Helioseismic and Magnetic Imager (SDO/HMI) data taken on January~1 of each year between 2016 and 2019. For each dataset, Dopplergrams spanning a 30-minute interval (00:00--00:30~UTC) are processed, and Fourier transforms are applied to extract root-mean-square (RMS) velocity amplitudes in the 3.3--5.5 mHz \textit{p}-mode frequency band. Table~\ref{tab:2} summarizes the calculated RMS values, which capture year-to-year variations in oscillation strength, in good agreement with solar-cycle trends during the declining phase of Solar Cycle~24.\par

\begin{table*}
	\caption{SDO/HMI Doppler velocity data.}
	\label{tab:2}
	\begin{ruledtabular}
		\begin{tabular}{ccccc}
			S.~No. & Date & Time range (UTC) & No.~of FITS files & RMS velocity amplitude (m\,s$^{-1}$) \\
			\hline
			1 & 01-01-2016 & 00:00--00:30 & 40 & 957.26 \\
			2 & 01-01-2017 & 00:00--00:30 & 40 & 767.27 \\
			3 & 01-01-2018 & 00:00--00:30 & 40 & 647.36 \\
			4 & 01-01-2019 & 00:00--00:30 & 40 & 727.00 \\
		\end{tabular}
	\end{ruledtabular}
\end{table*}

\begin{figure}
	\centering
	\begin{tabular}{c}
		\includegraphics[trim={5cm 0cm 5cm 0cm},clip,width=8cm]{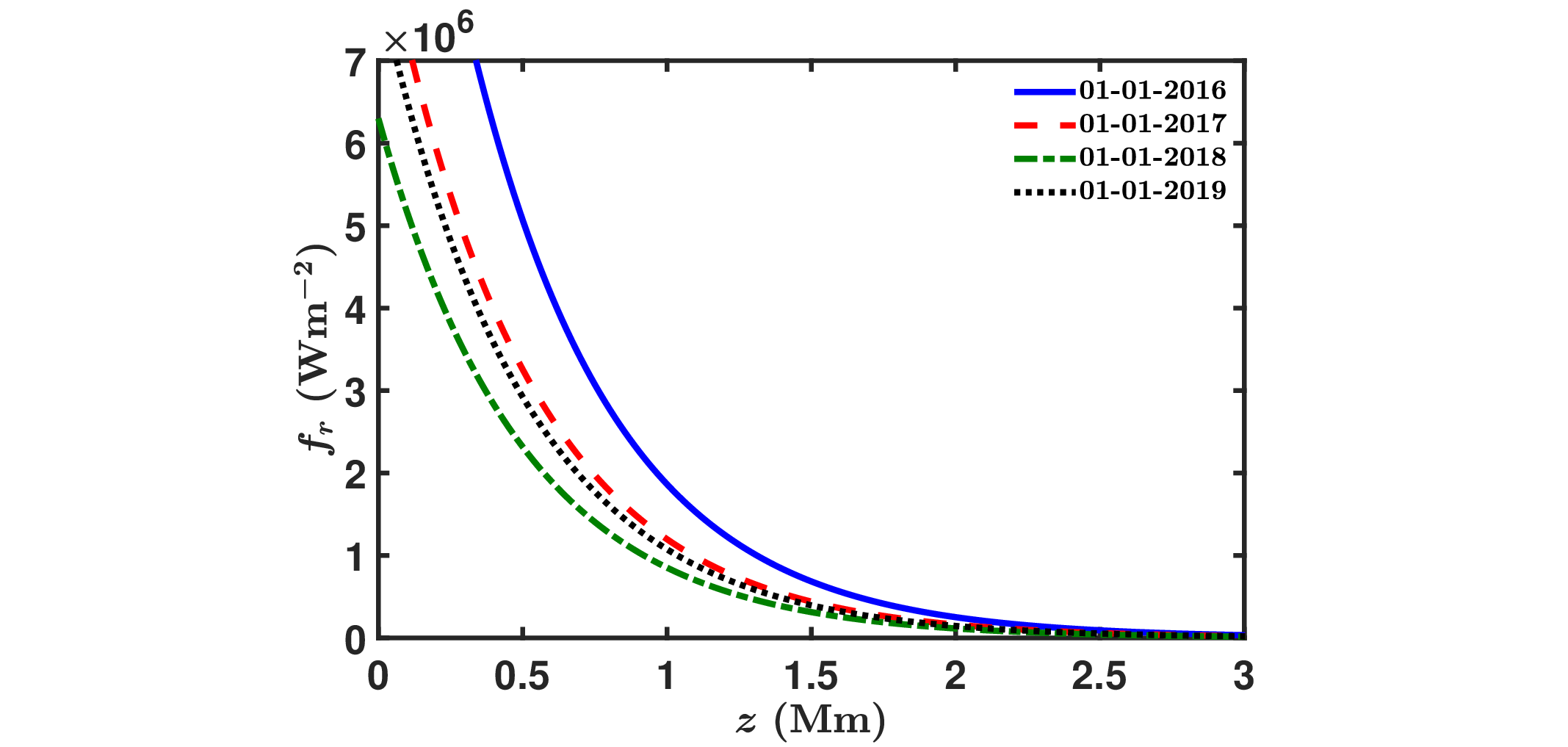}
	\end{tabular}
	\caption{Profile of the \textit{p}-mode--driven energy flux density $(f_r)$ as a function of height $(z)$ above the base of the photosphere, derived from four years (2016--2019) of SDO/HMI Doppler observations.}
	\label{fig:15}
\end{figure}

In Figure~\ref{fig:15}, we illustrate the resulting vertical profiles of the acoustic energy flux density $f_r(z)$ obtained from the relation
\[
	f_r(z) = \tfrac{1}{2}\,\rho(z) C_s v_{\mathrm{rms}}^2(z),
\]
where $\rho(z)$ is the mass density, $v_{\mathrm{rms}}$ denotes the root-mean-square vertical velocity fluctuation, and $C_s$ represents the ion-acoustic phase speed, modelled under an exponentially stratified atmosphere. The flux values start above $10^6$ W\,m$^{-2}$ at the photospheric base and decrease nearly exponentially with altitude, driven by stratification, mode conversion, and dissipative leakage. A gradual reduction in flux is observed over the years, with 2016 showing the strongest energy transport and 2019 the weakest, demonstrating how solar variability and damping processes regulate the efficiency of \textit{p}-mode energy propagation into the chromosphere.\par

\begin{figure}
	\centering
	\begin{tabular}{c}
		\includegraphics[trim={5cm 0cm 5cm 0cm},clip,width=8cm]{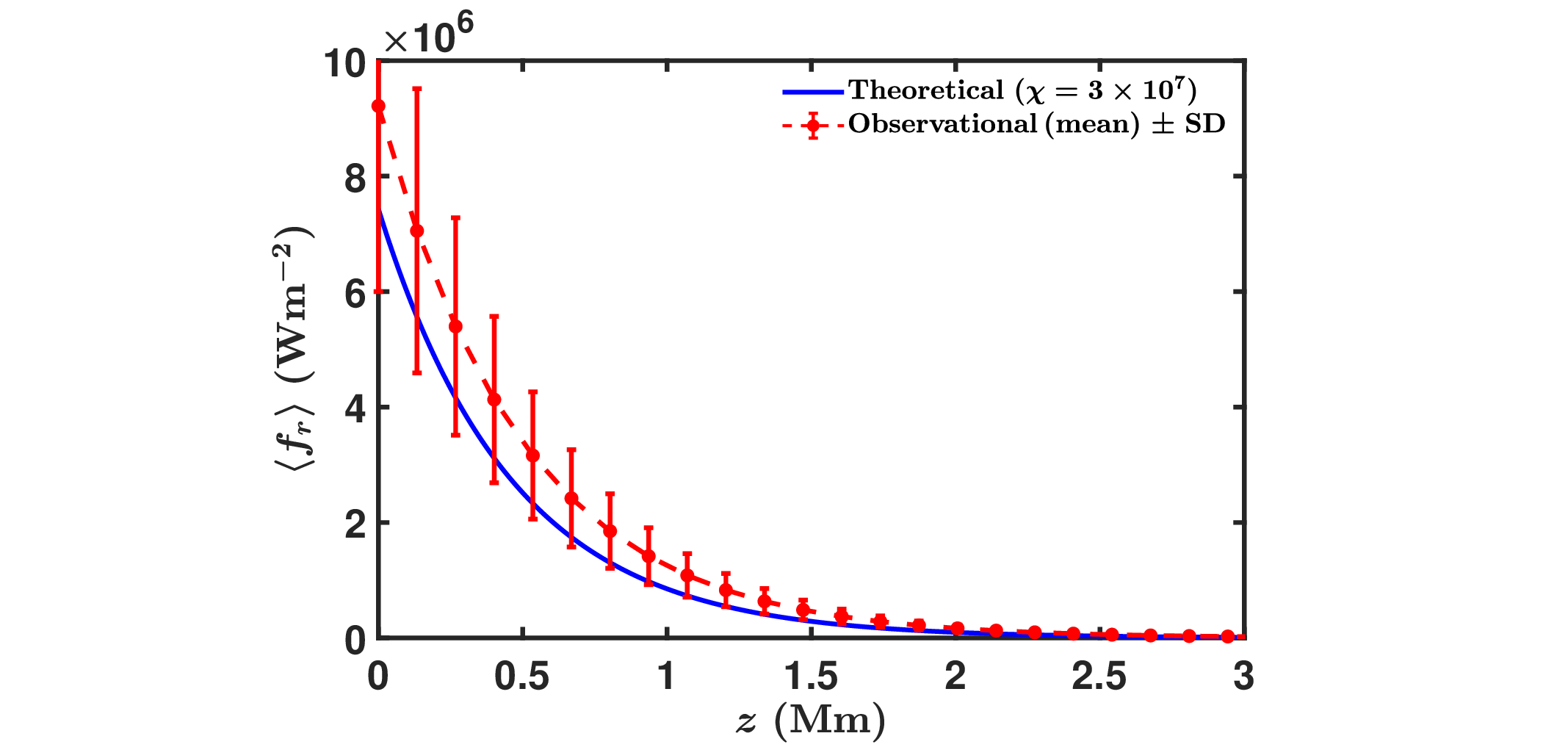}
	\end{tabular}
	\caption{Mean \textit{p}-mode--driven energy flux density $(\langle f_r\rangle)$ as a function of height $(z)$ above the base of the photosphere, comparing theoretical predictions for $\chi=3\times10^7$ m$^5$\,kg$^{-1}$\,s$^{-2}$ (solid blue curve) with averaged observational results from four years (2016--2019) of SDO/HMI Doppler data. The red dashed curve shows the observational mean, and vertical error bars represent the standard deviation (SD) across the datasets.}
	\label{fig:16}
\end{figure}

Figure~\ref{fig:16} provides a direct comparison between theoretical and observational mean profiles of \textit{p}-mode--driven energy flux density $\langle f_r \rangle$ as a function of height $z$ above the photospheric base. The theoretical curve, computed using the modified EiBI gravity parameter $\chi=3\times10^7$ m$^5$\,kg$^{-1}$\,s$^{-2}$, aligns closely with the averaged observational flux (red dashed curve) obtained from multi-year SDO/HMI Doppler velocity measurements. Both profiles reveal a rapid decline with altitude, indicating strong attenuation of upward-propagating acoustic energy due to stratification and dissipative processes. The vertical error bars quantify year-to-year variability, showing that while theoretical predictions fall within $15\%$ of observational uncertainties across most heights, deviations reach $\sim 20\%$ near the base where the flux peaks. Overall, the agreement underscores the robustness of the adopted theoretical model and suggests that nonlinear corrections introduced by the EiBI gravity provide a consistent framework for capturing the essential features of acoustic energy transport in the near-surface solar atmosphere.\par

Table~\ref{tab:3} summarizes the stabilizing and destabilizing influences of the EiBI gravity parameter ($\chi$) and relative polytropic sound speed ($\beta$) on solar plasma oscillations.\par

\begin{table*}
	\caption{Parametric influences on the solar plasma oscillations.}
	\label{tab:3}
	\begin{ruledtabular}
		\begin{tabular}{ccc}
			Parameter & Interior ($\xi=0.8$) & Surface ($\xi=2.25$) \\
			\hline
			$\chi > 0$ & Accelerating; neutral growth & Accelerating; destabilizing \\
			$\chi < 0$ & Decelerating; neutral growth & Decelerating; stabilizing \\
			$\beta$    & Accelerating; neutral growth & Accelerating; destabilizing \\
		\end{tabular}
	\end{ruledtabular}
\end{table*}

This parametric analysis establishes that EiBI gravity and thermodynamic polytropic conditions fundamentally reshape solar plasma stability, perturbation energetics, and outward energy transport, with the strongest imprints appearing in dense and stratified regions. By directly linking these effects to observable helioseismic signatures and benchmarking them against space-based Doppler data, we obtain the most stringent constraints to date on nonlinear gravity in the solar context. The quantified restructuring of energy partitioning and the characterization of flux attenuation further advance models of atmospheric heating and energy redistribution.

\section{Conclusions}
\label{sec:5}
This study presents a comprehensive theoretical and observational investigation of wave dynamics, stability, and energy transport in self-gravitating solar plasmas within the Eddington-inspired Born--Infeld (EiBI) gravity framework. Through systematic inclusion of EiBI gravity--modified Poisson equation in a polytropic, viscous, and turbulent solar plasma model, we rigorously perform a linear stability analysis on the Jeans-normalized basic governing equations. It yields a closed quadratic dispersion relation governing the propagation of Jeans-normalized parametric perturbations. We numerically explore the dispersion relation across a broad parameter space. The perturbative analysis reveals that the oscillatory modes remain linearly stable, with growth rates always negative, while the oscillation frequencies are significantly modulated by both the EiBI gravity parameter $\chi$ and the relative polytropic sound speed $\beta$.\par
Our systematic parametric analyses reveal transformative modifications to the solar plasma dynamics. Positive $\chi$ consistently elevate oscillation frequencies by 4--10\% and enhance phase speeds by up to 10\%. Similarly, larger $\beta$ values raise frequency by 20--55\% and increase phase speed by up to 55\%. Hence, positive $\chi$ and $\beta$ establish robust plasma stability and enhance wave propagating propensity. In contrast, negative $\chi$ strengthens gravitational binding and increases damping by up to 40\%, particularly at low wavenumbers ($K \leq 1$), demonstrating the bidirectional regulatory influence of nonlinear gravity corrections.\par
The perturbation energy partitioning analysis highlights that the EiBI corrections fundamentally restructure the kinetic--electrostatic--gravitational energy balance that has been assumed negligible in classical treatments. In the Newtonian limit, the energy budget is dominated by electrostatic and kinetic contributions, with gravity contributing less than 4\% to perturbation energetics. However, for nonzero $\chi$, the gravitational fraction becomes substantial---peaking at 28--32\% for negative $\chi$ and 7\% for positive $\chi$---before the released energy is efficiently channelled into kinetic modes at higher wavenumber. These transitions indicate a profound restructuring of plasma oscillation energetics caused by the nonlinear gravity corrections.\par
Furthermore, we show that the outward acoustic energy flux is directly regulated by EiBI gravity and thermodynamic polytropic conditions. Positive $\chi$ enhances the modal flux levels up to 10\%, while negative $\chi$ suppresses it by 4--10\% relative to Newtonian results. Larger $\beta$ values further boost energy flux by as much as 55\%, consistent with stronger polytropic pressure support enabling more efficient acoustic mode propagation. The height-dependent flux profiles indicate an exponential-like attenuation with altitude, consistent with leakage and mode conversion processes in the photosphere--chromosphere interface. Crucially, our theoretical predictions with $\chi=3\times10^7$ m$^5$\,kg$^{-1}$\,s$^{-2}$ show excellent agreement with the four-year SDO/HMI Doppler velocity observations, establishing the first empirical constraint on the solar EiBI gravity through helioseismology.\par
In summary, our investigated findings demonstrate that nonlinear EiBI gravity corrections yield rich, quantifiable, and observable imprints on solar plasma oscillation dynamics, collective stability thresholds, and energy transport mechanisms. Such signatures not only provide a direct pathway to test gravity beyond general relativity in the solar context, but also significantly advance the modelling accuracy of helioseismic and energy transfer phenomena. Future extensions incorporating more realistic plasma equations of state, magnetic fields, higher-order nonlinear couplings, and multi-modal nonlinear oscillation analyses will further refine these constraints and may uncover new avenues for probing fundamental gravity through the solar and stellar plasma oscillations.

\begin{acknowledgments}
The authors acknowledge the Department of Physics, Tezpur University, for providing institutional support throughout this work. The constructive discussions and encouragement from members of the Astrophysical Plasma and Nonlinear Dynamics Research Laboratory (APNDRL) are also duly appreciated. PKK gratefully acknowledges the Inter-University Centre for Astronomy and Astrophysics (IUCAA), Pune, India, for the award of an Associateship. SD acknowledges the Department of Science and Technology (DST), Government of India, for financial support through the DST-INSPIRE Fellowship (Grant No.: DST/INSPIRE Fellowship/2021/IF210234).
\end{acknowledgments}

\section*{Data Availability}
All data supporting the findings of this study are contained within this article \cite{ThisWork}.

\bibliography{Bibliography}

@article{Adams1994,
   author = {Fred C. Adams and Marco Fatuzzo and Richard Watkins},
   doi = {10.1086/174100},
   issn = {0004-637X},
   journal = {The Astrophysical Journal},
   month = {5},
   pages = {629},
   title = {General analytic results for nonlinear waves and solitons in molecular clouds},
   volume = {426},
   url = {http://adsabs.harvard.edu/doi/10.1086/174100},
   year = {1994}
}

@article{Afonso2019,
   author = {Victor I. Afonso and Gonzalo J. Olmo and Emanuele Orazi and Diego Rubiera-Garcia},
   doi = {10.1088/1475-7516/2019/12/044},
   issn = {1475-7516},
   issue = {12},
   journal = {Journal of Cosmology and Astroparticle Physics},
   month = {12},
   pages = {044-044},
   title = {New scalar compact objects in Ricci-based gravity theories},
   volume = {2019},
   year = {2019}
}

@article{Afonso2020,
   author = {Victor I. Afonso},
   doi = {10.1142/S0218271820410114},
   issn = {0218-2718},
   issue = {11},
   journal = {International Journal of Modern Physics D},
   month = {8},
   pages = {2041011},
   title = {Compact scalar field solutions in EiBI gravity},
   volume = {29},
   year = {2020}
}

@book{Ambastha2020,
   author = {Ashok Ambastha},
   city = {Boca Raton},
   doi = {10.1201/9781003005674},
   isbn = {9781003005674},
   month = {3},
   publisher = {CRC Press},
   title = {Physics of the Invisible Sun},
   url = {https://www.taylorfrancis.com/books/9781003005674},
   year = {2020}
}

@inbook{Aschwanden2014,
   author = {Markus J Aschwanden},
   city = {Boston},
   doi = {https://doi.org/10.1016/B978-0-12-415845-0.00011-6},
   edition = {Third Edit},
   editor = {Tilman Spohn and Doris Breuer and Torrence V Johnson},
   isbn = {978-0-12-415845-0},
   booktitle = {Encyclopedia of the Solar System (Third Edition)},
   pages = {235-259},
   publisher = {Elsevier},
   title = {Chapter 11 - The Sun},
   url = {https://www.sciencedirect.com/science/article/pii/B9780124158450000116},
   year = {2014}
}

@article{Avelino2012a,
	author = {P. P. Avelino},
	doi = {10.1103/PhysRevD.85.104053},
	issn = {1550-7998},
	issue = {10},
	journal = {Physical Review D},
	month = {5},
	pages = {104053},
	title = {Eddington-inspired Born-Infeld gravity: Astrophysical and cosmological constraints},
	volume = {85},
	url = {https://link.aps.org/doi/10.1103/PhysRevD.85.104053},
	year = {2012}
}

@article{Avelino2012b,
   author = {P.P Avelino},
   doi = {10.1088/1475-7516/2012/11/022},
   issn = {1475-7516},
   issue = {11},
   journal = {Journal of Cosmology and Astroparticle Physics},
   month = {11},
   pages = {022-022},
   title = {Eddington-inspired Born-Infeld gravity: nuclear physics constraints and the validity of the continuous fluid approximation},
   volume = {2012},
   url = {https://iopscience.iop.org/article/10.1088/1475-7516/2012/11/022},
   year = {2012}
}

@article{Banados2010,
   author = {M{\'a}ximo Ba{\~n}ados and Pedro G. Ferreira},
   doi = {10.1103/PhysRevLett.105.011101},
   issn = {0031-9007},
   issue = {1},
   journal = {Physical Review Letters},
   month = {7},
   pages = {011101},
   title = {Eddington’s Theory of Gravity and Its Progeny},
   volume = {105},
   url = {https://link.aps.org/doi/10.1103/PhysRevLett.105.011101},
   year = {2010}
}

@article{Banerjee2022,
   author = {Pritam Banerjee and Debojyoti Garain and Suvankar Paul and Rajibul Shaikh and Tapobrata Sarkar},
   doi = {10.3847/1538-4357/ac324f},
   issn = {0004-637X},
   issue = {1},
   journal = {The Astrophysical Journal},
   month = {1},
   pages = {20},
   title = {A Stellar Constraint on Eddington-inspired Born–Infeld Gravity from Cataclysmic Variable Binaries},
   volume = {924},
   year = {2022}
}

@article{Barnes2020,
	author = {Will T. Barnes and Monica G. Bobra and Steven D. Christe and Nabil Freij and Laura A. Hayes and Jack Ireland and Stuart Mumford and David Perez-Suarez and Daniel F. Ryan and Albert Y. Shih and Prateek Chanda and Kolja Glogowski and Russell Hewett and V. Keith Hughitt and Andrew Hill and Kaustubh Hiware and Andrew Inglis and Michael S. F. Kirk and Sudarshan Konge and James Paul Mason and Shane Anthony Maloney and Sophie A. Murray and Asish Panda and Jongyeob Park and Tiago M. D. Pereira and Kevin Reardon and Sabrina Savage and Brigitta M. Sip\H{o}cz and David Stansby and Yash Jain and Garrison Taylor and Tannmay Yadav and Rajul and Trung Kien Dang},
	doi = {10.3847/1538-4357/ab4f7a},
	issn = {0004-637X},
	issue = {1},
	journal = {The Astrophysical Journal},
	month = {2},
	pages = {68},
	title = {The SunPy Project: Open Source Development and Status of the Version 1.0 Core Package},
	volume = {890},
	year = {2020}
}

@article{Bessiri2021,
   author = {Ahmed Bessiri and Kamel Ourabah and Taha Houssine Zerguini},
   doi = {10.1088/1402-4896/ac1cd2},
   issn = {0031-8949},
   issue = {12},
   journal = {Physica Scripta},
   month = {12},
   pages = {125208},
   title = {Jeans instability in Eddington-inspired Born--Infeld (EiBI) gravity: a quantum approach},
   volume = {96},
   year = {2021}
}

@inbook{Braginskii1965,
   author = {S. I. Braginskii},
   city = {New York},
   editor = {M. A. Leontovich},
   booktitle = {Reviews of Plasma Physics},
   pages = {205-311},
   publisher = {Consultants Bureau},
   title = {Transport processes in a plasma},
   volume = {1},
   url = {https://ui.adsabs.harvard.edu/abs/1965RvPP....1..205B},
   year = {1965}
}

@article{Buresti2015,
   author = {Guido Buresti},
   doi = {10.1007/s00707-015-1380-9},
   issn = {0001-5970},
   issue = {10},
   journal = {Acta Mechanica},
   month = {10},
   pages = {3555-3559},
   title = {A note on Stokes’ hypothesis},
   volume = {226},
   year = {2015}
}

@article{Cally1995,
   author = {P. S. Cally},
   doi = {10.1086/176226},
   issn = {0004-637X},
   journal = {The Astrophysical Journal},
   month = {9},
   pages = {372},
   title = {Effects of Weak-to-Moderate Vertical Magnetic Fields on Solar f- and p-Modes},
   volume = {451},
   year = {1995}
}

@book{Carroll2019,
   author = {Sean M Carroll},
   doi = {10.1017/9781108770385},
   isbn = {978-0-8053-8732-2, 978-1-108-48839-6, 978-1-108-77555-7},
   month = {9},
   publisher = {Cambridge University Press},
   title = {Spacetime and Geometry: An Introduction to General Relativity},
   year = {2019}
}

@article{Casanellas2012,
   author = {Jordi Casanellas and Paolo Pani and Ilídio Lopes and Vitor Cardoso},
   doi = {10.1088/0004-637X/745/1/15},
   issn = {15384357},
   issue = {1},
   journal = {Astrophysical Journal},
   month = {1},
   pages = {15},
   title = {Testing alternative theories of gravity using the sun},
   volume = {745},
   url = {https://iopscience.iop.org/article/10.1088/0004-637X/745/1/15},
   year = {2012}
}

@article{Chandra2013,
   author = {Suresh Chandra},
   doi = {10.1007/s12648-012-0239-3},
   issn = {0973-1458},
   issue = {6},
   journal = {Indian Journal of Physics},
   month = {6},
   pages = {601-606},
   title = {Investigation of diffusivity and viscosity in solar plasma},
   volume = {87},
   url = {http://link.springer.com/10.1007/s12648-012-0239-3},
   year = {2013}
}

@book{Chen2016,
   author = {Francis F. Chen},
   city = {Cham},
   doi = {10.1007/978-3-319-22309-4},
   isbn = {978-3-319-22308-7},
   publisher = {Springer International Publishing},
   title = {Introduction to Plasma Physics and Controlled Fusion},
   year = {2016}
}

@book{Choudhuri1998,
   author = {Arnab Rai Choudhuri},
   doi = {10.1017/CBO9781139171069},
   isbn = {9780521555432},
   month = {11},
   publisher = {Cambridge University Press},
   title = {The Physics of Fluids and Plasmas},
   year = {1998}
}

@article{Christensen-Dalsgaard2002,
   author = {Jørgen Christensen-Dalsgaard},
   doi = {10.1103/RevModPhys.74.1073},
   issn = {00346861},
   issue = {4},
   journal = {Reviews of Modern Physics},
   month = {11},
   pages = {1073-1129},
   title = {Helioseismology},
   volume = {74},
   url = {https://link.aps.org/doi/10.1103/RevModPhys.74.1073},
   year = {2002}
}

@book{Clarke2007,
   author = {Cathie Clarke and Bob Carswell},
   doi = {10.1017/CBO9780511813450},
   isbn = {9780521853316},
   month = {3},
   publisher = {Cambridge University Press},
   title = {Principles of Astrophysical Fluid Dynamics},
   year = {2007}
}

@article{Community2015,
	author = {The SunPy Community and Stuart J Mumford and Steven Christe and David Pérez-Suárez and Jack Ireland and Albert Y Shih and Andrew R Inglis and Simon Liedtke and Russell J Hewett and Florian Mayer and Keith Hughitt and Nabil Freij and Tomas Meszaros and Samuel M Bennett and Michael Malocha and John Evans and Ankit Agrawal and Andrew J Leonard and Thomas P Robitaille and Benjamin Mampaey and Jose Iván Campos-Rozo and Michael S Kirk},
	doi = {10.1088/1749-4699/8/1/014009},
	issn = {1749-4699},
	issue = {1},
	journal = {Computational Science \& Discovery},
	month = {7},
	pages = {014009},
	title = {SunPy--Python for solar physics},
	volume = {8},
	year = {2015}
}

@article{Das2023,
   author = {Souvik Das and Ahmed Atteya and Pralay Kumar Karmakar},
   doi = {10.1093/mnras/stad1664},
   issn = {13652966},
   issue = {4},
   journal = {Monthly Notices of the Royal Astronomical Society},
   month = {6},
   pages = {5635-5660},
   title = {A theoretic analysis of magnetoactive {GES}-based turbulent solar plasma instability},
   volume = {523},
   url = {https://academic.oup.com/mnras/article/523/4/5635/7190640},
   year = {2023}
}

@article{Das2024,
   author = {Souvik Das and Pralay Kumar Karmakar},
   doi = {https://doi.org/10.1016/j.cjph.2024.05.008},
   issn = {0577-9073},
   journal = {Chinese Journal of Physics},
   pages = {157-166},
   title = {Solar {GES}-structure modified with {EiBI} gravity},
   volume = {91},
   url = {https://www.sciencedirect.com/science/article/pii/S0577907324001874},
   year = {2024}
}

@article{DeMartino2017,
   author = {Ivan De Martino and Antonio Capolupo},
   doi = {10.1140/epjc/s10052-017-5300-0},
   issn = {1434-6044},
   issue = {10},
   journal = {The European Physical Journal C},
   month = {10},
   pages = {715},
   title = {Kinetic theory of Jean instability in Eddington-inspired Born–Infeld gravity},
   volume = {77},
   year = {2017}
}

@article{DePontieu2007,
   author = {B. De Pontieu and S. W. McIntosh and M. Carlsson and V. H. Hansteen and T. D. Tarbell and C. J. Schrijver and A. M. Title and R. A. Shine and S. Tsuneta and Y. Katsukawa and K. Ichimoto and Y. Suematsu and T. Shimizu and S. Nagata},
   doi = {10.1126/science.1151747},
   issn = {0036-8075},
   issue = {5856},
   journal = {Science},
   month = {12},
   pages = {1574-1577},
   title = {Chromospheric Alfv{\'e}nic Waves Strong Enough to Power the Solar Wind},
   volume = {318},
   url = {https://www.science.org/doi/10.1126/science.1151747},
   year = {2007}
}

@article{Delsate2012,
   author = {T{\'e}rence Delsate and Jan Steinhoff},
   doi = {10.1103/PhysRevLett.109.021101},
   issn = {0031-9007},
   issue = {2},
   journal = {Physical Review Letters},
   month = {7},
   pages = {021101},
   title = {New Insights on the Matter-Gravity Coupling Paradigm},
   volume = {109},
   year = {2012}
}

@article{Demarque1999,
   author = {P. Demarque and D. B. Guenther},
   doi = {10.1073/pnas.96.10.5356},
   issn = {0027-8424},
   issue = {10},
   journal = {Proceedings of the National Academy of Sciences},
   month = {5},
   pages = {5356-5359},
   title = {Helioseismology: Probing the interior of a star},
   volume = {96},
   url = {https://pnas.org/doi/full/10.1073/pnas.96.10.5356},
   year = {1999}
}

@article{Deubner1984,
   author = {Franz-Ludwig Deubner and Douglas Gough},
   doi = {10.1146/annurev.aa.22.090184.003113},
   issn = {0066-4146},
   issue = {1},
   journal = {Annual Review of Astronomy and Astrophysics},
   month = {9},
   pages = {593-619},
   title = {Helioseismology: Oscillations as a Diagnostic of the Solar Interior},
   volume = {22},
   year = {1984}
}

@article{Fontenla1993,
   author = {J. M. Fontenla and D. Rabin and D. H. Hathaway and R. L. Moore},
   doi = {10.1086/172408},
   issn = {0004-637X},
   journal = {The Astrophysical Journal},
   month = {3},
   pages = {787},
   title = {Measurement of p-mode energy propagation in the quiet solar photosphere},
   volume = {405},
   year = {1993}
}

@article{Gough1996,
   author = {D. O. Gough and A. G. Kosovichev and J. Toomre and E. Anderson and H. M. Antia and S. Basu and B. Chaboyer and S. M. Chitre and J. Christensen-Dalsgaard and W. A. Dziembowski and A. Eff-Darwich and J. R. Elliott and P. M. Giles and P. R. Goode and J. A. Guzik and J. W. Harvey and F. Hill and J. W. Leibacher and M. J. P. F. G. Monteiro and O. Richard and T. Sekii and H. Shibahashi and M. Takata and M. J. Thompson and S. Vauclair and S. V. Vorontsov},
   doi = {10.1126/science.272.5266.1296},
   issn = {0036-8075},
   issue = {5266},
   journal = {Science},
   month = {5},
   pages = {1296-1300},
   title = {The Seismic Structure of the Sun},
   volume = {272},
   year = {1996}
}

@book{Hansen2004,
   author = {Carl J. Hansen and Steven D. Kawaler and Virginia Trimble},
   city = {New York, NY},
   doi = {10.1007/978-1-4419-9110-2},
   isbn = {978-1-4612-6497-2},
   publisher = {Springer New York},
   title = {Stellar Interiors},
   url = {http://link.springer.com/10.1007/978-1-4419-9110-2},
   year = {2004}
}

@book{Hartle2003,
   author = {J B Hartle},
   isbn = {978-0-8053-8662-2},
   title = {Gravity: An introduction to Einstein's general relativity},
   url = {https://ui.adsabs.harvard.edu/abs/2003gieg.book.....H},
   year = {2003}
}

@article{Hendry1993,
   author = {Archibald W. Hendry},
   doi = {10.1119/1.17362},
   issn = {0002-9505},
   issue = {10},
   journal = {American Journal of Physics},
   month = {10},
   pages = {906-910},
   title = {A polytropic model of the Sun},
   volume = {61},
   year = {1993}
}

@article{Hollweg1985,
   author = {Joseph V. Hollweg},
   doi = {10.1029/JA090iA08p07620},
   issn = {0148-0227},
   issue = {A8},
   journal = {Journal of Geophysical Research: Space Physics},
   month = {8},
   pages = {7620-7622},
   title = {Viscosity in a magnetized plasma: Physical interpretation},
   volume = {90},
   year = {1985}
}

@book{Horedt2004,
   author = {G.P. Horedt},
   city = {Dordrecht},
   doi = {10.1007/1-4020-2351-0},
   isbn = {1-4020-2350-2},
   publisher = {Kluwer Academic Publishers},
   title = {Polytropes},
   volume = {306},
   url = {http://link.springer.com/10.1007/1-4020-2351-0},
   year = {2004}
}

@article{Kim2014,
   author = {Hyeong-Chan Kim},
   doi = {10.1103/PhysRevD.89.064001},
   issn = {1550-7998},
   issue = {6},
   journal = {Physical Review D},
   month = {3},
   pages = {064001},
   title = {Physics at the surface of a star in Eddington-inspired Born-Infeld Gravity},
   volume = {89},
   year = {2014}
}

@book{Landau1987,
   author = {L.D. Landau and E.M. Lifshitz},
   doi = {10.1016/C2013-0-03799-1},
   edition = {2},
   isbn = {9780080339337},
   publisher = {Elsevier},
   title = {Fluid Mechanics},
   volume = {6},
   year = {1987}
}

@article{Larson1981,
   author = {R. B. Larson},
   doi = {10.1093/mnras/194.4.809},
   issn = {0035-8711},
   issue = {4},
   journal = {Monthly Notices of the Royal Astronomical Society},
   month = {4},
   pages = {809-826},
   title = {Turbulence and star formation in molecular clouds},
   volume = {194},
   url = {https://academic.oup.com/mnras/article-lookup/doi/10.1093/mnras/194.4.809},
   year = {1981}
}

@article{Lizano1989,
   author = {Susana Lizano and Frank H. Shu},
   doi = {10.1086/167640},
   issn = {0004-637X},
   journal = {The Astrophysical Journal},
   month = {7},
   pages = {834},
   title = {Molecular cloud cores and bimodal star formation},
   volume = {342},
   url = {http://adsabs.harvard.edu/doi/10.1086/167640},
   year = {1989}
}

@book{Misner1973,
   author = {Charles W Misner and K S Thorne and J A Wheeler},
   city = {San Francisco},
   isbn = {978-0-7167-0344-0, 978-0-691-17779-3},
   publisher = {W. H. Freeman},
   title = {Gravitation},
   url = {https://ui.adsabs.harvard.edu/abs/1973grav.book.....M},
   year = {1973}
}

@article{Pani2011,
   author = {Paolo Pani and Vitor Cardoso and T{\'e}rence Delsate},
   doi = {10.1103/PhysRevLett.107.031101},
   issn = {0031-9007},
   issue = {3},
   journal = {Physical Review Letters},
   month = {7},
   pages = {031101},
   title = {Compact Stars in Eddington Inspired Gravity},
   volume = {107},
   url = {https://link.aps.org/doi/10.1103/PhysRevLett.107.031101},
   year = {2011}
}

@article{Pani2012a,
   author = {Paolo Pani and T{\'e}rence Delsate and Vitor Cardoso},
   doi = {10.1103/PhysRevD.85.084020},
   issn = {1550-7998},
   issue = {8},
   journal = {Physical Review D},
   month = {4},
   pages = {084020},
   title = {Eddington-inspired Born-Infeld gravity: Phenomenology of nonlinear gravity-matter coupling},
   volume = {85},
   url = {https://link.aps.org/doi/10.1103/PhysRevD.85.084020},
   year = {2012}
}

@article{Pani2012b,
   author = {Paolo Pani and Thomas P. Sotiriou},
   doi = {10.1103/PhysRevLett.109.251102},
   issn = {0031-9007},
   issue = {25},
   journal = {Physical Review Letters},
   month = {12},
   pages = {251102},
   title = {Surface Singularities in Eddington-Inspired Born-Infeld Gravity},
   volume = {109},
   year = {2012}
}

@article{Prasetyo2021,
   author = {I. Prasetyo and H. Maulana and H. S. Ramadhan and A. Sulaksono},
   doi = {10.1103/PhysRevD.104.084029},
   issn = {2470-0010},
   issue = {8},
   journal = {Physical Review D},
   month = {10},
   pages = {084029},
   title = {$2.6\text{ }\text{ }{M}_{\ensuremath{\bigodot}}$ compact object and neutron stars in Eddington-inspired Born--Infeld theory of gravity},
   volume = {104},
   year = {2021}
}

@book{Priest2014,
   author = {Eric Priest},
   doi = {10.1017/cbo9781139020732},
   isbn = {9780521854719},
   journal = {Magnetohydrodynamics of the Sun},
   month = {4},
   publisher = {Cambridge University Press},
   title = {Magnetohydrodynamics of the Sun},
   url = {https://www.cambridge.org/core/product/identifier/9781139020732/type/book},
   year = {2014}
}

@article{Roberts1986,
   author = {B. Roberts and W. R. Campbell},
   doi = {10.1038/323603a0},
   issn = {0028-0836},
   issue = {6089},
   journal = {Nature},
   month = {10},
   pages = {603-605},
   title = {Magnetic field corrections to solar oscillation frequencies},
   volume = {323},
   year = {1986}
}

@article{Rodrigues2008,
   author = {Davi C. Rodrigues},
   doi = {10.1103/PhysRevD.78.063013},
   issn = {1550-7998},
   issue = {6},
   journal = {Physical Review D},
   month = {9},
   pages = {063013},
   title = {Evolution of anisotropies in Eddington-Born-Infeld cosmology},
   volume = {78},
   year = {2008}
}

@article{Sham2014,
   author = {Y.-H. Sham and L.-M. Lin and P. T. Leung},
   doi = {10.1088/0004-637X/781/2/66},
   issn = {0004-637X},
   issue = {2},
   journal = {The Astrophysical Journal},
   month = {1},
   pages = {66},
   title = {TESTING UNIVERSAL RELATIONS OF NEUTRON STARS WITH A NONLINEAR MATTER-GRAVITY COUPLING THEORY},
   volume = {781},
   year = {2014}
}

@article{Sham2013,
   author = {Y.-H. Sham and P. T. Leung and L.-M. Lin},
   doi = {10.1103/PhysRevD.87.061503},
   issn = {1550-7998},
   issue = {6},
   journal = {Physical Review D},
   month = {3},
   pages = {061503},
   title = {Compact stars in Eddington-inspired Born-Infeld gravity: Anomalies associated with phase transitions},
   volume = {87},
   year = {2013}
}

@article{Shoda2018,
   author = {Munehito Shoda and Takaaki Yokoyama},
   doi = {10.3847/1538-4357/aaa54f},
   issn = {0004-637X},
   issue = {1},
   journal = {The Astrophysical Journal},
   month = {2},
   pages = {9},
   title = {High-frequency Spicule Oscillations Generated via Mode Conversion},
   volume = {854},
   url = {https://iopscience.iop.org/article/10.3847/1538-4357/aaa54f},
   year = {2018}
}

@book{Stix2002,
   author = {Michael Stix},
   city = {Berlin, Heidelberg},
   doi = {10.1007/978-3-642-56042-2},
   isbn = {978-3-642-62477-3},
   publisher = {Springer Berlin Heidelberg},
   title = {The Sun},
   url = {http://link.springer.com/10.1007/978-3-642-56042-2},
   year = {2002}
}

@inbook{Stokes1845,
   author = {G.G. Stokes},
   doi = {10.1190/1.9781560801931.ch3e},
   booktitle = {Trans. Cambridge Philos. Soc.},
   pages = {287-319},
   publisher = {Cambridge Philosophical Society},
   title = {On the theories of the internal friction of fluids in motion, and of the equilibrium and motion of elastic  solids.},
   volume = {8},
   year = {1845}
}

@article{Ulrich1970,
   author = {Roger K. Ulrich},
   doi = {10.1086/150731},
   issn = {0004-637X},
   journal = {The Astrophysical Journal},
   month = {12},
   pages = {993},
   title = {The Five-Minute Oscillations on the Solar Surface},
   volume = {162},
   url = {http://adsabs.harvard.edu/doi/10.1086/150731},
   year = {1970}
}

@article{Yang2023,
   author = {Qiaoyun Yang and Ling Tan and Hui Chen and Sanqiu Liu},
   doi = {10.1016/j.newast.2022.101947},
   issn = {13841076},
   journal = {New Astronomy},
   month = {2},
   pages = {101947},
   title = {Jeans instability analysis of viscoelastic astrofluids in Eddington-Inspired-Born–Infeld(EiBI) gravity},
   volume = {99},
   url = {https://linkinghub.elsevier.com/retrieve/pii/S1384107622001294},
   year = {2023}
}

@article{ThisWork,
	author  = {Souvik Das and Pralay Kumar Karmakar},
	title   = {Stability and wave dynamics in polytropic {EiBI} gravitating solar plasmas},
	journal = {Physical Review E},
	year    = {2025}
}

\end{document}